\documentclass[a4paper,twocolumn,11pt,accepted=2024-02-06]{quantumarticle}
\pdfoutput=1
\usepackage[utf8]{inputenc}
\usepackage[english]{babel}
\usepackage[T1]{fontenc}
\usepackage{hyperref}
\usepackage{graphicx}
\usepackage[dvipsnames]{xcolor}
\PassOptionsToPackage{compress}{natbib}
\usepackage[numbers]{natbib}
\usepackage{float}
\usepackage{bbm}
\usepackage{amsmath}
\usepackage{amsfonts}
\usepackage{physics}
\usepackage{mathtools}
\usepackage{tikz}
\usepackage{appendix}
\usetikzlibrary{arrows.meta}
\usepackage{dsfont}
\usepackage{caption}
\usepackage{subcaption}
\usepackage{placeins}
\usetikzlibrary{arrows,shapes}
\definecolor{mypink}{RGB}{228, 83, 139}
\definecolor{mymagneta}{RGB}{190, 74, 193}
\definecolor{mypurple}{RGB}{138, 63, 252}
\definecolor{mygreyblue}{RGB}{132, 167, 214}
\definecolor{myblue}{RGB}{54, 95, 148}
\definecolor{mynavyblue}{RGB}{0, 32, 96}
\tikzset{font={\fontsize{12pt}{12}\selectfont}}
\usepackage{mathtools}
\tikzstyle{vertex}=[circle, fill=mypink, minimum size=30pt, inner sep=0pt, text=white]
\tikzstyle{vertex_mt}=[circle, fill=mypink, minimum size=15pt, inner sep=0pt, text=white]
\tikzstyle{vertexb}=[circle, fill=myblue, minimum size=30pt, inner sep=0pt, text=white]
\tikzstyle{vertexblank}=[circle, fill=white, minimum size=1pt, inner sep=0pt, text=white]
\tikzstyle{vertex_small}=[circle, fill=mypink, minimum size=10pt, inner sep=0pt, text=white]
\tikzstyle{edge} = [draw,ultra thick,-]
\tikzstyle{dedge} = [draw,ultra thick, dashed]
\begin{document}
\title{Rapid quantum approaches for combinatorial optimisation inspired by optimal state-transfer}

\author{Robert J.\ Banks}
\email{robert.banks.20@ucl.ac.uk}
\affiliation{London Centre for Nanotechnology, UCL, London WC1H 0AH, UK}
\author{Dan E.\ Browne}
\affiliation{Department of Physics and Astronomy, UCL, London WC1E 6BT, UK}
\author{P.\ A.\ Warburton}
\affiliation{London Centre for Nanotechnology, UCL, London WC1H 0AH, UK}
\affiliation{Department of Electronic \& Electrical Engineering, UCL, London WC1E 7JE, UK}

\begin{abstract}
We propose a new design heuristic to tackle combinatorial optimisation problems, inspired by Hamiltonians for optimal state-transfer. The result is a rapid approximate optimisation algorithm. We provide numerical evidence of the success of this new design heuristic. We find this approach results in a better approximation ratio than the Quantum Approximate Optimisation Algorithm at lowest depth for the majority of problem instances considered, while utilising comparable resources. This opens the door to investigating new approaches for tackling combinatorial optimisation problems, distinct from adiabatic-influenced approaches. 
\end{abstract}
\maketitle

\section{Introduction}
In a combinatorial optimisation problem, such as MAX-CUT or the TRAVELLING-SALESPERSON \cite{Pap81}, the aim is to evolve from some initial state to a final state that encodes the solution to the optimisation problem. One approach might be to evolve adiabatically, encoding each of the initial and final states as the ground state of some Hamiltonian and interpolating sufficiently slowly between them. In practice this approach is limited by the minimum spectral gap of the interpolating Hamiltonian \cite{Ami09, Rei04}. This approach is known as adiabatic quantum optimisation (AQO) \cite{Apo89,Far00, Kad98,Fin94,Alb18}.

In the absence of mature hardware, AQO has relied on the adiabatic principle as a guiding design tenet. In turn AQO has lead to Quantum Annealing (QA). Similar to AQO, QA attempts to interpolate continuously between the initial and final Hamiltonians. QA denotes a broader approach than AQO to find (or approximate) the solution of the optimisation problem. This could include: thermal effects \cite{Dic13}; intentionally operating diabatically \cite{Cro21,Fry21,Som12}; or adding new terms to the Hamiltonian \cite{Far11, Zen16, Far02, Fein22, Cro14, Cho21}. QA has gone on to help inspire a number of other approaches, such as: the gate-based Quantum Approximate Optimisation Algorithm (QAOA) \cite{Far14};  continuous-time quantum walks (QW) for optimisation \cite{Cal19,Ken20}; and combinations of different approaches \cite{Cal21, Mor19}.  In short, the adiabatic principle has been a successful design-principle for developing new heuristic quantum algorithms for optimisation. 

An alternative approach for evolving from an initial state to the final state might be via Hamiltonians for optimal state transfer \cite{Bro06}. These are Hamiltonians that transfer the system from the initial state to the final state in the shortest possible time. In this paper we use these Hamiltonians as our underlying design principle. Such Hamiltonians typically require knowledge of the final state (i.e., the solution to the optimisation problem). In the absence of this information we therefore focus on how the behaviour of these Hamiltonians might be approximated to find approximate solutions to optimisation problems. The result is a rapid continuous-time approach, with a single variational parameter. 

The framework of this paper is as follows: Sections \ref{sec:QAframe}, \ref{sec:hamdes}, and \ref{sec:prob} introduce the requisite background material and introduce the Hamiltonians considered in the rest of the paper (Sec.\ \ref{sec:hamdes}). Sections \ref{sec:H1}, \ref{sec:QZ}, and \ref{sec:lpa} provide analytical and numerical evidence for the performance of these Hamiltonians. Throughout the paper we adopt the convention $\hbar=1$. The corresponding Pauli matrices are denoted by $X,Y$ and $Z$. The identity is denoted by $I$. The commutator is denoted by $[\cdot,\cdot]$. The simulations make use of the Python packages QuTiP \cite{Joh12,Joh13} and Qiskit \cite{Qiskit}. Details about the numerical studies, including discussion about the presentation of numerical results can be found in Appendix \ref{app:num}.

\section{The QA-framework}
\label{sec:QAframe}
Before elaborating on our new approach, in this section we provide a very brief overview of the adiabatic influenced algorithms mentioned in the introduction.  

In QA, typically the optimisation problem is encoded as finding the ground-state, $\ket{\psi_f}$, of an Ising Hamiltonian, $H_f$. The system is initialised in the ground-state, $\ket{\psi_i}$, of an easy to prepare Hamiltonian $H_i$. Usually, $H_i$ is taken to be the transverse-field Hamiltonian. The optimisation problem is solved by evolving the system to the ground-state of $H_f$. We refer to this set-up as the QA-framework. We utilise this encoding of the optimisation problem for our new approach. 

For the adiabatic influenced approaches (AQO, QA, QAOA and QW) the system is subjected to the following Hamiltonian: 
\begin{equation}
\label{eq:QAham}
    H(t)=A(t) H_i + B(t) H_f,
\end{equation}
where $t\in[0,T]$ denotes time. The design philosophy behind how the schedules $A(t)$ and $B(t)$ are chosen, is what distinguishes these near-term intermediate-scale quantum (NISQ) \cite{Pre18} approaches. For AQO, the schedules are chosen to interpolate adiabatically between the two ground states \cite{Apo89,Far00, Kad98,Fin94}. In QA, this restriction is loosened to typically continuous, monotonically decreasing for $A(t)$ and increasing for $B(t)$, schedules \cite{Cro21,Hau20}. In QW $A(t)$ and $B(t)$ are taken to be constants over the whole evolution \cite{Cal19,Ken20}. 

QAOA \cite{Far14} is a gate-based design-philosophy for determining the schedules. In QAOA either $A(t)=1$ and $B(t)=0$, or $A(t)=0$ and $B(t)=1$. The switching parameter, $p$, controls the number of times the schedules alternate between the two parameter settings. The duration between switching is either determined prior to the evolution by a classical computer or by using a classical variational outer-loop attempting to minimise $\langle H_f \rangle$ by varying $2p$ free-parameters. QAOA again relies on the adiabatic principle to provide a guarantee of finding the ground-state in the limit of infinite $p$. However, far from this limit, the variational method allows QAOA to exploit non-adiabatic evolution \cite{Zho20}. QAOA has been further generalised, retaining its switching framework to the `Quantum Alternating Operator Ansatz' \cite{Had19}. It has also become a popular choice for benchmarking the performance of quantum hardware \cite{Har21, Gra22, Ott17}.

Recently Brady et al.\ applied optimal control-theory to help design and investigate schedules \cite{Bra21_1,Bra21,fei22}. They demonstrated for $0\leq A(t) \leq1$ and $B(t)=1-A(t)$, that the optimal schedule starts and finishes with a `bang'  (i.e., one of the controls takes its maximum value)\cite{Bra21_1}. This claim was further investigated in \cite{Venuti_2021} and applied to open quantum systems.

Clearly there are many design-philosophies for tacking problems within the QA-framework. Although, some have been inspired by the adiabatic theorem, practical implementation might have little resemblance to this original idea. The next section provides a brief introduction to Hamiltonians for optimal state-transfer as well as the motivation for the choice of Hamiltonians used in this paper.

\section{Hamiltonian design}
\label{sec:hamdes}

The aim of optimal state-transfer is to find a Hamiltonian that transfers the system from an initial state (i.e., $\ket{\psi_i}$) to a known final state (i.e., $\ket{\psi_f}$) in the shortest possible time. We shall refer to this Hamiltonian as the optimal Hamiltonian (although it is by no means unique).

One notable approach to finding the optimal Hamiltonian comes from Nielsen et al. \cite{Nie05,Nie06,Dow07} who investigated the use of differential geometry, at the level of unitaries, to find geodesics connecting the identity to the desired unitary. The length of the geodesic was linked to the computational complexity of the problem \cite{Nie06}. A second approach comes from Carnali et al.. Inspired by the brachistochrone problem, they developed a variational approach at both the level of state-vectors \cite{Car06} and unitaries \cite{Car07}. The quantum brachistochrone problem has generated a considerable amount of literature and interest \cite{Rez09,Wan15,Wak20,Wan21, San21, Yan22}. 

Both approaches allow for constraints to be imposed on the Hamiltonian. Both concluded that if the only constraint is on the total energy of the Hamiltonian, the optimal Hamiltonian is constant in time. In the rest of this section we outline the geometric argument put forward by Brody et al.\ \cite{Bro06} to find the optimal Hamiltonian in this case. The reader is invited to refer to the original work for the explicit details. 

The optimal Hamiltonian will generate evolution in a straight line in the (complex-projective) space in which the states live.  Intuitively, a line in this space between $\ket{\psi_i}$ and $\ket{\psi_f}$ consists of superpositions of the two states. Hence, the line in the complex-projective space can be represented on the Bloch sphere.

If, without loss of generality, $\ket{\psi_i}$ and $\ket{\psi_f}$ are placed in the traditional $z-x$ plane of the Bloch sphere, it is clear that the optimal Hamiltonian generates rotations in this plane. The optimal Hamiltonian is then (reminiscent of the cross-product):
\begin{equation}
    H_{opt}=-i \left(\ket{\psi_i}\bra{\psi_f}-\ket{\psi_f}\bra{\psi_i}\right).
    \label{eq:optHam}
\end{equation}
This can be scaled to meet the condition on the energy of the Hamiltonian. It then remains to calculate the time required to transfer between the two states. This will depend on how far apart the states are and how fast the evolution is. This is encapsulated in the Anandan-Aharonov relationship \cite{Ana90}:
\begin{equation}
    \frac{ds}{dt}=2\delta E(t).
\end{equation}

The left-hand-side denotes the speed of the state, $\ket{\psi(t)}$, where $ds$ is the infinitesimal distance between $\ket{\psi\left(t+dt\right)}$ and $\ket{\psi\left(t\right)}$ \footnote{The distance is measured by the Fubini-study metric, $ds^2=4\left(1-\abs{\bra{\psi\left(t\right)}\ket{\psi\left(t+dt\right)}}^2\right)$, on the complex-projective space.}.  Under evolution by the Schr\"odinger equation, with Hamiltonian $H$, the instantaneous speed of the evolution is given by the uncertainty in the energy, $\delta E(t)^2=\bra{\psi(t)}H(t)^2\ket{\psi(t)}-\bra{\psi(t)}H(t)\ket{\psi(t)}^2$.

Since the optimal Hamiltonian is constant in time, $\delta E$ can be evaluated using the initial state.  Therefore, the time of evolution is:
\begin{equation}
T=\frac{\arccos{\abs{\bra{\psi_f}\ket{\psi_i}}}}{ \sqrt{1-\abs{\bra{\psi_f}\ket{\psi_i}}^2}}.
\end{equation}

 In summary, $e^{-i H_{opt} T}\ket{\psi_i}$ generates the state $\ket{\psi_f}$. The goal of the rest of this paper is to harness some of the physics behind this expression for computation.  To this end we primarily focus on Hamiltonians which are constant in time. Appendix \ref{sec:opHamQa} contains more details on what optimal Hamiltonians might look like within the QA-framework. 
 
 In the style of a variational quantum eigensolver (VQE) \cite{Per14,Fed21}, we allow $T$ to be a variational parameter that needs to be optimised in our new approach. In this paper we select the $T$ that minimises the final value of $\langle H_f \rangle$. There is scope to extend this to different metrics \cite{Li20,Bar20}. As $T$ is a variational parameter the Hamiltonian is only important up to some constant factor. Rewriting Eq.\ \ref{eq:optHam} up to some constant gives:
\begin{equation}
    \label{eq:optHamCom}
    H_{opt}\propto\frac{1}{2i}\left[\ket{\psi_i}\bra{\psi_i},\ket{\psi_f}\bra{\psi_f}\right],
\end{equation}
assuming $\ket{\psi_i}$ and $\ket{\psi_f}$ have a non-zero overlap (this is a given in the standard QA-framework). This equation (Eq.\ \ref{eq:optHamCom}) provides the starting point for all the Hamiltonians considered in this paper. 

The optimal Hamiltonian (i.e., Eq.\ \ref{eq:optHamCom}) requires knowledge of the final state. In practice, when attempting to solve an optimisation problem, one doesn't have direct access to $\ket{\psi_f}$.  Instead one has easy access to $H_i$ and $H_f$. Therefore, we make the pragmatic substitutions $\ket{\psi_i}\bra{\psi_i} \rightarrow H_i$ and $\ket{\psi_f}\bra{\psi_f} \rightarrow H_f$ into Eq.\ \ref{eq:optHamCom}

\begin{align}
H_{opt}\propto\frac{1}{2i}&\left[\ket{\psi_i}\bra{\psi_i},\ket{\psi_f}\bra{\psi_f}\right] \nonumber \\
    & \hspace{0.6cm}\downarrow\hspace{1.3cm}\downarrow \nonumber\\
    H_1=\frac{1}{2i}&\left[\hspace{0.4cm}H_i\hspace{0.4cm},\hspace{0.6cm}H_f\hspace{0.4cm}\right].
\end{align}
This Hamiltonian is the most amenable to NISQ implementation of all the Hamiltonians considered in this paper, therefore the bulk of the paper is devoted to demonstrating its performance. The results can be seen in Sec.\ \ref{sec:H1}.

The substitutions $\ket{\psi_i}\bra{\psi_i}\rightarrow H_i$ and $\ket{\psi_f}\bra{\psi_f}\rightarrow H_f$ introduce errors, such that the evolution under $H_1$ no longer closely follows the evolution under $H_{opt}$. In Sec.\ \ref{sec:QZ} we try to correct for this error by adding a new term to the Hamiltonian
\begin{equation}H_{1,improved}=H_1+H_{QZ}.
\end{equation}
The proposed form of $H_{QZ}$ is motivated by the quantum Zermello problem \cite{Bro15,Brody_2015,Rus14,Rus15}.

Finally, in Sec.\ \ref{sec:lpa}  we exploit our knowledge of the initial state and propose the substitution $\ket{\psi_f}\bra{\psi_f}\rightarrow f(H_f)$, where $f(\cdot)$ is some real function:

\begin{align}
    H_{opt}\propto\frac{1}{2i}&\left[\ket{\psi_i}\bra{\psi_i},\ket{\psi_f}\bra{\psi_f}\right] \nonumber\\
    & \hspace{2.1cm}\downarrow \nonumber\\
    H_{\psi_i}=\frac{1}{2i}&\left[\ket{\psi_i}\bra{\psi_i},\hspace{0.2cm}f\left(H_f\right)\hspace{0.1cm}\right].
\end{align}

\section{Problems considered}
\label{sec:prob}
To assess the performance of the Hamiltonians proposed in Sec.\ \ref{sec:hamdes}, we apply them to the combinatorial optimisation problem known as MAX-CUT. MAX-CUT can be encoded as finding the ground-state of the following Ising Hamiltonian:
\begin{equation}
    H_f=\sum_{(i,j) \in E} Z_i Z_j,
\end{equation}
where $E$ denotes the set of edges in a graph $G=(V,E)$. Through-out this paper we examine three different types of graph: 2-regular, 3-regular and random graphs. For the random graphs each edge is selected with probability 2/3. More details about these graphs in the context of the quantum algorithms discussed in Sec.\ \ref{sec:QAframe} can be found in Appendix \ref{app:prob_plus}.

To measure the performance of the different approaches we look at two  metrics: the ground-state probability and approximation-ratio. We define the approximation ratio to be $\langle H_f \rangle/E_{min}$ where $E_{min}$ is the energy of the ground-state solution of $H_f$ and the expectation is with respect to the final state. Further justification of these choice of metrics can be found in Appendix \ref{app:met}.

Of crucial interest for NISQ implementations is the duration of each run. For continuous-time approaches this is simply the time of each run. We would like to compare this approach to QAOA, which has the following ansatz:
\begin{equation}
    \label{eq:QAOA}
    \ket{\psi_{QAOA}}=\prod_{k=1}^p \left(e^{-i \beta_k H_i}e^{-i \gamma_k H_f}\right) \ket{+}
\end{equation}
where the $\beta_k$s and $\gamma_k$s are the variational parameters. We take the time of a QAOA run to be the sum of the variational parameters. In Sec.\ \ref{sec:QAOA} we compare the optimal time of QAOA $p=1$ and $H_1$, the optimal time being the time that maximises the approximation ratio. To make a fair comparison between the two approaches with different problem sizes, we fix the energy of the Hamiltonians in Sec.\ \ref{sec:QAOA} to be: 
\begin{equation}
     \label{eq:norm}
    \frac{1}{2^n}\Tr{H_*^2}=n,
\end{equation}
for $*=i,f,1$, where $n$ is the number of qubits.

As mentioned this paper will focus on MAX-CUT but to illustrate the wide applicability of these model we also apply it to a Sherrington-Kirpatrick inspired model, the details of which can be found in Appendix \ref{app:skm}.

\section{Taking the commutator between the initial and final Hamiltonian}
\label{sec:H1}
In Sec.\ \ref{sec:hamdes}. we motivated the Hamiltonian
\begin{equation}
    \label{eq:sqham}
    H_1=\frac{1}{2i}\left[H_i,H_f\right]
\end{equation}
by substituting out the projectors in Eq.\ \ref{eq:optHamCom} for easily accessible Hamiltonians. In this section we explore the effectiveness of these substitutions. We begin by demonstrating that Eq.\ref{eq:sqham} generates the optimal rotation for a single qubit (Sec.\ \ref{sec:singleQ}). In Sec.\ \ref{sec:SERand} we show that $H_1$ has the potential to outperform random guessing within the QA-framework. The rest of the section analyses the performance of $H_1$ on the problems outlined in Sec.\ \ref{sec:prob}.

\subsection{The optimal approach for a single qubit}
\label{sec:singleQ}
\begin{figure}
\centering
\begin{tikzpicture}[line cap=round, line join=round, >=Triangle, scale=0.8]
  \clip(-2.19,-2.49) rectangle (2.66,2.8);

  \draw [shift={(0,0)}, lightgray, fill, fill opacity=0.1] (0,0) -- (-135.7:0.4) arc (-135.7:-41:0.4) -- cycle;
  \draw(0,0) circle (2cm);
  \draw [rotate around={0.:(0.,0.)},dash pattern=on 3pt off 3pt] (0,0) ellipse (2cm and 0.9cm);

  \draw [->] (0,0) -- (0,2);
  \draw [->] (0,0) -- (-0.81,-0.79);
  \draw [->] (0,0) -- (0.90,-0.8);
  \draw [->, style=dashed] (0,0) -- (0.81,0.79);
  \draw [->, style=dashed] (0,0) -- (-0.9,0.8);

  \draw[ ->] (0.5,1.3);
  \draw (0,1.2) ellipse (0.5cm and 0.225cm);
  \draw (-0.08,-0.3) node[anchor=north west] {$\theta$};
 
  \draw (-1.05,-0.75) node[anchor=north west] {$\mathbf {\hat{m}}$};
  \draw (0.8,-0.7) node[anchor=north west] {$\mathbf {\hat{n}}$};
  \draw (-0.5,2.8) node[anchor=north west] {$\mathbf {\hat{k}}=\mathbf {\hat{m}}\cross\mathbf {\hat{n}}$};
\end{tikzpicture}
\caption{The geometric intuition behind finding the Hamiltonian for optimally transferring between the ground-states of $H_i$ and $H_f$ on the Bloch sphere. The vectors $\pm \hat{m}$ ($\pm \hat{n}$) are the eigenvectors of $H_i$ ($H_f$). The aim is to generate a rotation of $\theta$ around $\hat{k}$ to map $\pm \hat{m}$ to $\pm \hat{n}$. The handedness of the cross-product takes into account the direction.}
\label{fig:bloch}
\end{figure}
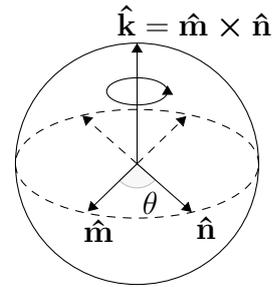

Here we outline a simple geometric argument which shows that $H_1$ generates the optimal rotation for a single qubit (hence the name $H_1$). The eigenstates of $H_i$ and $H_f$ can be represented as points on the surface of the Bloch sphere, see Fig.\ \ref{fig:bloch}. Since these points lie in a plane, the aim is to write down a Hamiltonian that generates rotation in this plane. By writing the Hamiltonians in the Pauli basis, it is clear the Hamiltonian that generates the correct rotation is:
\begin{equation*}
    H_1=\frac{1}{2i}\left[H_i,H_f\right].
\end{equation*}
For the full details see Appendix \ref{app:sq}. Again, by using the Anandan-Aharonov relationship:
\begin{equation}
    \frac{d\theta}{dt}=2\delta E,
\end{equation}
where $\delta E$ is the uncertainty in the energy and $\theta$ is the distance between the desired states (Fig.\ \ref{fig:bloch}), we can calculate the time required to transfer between the two ground-states. 

Having established $H_1$ as the optimal Hamiltonian for a single qubit, the next section investigates its performance on larger problems. 

\subsection{Application to larger problems}
\subsubsection{Outperforming random guessing for short times}
\label{sec:SERand}

In this section we demonstrate that $H_1$ can always do better than random guessing within the QA-framework. Starting with the time-dependent Schr\"odinger equation:
\begin{equation*}
    \ket{\dot{\psi}(t)}=-i H_1 \ket{\psi(t)},
\end{equation*}
we expand $\ket{\psi(t)}$ in terms of the eigenbasis of $H_f$, so $\ket{\psi(t)}=\sum_k c_{k}(t) \ket{k}$ where $\ket{k}$ are the eigenvectors of $H_f$ with associated eigenvalue $E_k$. The eigenvalues are ordered such that $E_0\leq E_1 \leq E_2 \dots$. Substituting this into the Schr\"odinger equation gives:
\begin{align*}
    \sum_k \dot{c}_k(t) \ket{k}&=-\frac{i}{2i}\sum_k c_k(t) \left(H_i H_f -H_f H_i \right)\ket{k}\\
    &=-\frac{1}{2}\sum_k c_k(t) E_k \left(H_i-H_f H_i\right)\ket{k}
\end{align*}

Acting with $\bra{j}$ on each side, to find $\dot{c}_j(t)$, gives:
\begin{equation}
    \dot{c}_j(t)=-\frac{1}{2}\sum_k c_k(t) \underbrace{\left(E_k-E_j\right)}_{\text{"Velocity"}}\overbrace{\bra{j}H_i\ket{k}}^{\substack{\text{How the basis}\\ \text{states of }H_f \\ \text{are connected}}}.
    \label{eq:TDSEcoef}
\end{equation}

In the standard QA-framework $H_i=-\sum_k^n X_k$, $c_k(0)=1/\sqrt{2^n}$, for all $k$, and the basis states (e.g. $\ket{k}$), correspond to computational basis states. Accordingly, $H_1$ connects computational basis states which are a Hamming-distance of one away.

The difference in energy of the computational basis states intuitively provides something akin to a velocity, with greater rates of change between states which are further apart in energy.

Focusing on the derivative of the ground-state amplitude at $t=0$ we have:
\begin{equation}
    \dot{c}_0(0)=-\frac{1}{2}\sum_k \underbrace{c_k(0)}_{>0}\underbrace{\left(E_k-E_0\right)}_{\geq 0}\underbrace{\bra{0}-\sum_j X_j\ket{k}}_{=0 \text{ or }-1},
\end{equation}
so $\dot{c}_0(0)\geq0$, with equality if all states in a Hamming-distance of one have the same energy as $\ket{0}$. In this case the above logic can be repeated for these states. Hence, at $t=0$, the ground state amplitude is increasing - meaning $H_1$ can do better than random guessing by measuring on short times. This is evidence that $H_1$ is capturing something of the of the optimal Hamiltonian for short times. Indeed, for short times all the amplitudes flow from higher-energy states to lower-energy states. 

The above logic can be extended to the case where $H_i$ is any stoquastic Hamiltonian in the computational basis \cite{Bra10,Alb18} and $\ket{\psi_i}$ the corresponding ground-state.  That is to say, we require $H_i$ to have non-positive off-diagonal elements in the computational basis (i.e. stoquastic) and as a consequence we can can write the ground-state of $H_i$  with real non-negative amplitudes \cite{Alb18}. Consequently, for any stoquastic choice of $H_i$ the ground-state amplitude is increasing at $t=0$ and can do better than the initial value of $c_0$ at short times. 

We could also take a generalised version of $H_1$:
\begin{equation}
    H_{1,gen}=\frac{1}{2i}\left[f\left(H_i\right),g\left(H_f\right)\right],
\end{equation}
where $f$ and $g$ are real functions. Eq.\ \ref{eq:TDSEcoef} becomes:
\begin{equation}
    \dot{c}_j(t)=-\frac{1}{2}\sum_k c_k(t) \left(g\left(E_k\right)-g\left(E_j\right)\right)\bra{j}f\left(H_i\right)\ket{k}.
\end{equation}

The function acting on $H_i$ (i.e., $f(\cdot)$) controls how the computational-basis states are connected, while the function acting on $H_f$ (i.e., $g(\cdot)$) controls the velocity between computational basis states. If $f(\cdot)$ is the identity and $H_i$ stoquastic, then any monotonic function for $g(\cdot)$ (e.g., $H_f^3$, $H_f^5$, $\exp{H_f}$,...) will do better than $c_0(0)$ for short $t$. Taking $H_i$ to be the transverse-field Hamiltonian, that is better than random-guessing. 

The above analysis demonstrates that $H_1$ has potential for tackling generic problems within the QA-framework. The next sections apply $H_1$ to specific examples in an attempt to quantify the success of this approach. For the rest of this paper we take $H_i$ to be the transverse-field Hamiltonian.

\subsubsection{MAX-CUT on two-regular graphs}
\label{sec:RoD}

\begin{figure}
    \centering
\includegraphics[width=0.48\textwidth]{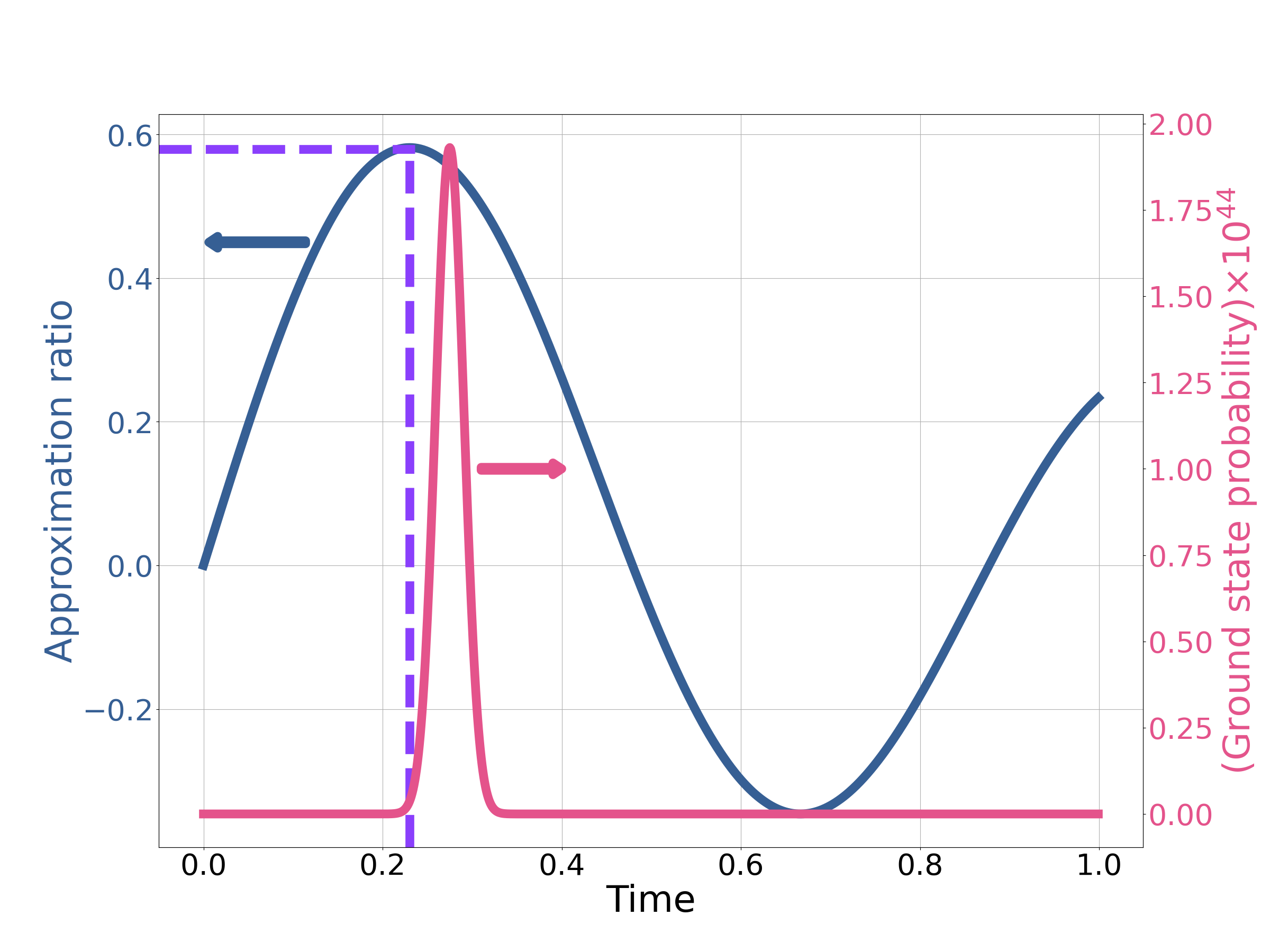}
    \caption{A time-domain plot of the ground-state probability (in pink) and approximation ratio (in blue) for $H_1$ applied to MAX-CUT on a 2-regular graph with 400 qubits. Random guessing corresponds to a ground-state probability of $2^{-399}\approx 10^{-120}$. The dashed purple line shows the location of the optimal time, corresponding to the maximum in approximation ratio.}
    \label{fig:RoDtd}
\end{figure}

Here we study the performance of $H_1$ on MAX-CUT with two-regular graphs. The explicit form of $H_1$ is then:
\begin{equation}
    H_1=\sum_{j=1}^n \left(Y_jZ_{j+1}+Z_jY_{j+1}\right).
\end{equation}
This can be solved analytically by mapping the problem onto free fermions via the Jordan-Wigner transformation. Details can be found in Appendix \ \ref{app:ferm}.

A time domain plot for the approximation ratio and ground-state probability is shown in Fig.\ \ref{fig:RoDtd} for 400 qubits. The peak in approximation ratio corresponding to the optimal time. As expected from the previous section (Sec.\ \ref{sec:SERand}) the approximation ratio is increasing at $t=0$.  There is a clear peak in ground-state probability at a time of $t\approx 0.275$. This peak remains present for larger problem sizes too. The peak also occurs at a later time than the peak in approximation ratio. Further insight into this phenomena may be found in Sec.\ \ref{sec:lpa}.

\begin{figure*}
    \centering
    \begin{subfigure}[t]{0.3\textwidth}
        \centering
         \includegraphics[width=\textwidth]{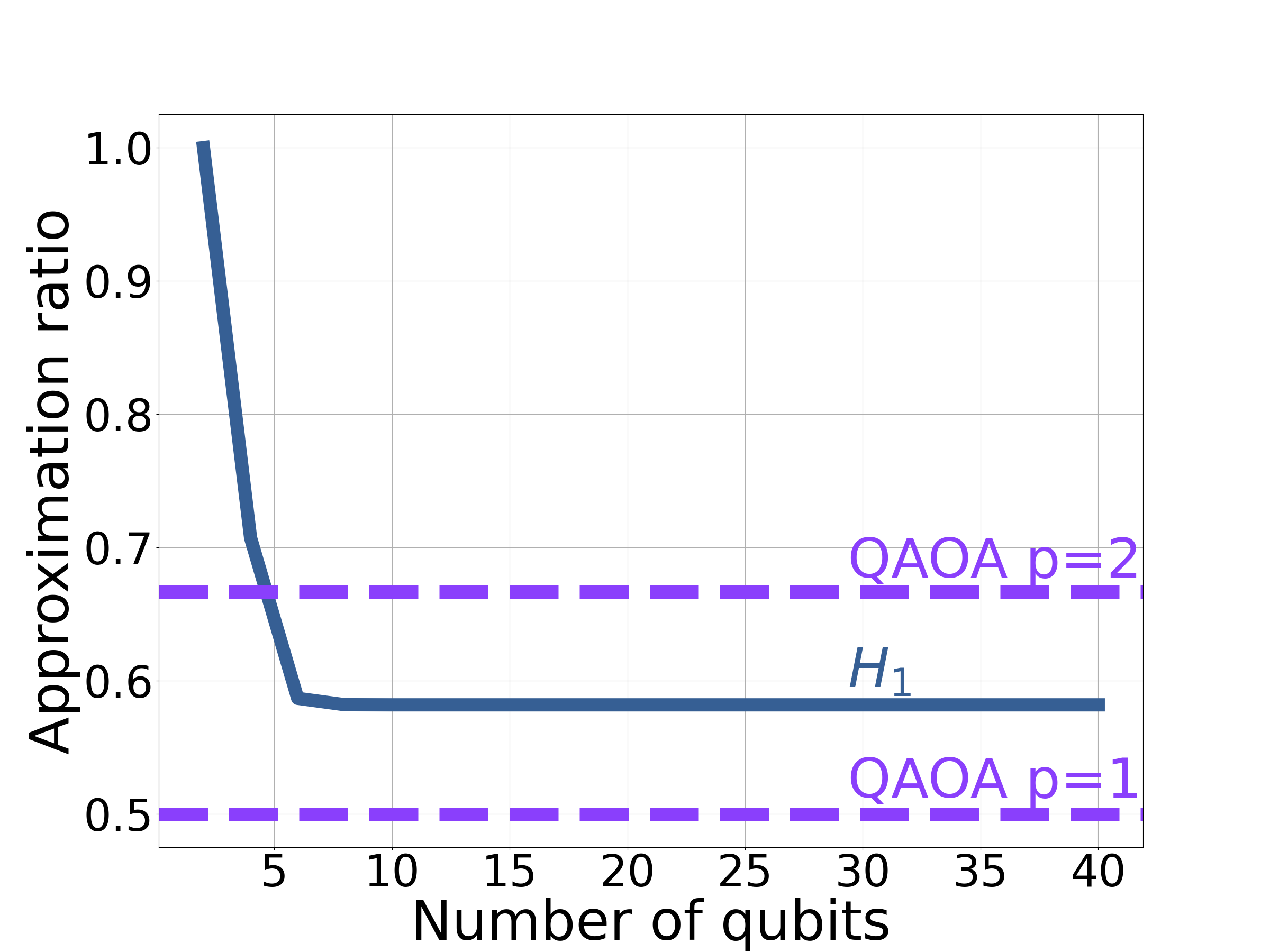}
         \caption{The approximation ratio for $H_1$ applied to MAX-CUT on two-regular graphs (solid line). The dashed lines show the performance of QAOA for this problem when $p < n/2$ \cite{Far14,mbe19}. The corresponding optimal times can be found in Fig. \ref{fig:RoDts}. The approximation ratio for $H_1$ freezes out at $0.5819$, with a corresponding time of $0.2301$.}
         \label{fig:RodeC}
    \end{subfigure}
    \hfill
    \begin{subfigure}[t]{0.3\textwidth}
    \centering
    \includegraphics[width=\textwidth]{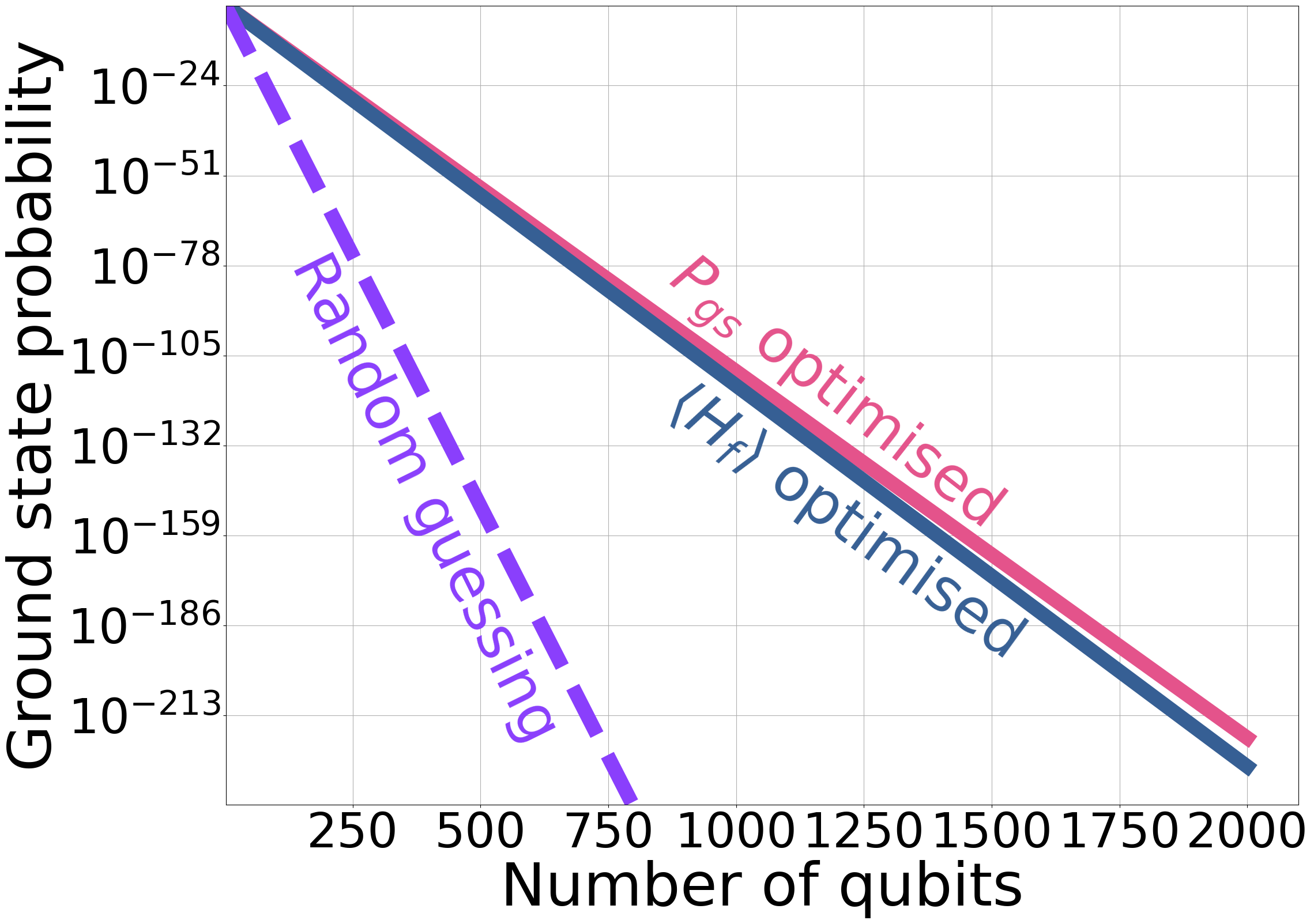}
    \caption{The ground-state probability for different problem sizes under the evolution of $H_1$. The blue line shows the ground-state probability for times that maximise the approximation ratio. The optimised ground-state probability is shown in pink. The dashed purple line shows the probability of randomly guessing the ground-state.}
    \label{fig:RoDegsp}
    \end{subfigure}
    \hfill
    \begin{subfigure}[t]{0.3\textwidth}
    \centering
    \includegraphics[width=\textwidth]{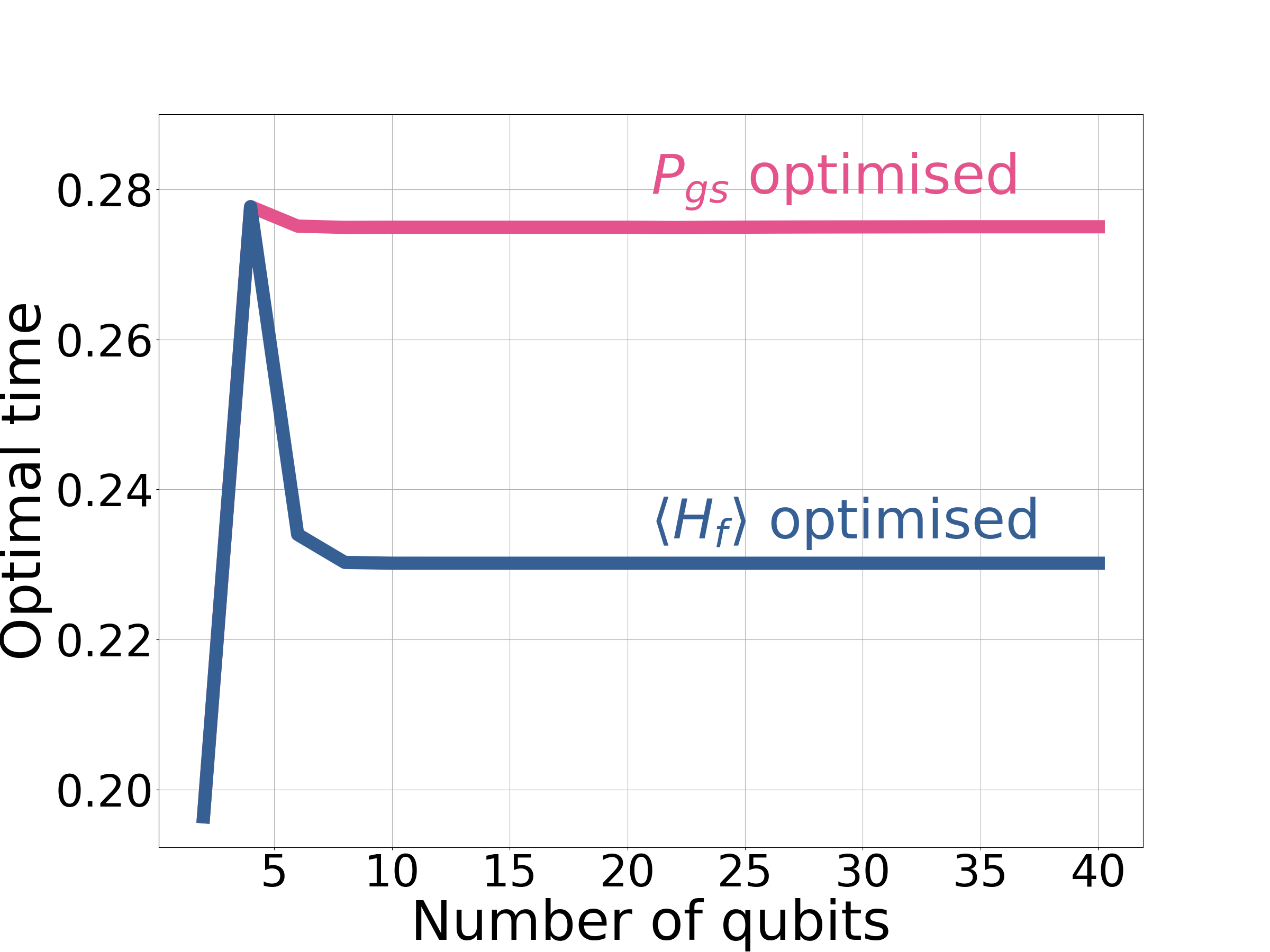}
    \caption{The optimal times for MAX-CUT on two-regular graphs. The blue (pink) line shows the time that optimises the approximation ratio (ground state probability).}
    \label{fig:RoDts}
    \end{subfigure}
    \caption{Performance of $H_1$ on MAX-CUT with 2-regular graphs. Only even numbers of qubits are plotted. }
\end{figure*}

The key result of this section is shown in Fig.\ \ref{fig:RodeC}, showing the optimal approximation ratio versus problem size for even numbers of qubits only. The corresponding plot for odd numbers of qubits can be found in Appendix \ref{app:ferm}. Notably the approximation ratio saturates for large problem sizes, achieving an approximation ratio of $0.5819$, compared to $0.5$ for QAOA $p=1$. This is despite QAOA $p=1$ having two variational parameters, compared to the single variational parameter for $H_1$. This behaviour is suggestive of $H_1$ optimising locally, since its approximation ratio is largely independent of problem size.

Despite MAX-CUT on two-regular graphs being an easy problem, the ground-state probability scales exponentially with problem size (Fig.\ \ref{fig:RoDegsp}). This is not necessarily a problem, as we present $H_1$ as an approximate approach only. Optimising the performance to give the best ground-state probability provides a small gain in performance but does not change the overall exponential scaling. The optimal times for both approximation ratio and ground-state probability can be found in Fig.\ \ref{fig:RoDts}. Both times freeze out at constant values for problem sizes greater than 10 qubits.

We have demonstrated with MAX-CUT on two-regular graphs, at large problem sizes, that $H_1$ can provide a better approximation ratio than QAOA $p=1$. We explore this comparison with QAOA $p=1$ further in Sec.\ \ref{sec:QAOA}.

\subsubsection{Performance on MAX-CUT problems with three-regular graphs}
\label{sec:3reg}

\begin{figure*}
    \centering
    \begin{subfigure}[t]{0.48\textwidth}
    \centering
    \includegraphics[width=\textwidth]{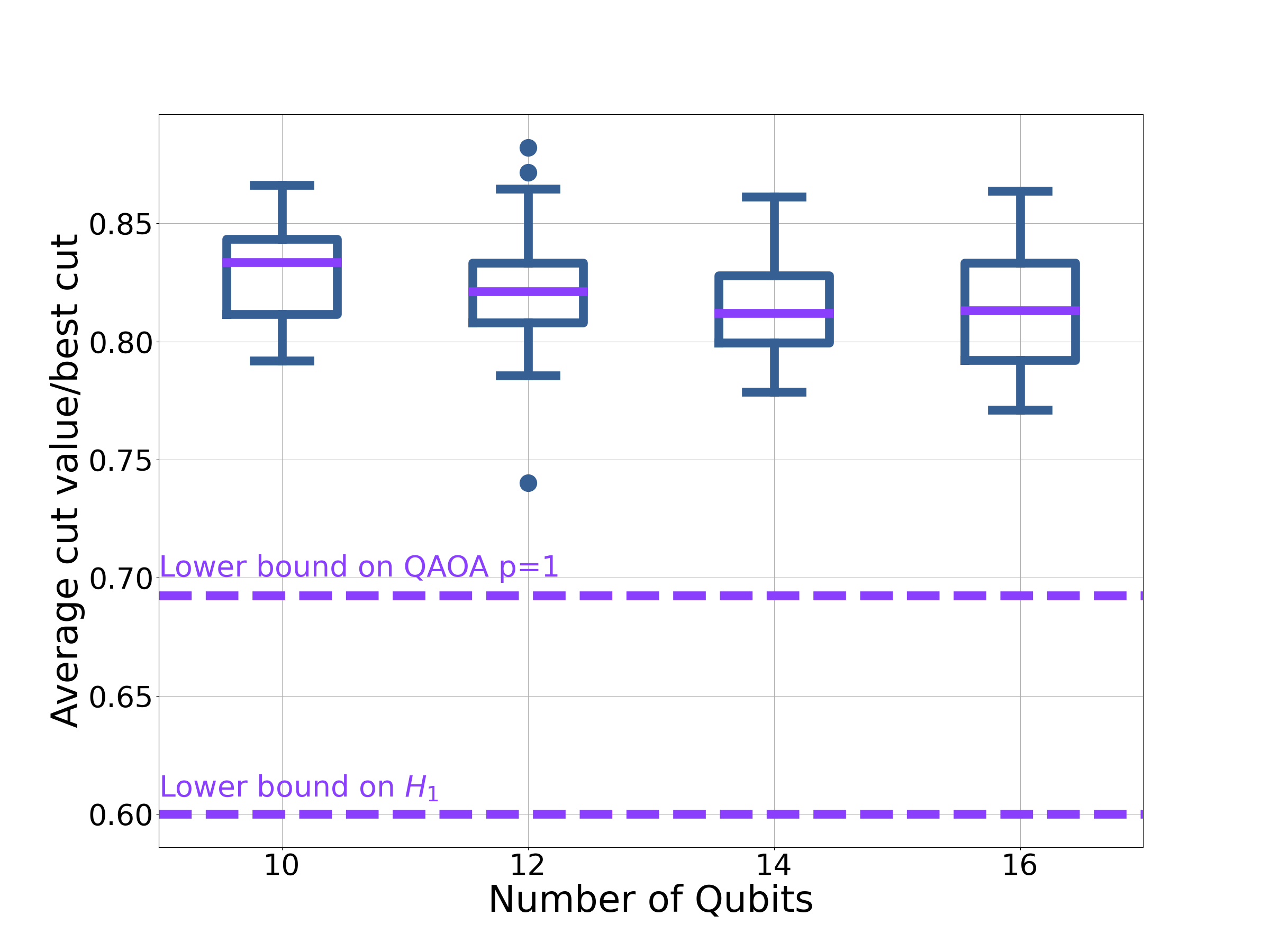}
    \caption{The y-axis shows the average cut-value from sampling $H_1$. The final time has been numerically optimised to give the best approximation ratio.
    }
    \label{fig:MC3Reg}
    \end{subfigure}
    \hfill
    \begin{subfigure}[t]{0.48\textwidth}
    \includegraphics[width=\textwidth]{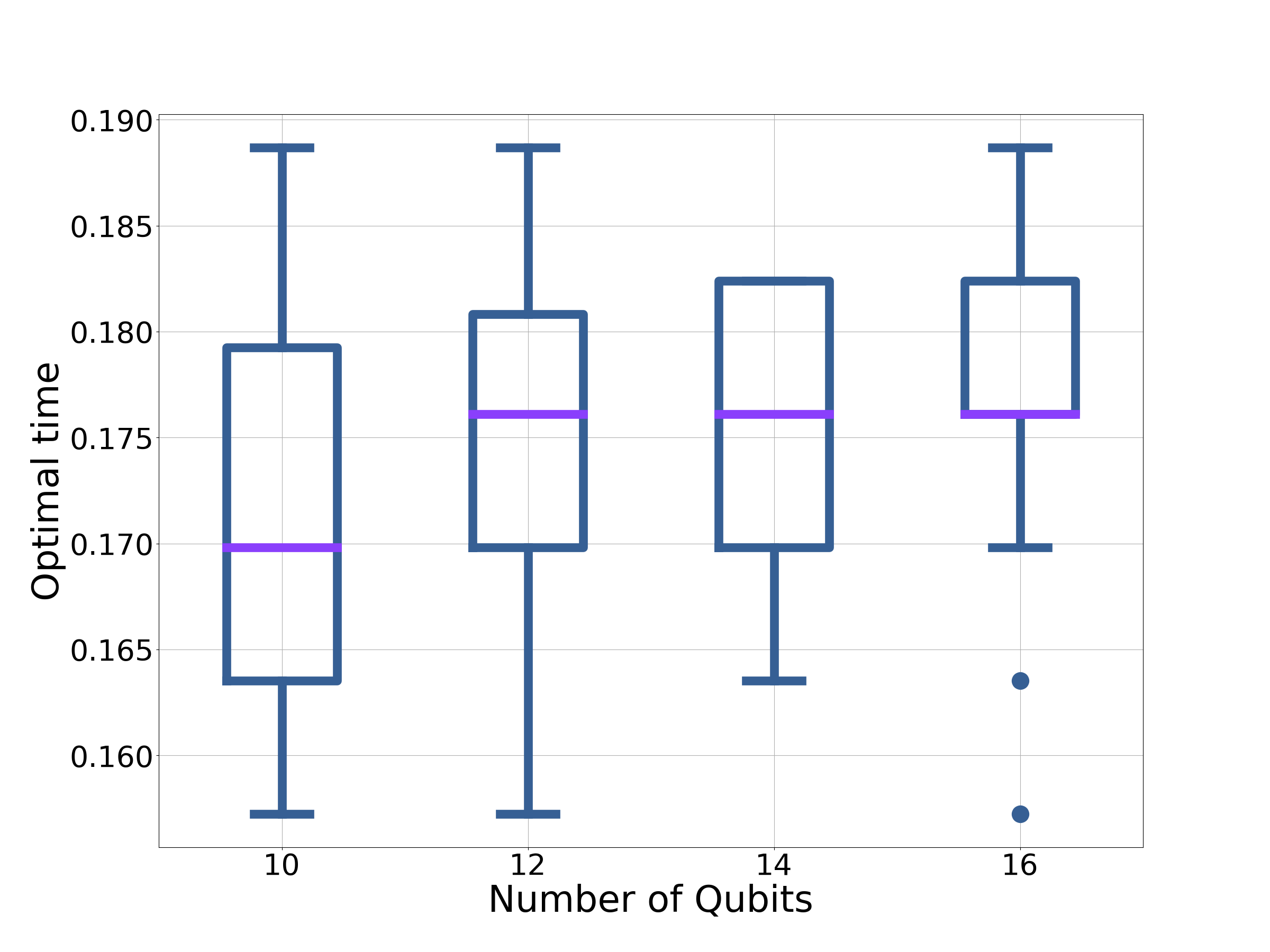}
    \caption{The optimal time for the three-regular MAX-CUT instances considered in Fig.\ \ref{fig:MC3Reg}. The optimal time was found by dividing the interval $[0,2 \pi] $ into 1000 time steps.}
    \label{fig:3RegOt}
    \end{subfigure}
    \caption{The performance of $H_1$ on randomly generated instances of three-regular graphs. For each problem size, 100 instances were generated. After accounting for graph isomorphisms the number of samples in order of ascending problem size were $[15,46,87,97]$. Disconnected graphs were allowed.}
\end{figure*}
\begin{figure*}
    \centering
    \begin{subfigure}[t]{0.3\textwidth}
        \centering
        \includegraphics[width=\textwidth]{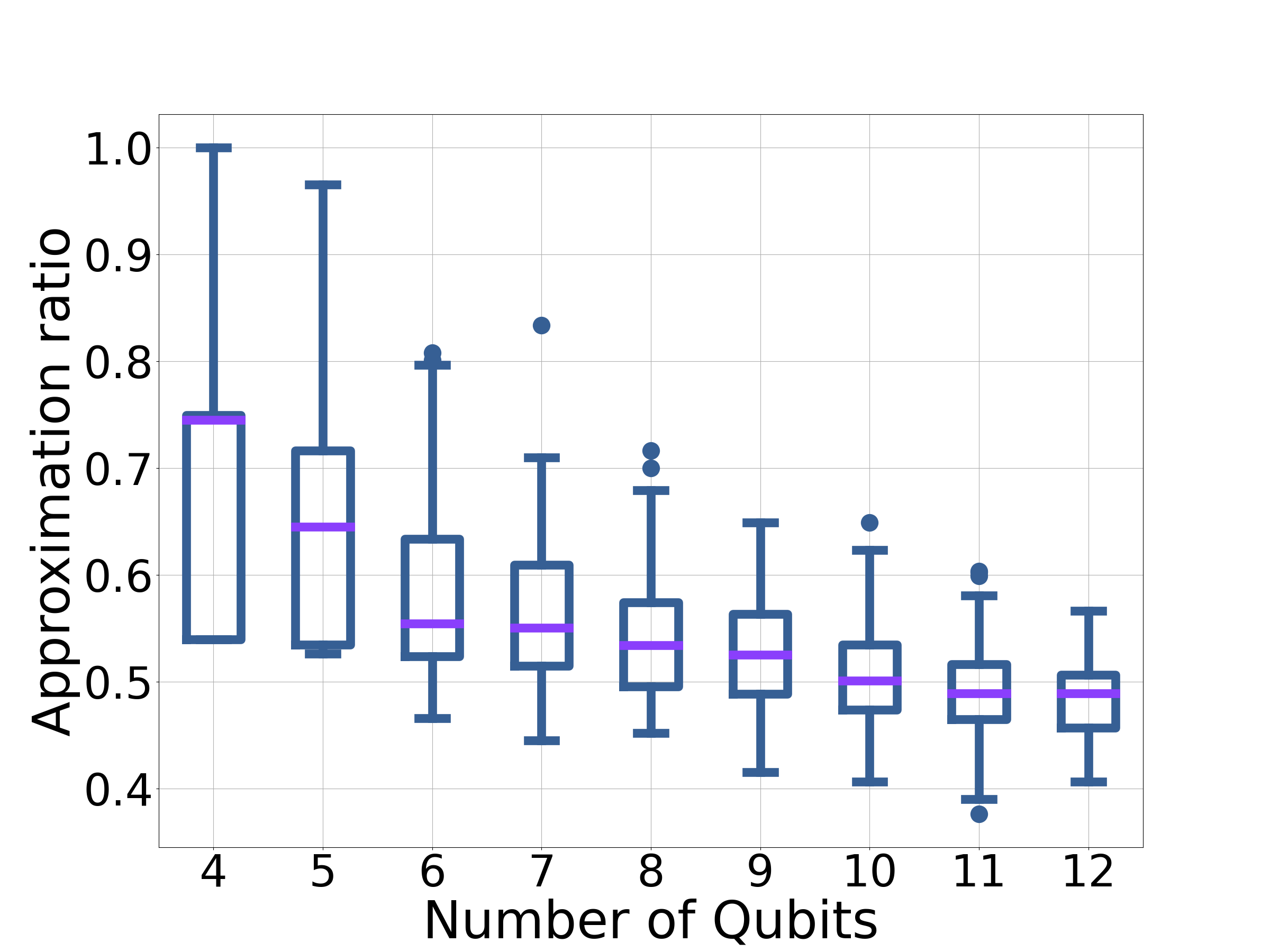}
        \caption{Approximation ratio for randomly generated MAX-CUT.}
        \label{fig:MCRandc}
    \end{subfigure}
    \hfill
    \begin{subfigure}[t]{0.3\textwidth}
        \centering
        \includegraphics[width=\textwidth]{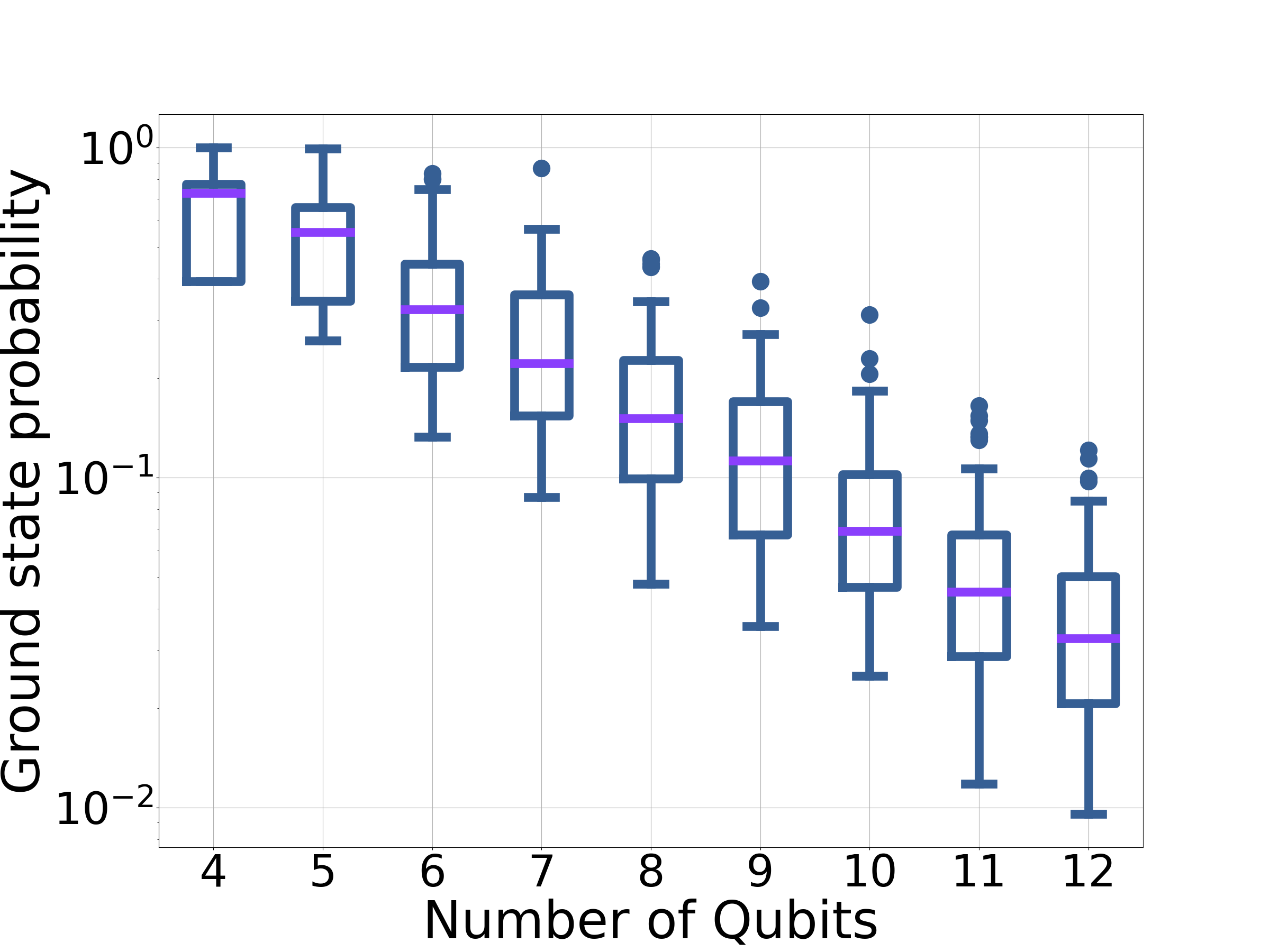}
        \caption{Ground-state probability for randomly generated MAX-CUT.}
        \label{fig:MCRandgsp}
    \end{subfigure}
    \hfill
    \begin{subfigure}[t]{0.3\textwidth}
        \centering
        \includegraphics[width=\textwidth]{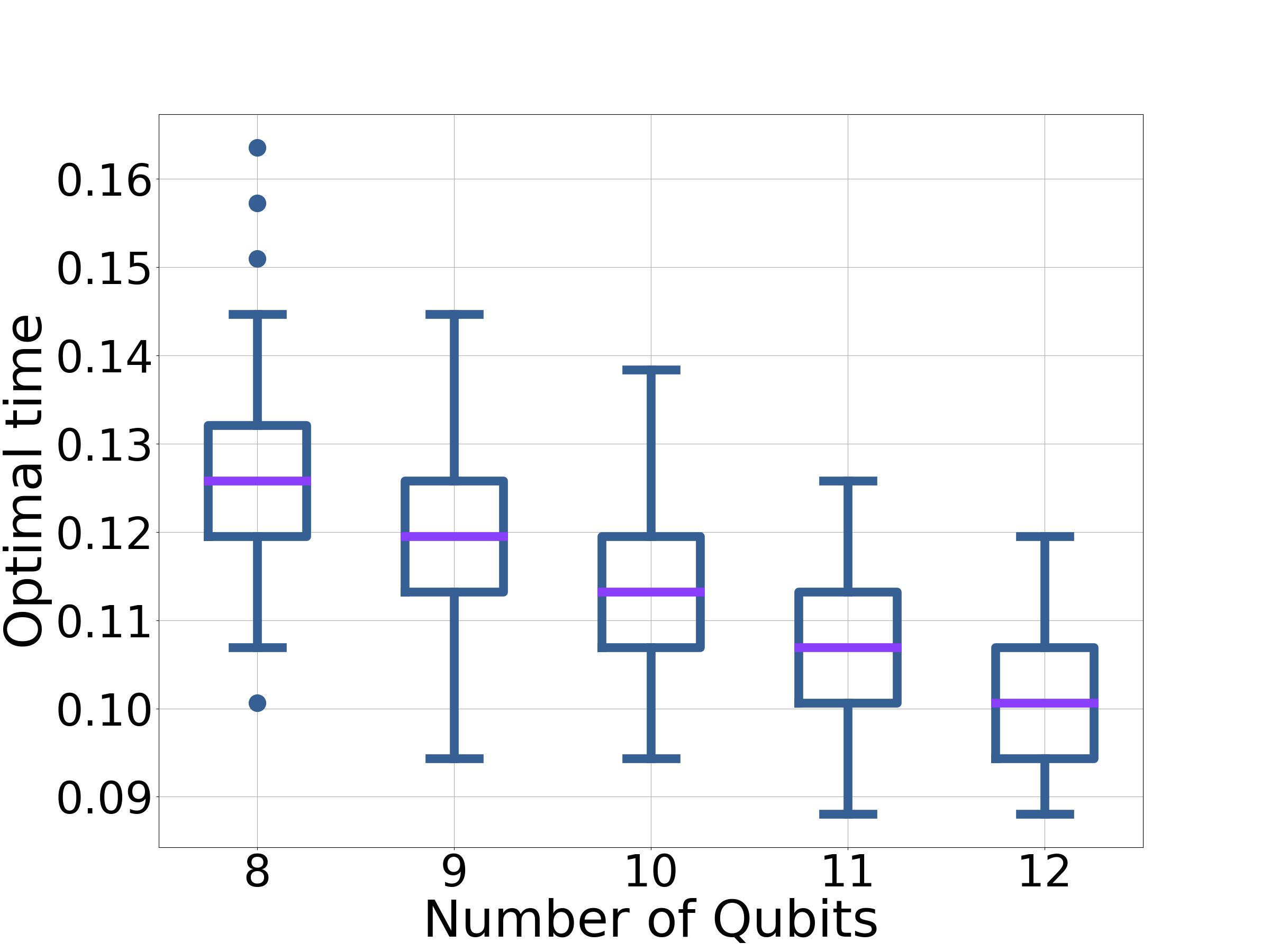}
        \caption{Optimal times for MAX-CUT on randomly generated graphs. The optimal time was found by dividing the interval $[0,2 \pi] $ into 1000 time steps.}
        \label{fig:MCRandOt}
    \end{subfigure}
    \caption{Performance of $H_1$ on 100 randomly-generated instances of MAX-CUT.}
    \label{fig:NumPer}
\end{figure*}

By exploiting locality in QA with short run-times, Braida et al.\ \cite{Bra22} were able to prove bounds on QA on MAX-CUT with three-regular graphs. Here we apply this approach to $H_1$. For details of the method the reader is referred to \cite{Bra22} and for explicit details of this computation to Appendix \ref{app:LRB}. We find that that $H_1$ finds at least $0.6003$ times the best cut. Hence $H_1$ has a marginally better worst-case than QA (which is 0.5933 times the best cut \cite{Bra22}), when this method is applied. Both bounds are not necessarily (and unlikely to be) tight. Resorting to numerical simulation gives Fig.\ \ref{fig:MC3Reg}. Here we can see, for the random instances considered, that $H_1$ never does worse than the QAOA $p=1$ worst bound. Direct comparisons to QAOA $p=1$ can be found in Appendix\ \ref{app:regQAOA}. Fig.\ \ref{fig:MC3Reg} also shows that the approximation ratio of $H_1$ on three-regular graphs has little dependence on the problem size, again suggesting that it is optimising locally.

\subsubsection{Numerical simulations on random instances of MAX-CUT}
\label{sec:H1Num}

In the previous two sections we established the performance of $H_1$ on problems with a high degree of structure, allowing for more  analytical investigation. In this section we explore the performance of $H_1$ numerically on MAX-CUT with random graphs. Consequently, we are restricted to exploring problem sizes that can be simulated classically. Further details about the numerics can be found in Appendix \ref{app:num}.

For each problem size we consider 100 randomly-generated instances. The results can be seen in Fig.\ \ref{fig:NumPer}. For each instance the time has been numerically optimised to maximise the approximation ratio within the interval $[0,2\pi)$. From Fig.\ \ref{fig:NumPer} we draw some conclusions from the simulations, with the caveat that either much larger sizes need to be simulated and/or analytic work is required to fully substantiate the claims. On all the problems considered $H_1$ performed better than random guessing (which results in an approximation ratio of 0). 

Focusing first on Fig.\ \ref{fig:MCRandc}: there appears to be some evidence that the distribution of approximation ratios is becoming smaller as the problem size is increased. In addition the approximation ratio tends to a constant value, independent of the problem size. From analysing the regular graphs, it is reasonable to assume that $H_1$ is optimising locally. Therefore, we would expect the performance of $H_1$ to depend on the subgraphs in the problem. If the performance is limited by one, or a certain combination of subgraphs, then that would explain the constant approximation ratio. As the problem size is increased the chance of having an atypical combination of subgraphs is likely to decrease, resulting in a smaller distribution.
 
The ground-state probability shows a clear exponential dependence on problem size (Fig.\ \ref{fig:MCRandgsp}).

\subsubsection{Estimating the optimal time}
Presenting the application of $H_1$ as a variational approach begs the question of how to find good initial guesses for the time, $T$, at which to measure the system. As previously noted the optimal time corresponds to the maximum approximation-ratio.

For our method we are not interested in finding the true maximum in approximation ratio. Sampling from a local maximum, close to $t=0$ is more achievable and reduces the time the system needs to be coherent. From Sec.\ \ref{sec:SERand} we expect the first turning point in approximation ratio after $t=0$ to be a local maximum. This is shown by Eq. \ref{eq:TDSEcoef}, with all amplitudes flowing from higher energy states to lower energy states a Hamming distance of one away at $t=0$. Further to this, throughout the work so far we have seen evidence of $H_1$ behaving locally. This local behaviour will allow us to  motivate good initial guesses for $H_1$.  As $H_1$ is optimising locally, its performance does not depend on the graph as a whole, but only on subgraphs.

\begin{figure}
\centering
\begin{subfigure}[t]{0.23\textwidth}
\centering
\begin{tikzpicture}[scale=0.8, auto,swap]

\foreach \pos/\name in { {(-1,0)/0}, {(0,0)/1}, {(1,0)/2}, {(2,0)/3}}
        \node[vertex_mt] (\name) at \pos {$\name$};

\foreach \source/ \dest  in {0/1, 1/2, 2/3}
        \path[edge] (\source) -- node {}(\dest);

\end{tikzpicture}
\caption{A subgraph of two-regular graphs}
\label{fig:subgraph_RD}
\end{subfigure}
\hfill
\begin{subfigure}[t]{0.23\textwidth}
\centering
\begin{tikzpicture}[scale=0.8, auto,swap]

\foreach \pos/\name in { {(-1,1)/0}, {(-1,-1)/1}, {(0,0)/2}, {(2,0)/3},{(3,1)/4},{(3,-1)/5}}
        \node[vertex_mt] (\name) at \pos {$\name$};

\foreach \source/ \dest  in {0/2, 1/2, 2/3,3/4,3/5}
        \path[edge] (\source) -- node {}(\dest);

\end{tikzpicture}
\caption{A small subgraph contained in three-regular graphs which is prevalent in graphs that $H_1$ performs worse on.}
\label{fig:subgraph_3Reg}
\end{subfigure}
\caption{Example subgraphs}
\end{figure}
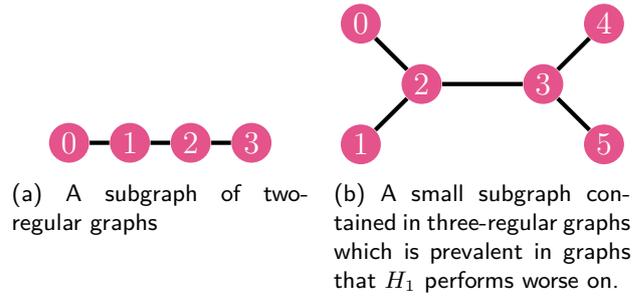

The optimal time for MAX-CUT on two-regular graphs was $T=0.23$. Under the assumption that $H_1$ is behaving locally we can estimate the optimal times by considering a smaller subgraph. The subgraph in question is shown in Fig.\ \ref{fig:subgraph_RD}. By numerically optimising $\langle Z_1Z_2 \rangle$ for this subgraph with $H_f=Z_0Z_1+Z_1Z_2+Z_2Z_3$, we can find good estimates for the optimal $T$. Optimising this subgraph, within the interval $T\in[0,1)$, gives $T=0.22$. This estimate matches the optimal time well. Choosing a larger subgraph will give a better estimate on the time.

Fig.\ \ref{fig:3RegOt} shows the optimal time for the larger instances of three-regular graphs considered in Fig.\ \ref{fig:MC3Reg}. The range of optimal times varied very little between problem instances and problem sizes, centered around $T=0.176$. As with the two-regular case we can examine subgraphs. In this case we consider the subgraph shown in Fig.\ \ref{fig:subgraph_3Reg}. This is the subgraph that saturates the Lieb-Robinson bound (Appendix \ref{app:LRB}).  Numerically optimising $\langle Z_2Z_3 \rangle$ for this subgraph with $H_f=Z_0Z_2+Z_1Z_2+Z_2Z_3+Z_3Z_4+Z_3Z_5$ gives a time of $T=0.19$. This again is a good estimate of the optimal time.

For problems with well understood local structure, such as regular graphs, we have shown that we can exploit this knowledge to provide reasonable estimates of the optimal times. These subgraphs are also of the same size used in finding the optimal time in QAOA $p=1$ \cite{Far14}.

For the MAX-CUT problems in Sec.\ \ref{sec:H1Num}, the optimal times can be found in Fig.\ \ref{fig:MCRandOt}. It appears that the optimal time tends to a constant value (or a small range of values), with $T<1$. The optimal times are clustered together, suggesting good optimal times might be transferable between problem instances. This approach is common within the QAOA literature \cite{Gal21}.

So far we have explored the performance of $H_1$. We have demonstrated that $H_1$ can provide a better approximation ratio for MAX-CUT on two-regular graphs. The intuition gained by studying QAOA $p=1$ has been transferable to the understanding of $H_1$. In the final part of this section we make some direct comparisons between QAOA $p=1$ and $H_1$.

\subsection{Direct numerical comparisons to QAOA p=1}
\label{sec:QAOA}

We have established in the previous sections that $H_1$ operates in a local fashion. Calculating the optimal time for sub-graphs the same size as those involved in QAOA $p=1$ gave good estimates for the optimal time for larger problem sizes. Therefore it is reasonable to assume that $H_1$ sees a similar proportion of the graph as QAOA $p=1$. Both approaches are variational with short run-times too. Since both approaches are using comparable resources, in this section we attempt to compare the two. Again we focus on the problems outlined in Sec.\ \ref{sec:prob}. 

For two-regular graphs, the optimal time for QAOA $p=1$ is 2.4 times longer than the optimal time for $H_1$ for large problem sizes, despite providing a poorer approximation ratio.
\begin{figure}[H]
    \begin{subfigure}[t]{0.48\textwidth}
        \centering
        \includegraphics[width=\textwidth]{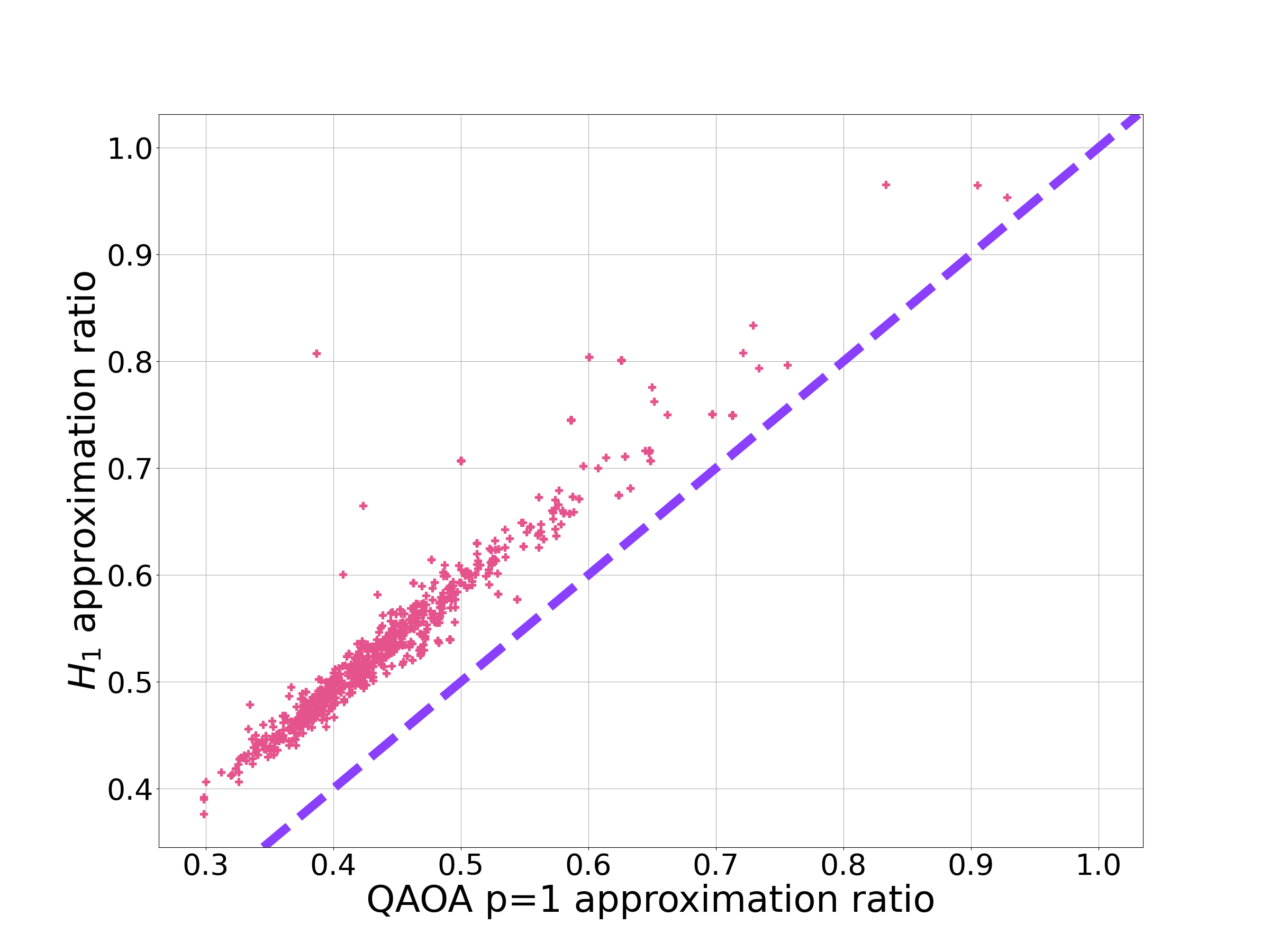}
        \caption{Approximation ratio comparison for MAX-CUT on randomly generated graphs.}
        \label{fig:comprandc}
    \end{subfigure}
    \hfill
    \\
    \begin{subfigure}[t]{0.48\textwidth}
        \centering
        \includegraphics[width=\textwidth]{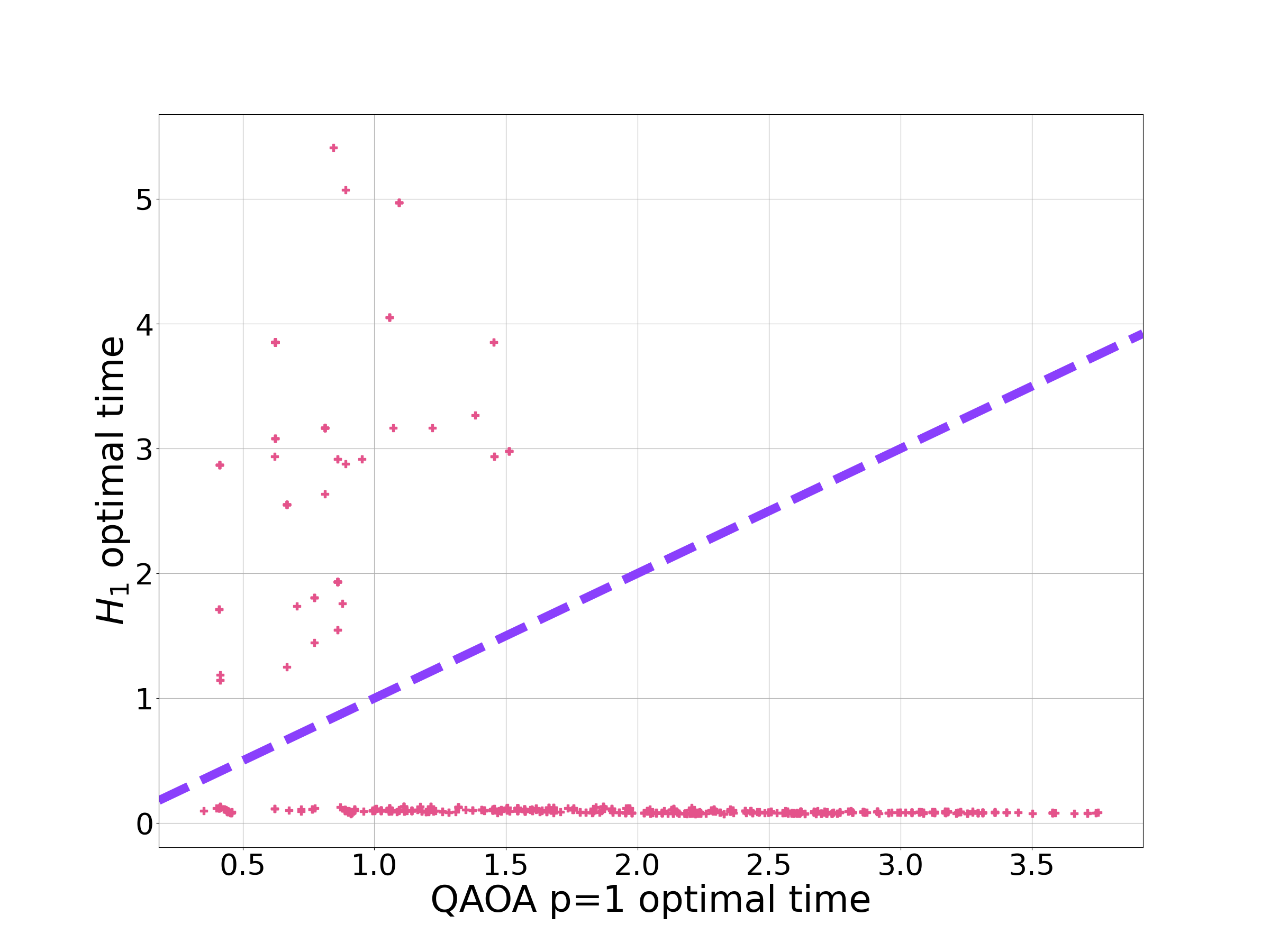}
        \caption{Optimal time comparison for MAX-CUT on randomly generated graphs.}
        \label{fig:comprandt}
    \end{subfigure}
    \caption{Comparison of $H_1$ (y-axis on the above plots) with QAOA $p=1$ (x-axis on the above plots)  for the problem instances considered in Sec.\ \ref{sec:H1Num}. The top line of figures compares approximation ratios. The bottom line of figures compares the optimal times (i.e.\ the time that maximises the approximation ratios) of the two approaches. The dashed purple line corresponds to equal performance.}
    \label{fig:comph1QAOA}
\end{figure}
The results comparing  $H_1$ and QAOA $p=1$, for all the problem instances considered in Sec.\ \ref{sec:H1Num}, are shown in Fig.\ \ref{fig:comph1QAOA}. 

The approximation ratios for each approach are largely correlated, suggesting in general that harder problems for QAOA $p=1$ corresponded to harder problems for $H_1$.For all instances considered, $H_1$ gave a greater than or equal to approximation ratio compared to QAOA $p=1$ (Fig.\ \ref{fig:comprandc}).

Turning now to the optimal time, $H_1$ had in the majority of cases the shorter optimal time (Fig.\ \ref{fig:comprandt}). In Appendix \ref{app:QAOA_better} we elaborate further on the exceptions, that is the MAX-CUT problems that have longer run-times than QAOA p=1.

This section has numerically demonstrated that $H_1$ provides a better approximation ratio than QAOA $p=1$ in a significantly shorter time for the majority of instances considered, justifing our description of this this approach as rapid, which is crucial for NISQ implementation \cite{Pre18}. Given that, $H_1$ tends to provide a better approximation ratio, in a shorter time, with fewer variational parameters, it raises the question - does QAOA $p=1$, the foundation of any QAOA circuit, make effective use of its afforded resources?

\FloatBarrier
\section{An improvement inspired by the Quantum Zermello problem}
\label{sec:QZ}
\subsection{The approach}
\label{sec:QZapp}

With QAOA it is clear how to get better approximation ratios, that is by increasing $p$. It is less clear how to do this with $H_1$. One suggestion might be to append this Hamiltonian to a QAOA circuit. However, the aim of this paper is to explore how Hamiltonians for optimal state-transfer can provide a guiding design principle.  Therefore, in this section we explore adding another term, motivated by this new design principle, to $H_1$ to improve the approximation ratio.

In Sec.\ \ref{sec:hamdes} we motivated $H_1$ from the optimal Hamiltonian by making the pragmatic substitutions $\ket{\psi_i}\bra{\psi_i} \rightarrow H_i$ and $\ket{\psi_f}\bra{\psi_f} \rightarrow H_f$. Subsequently, we demonstrated that $H_1$ provides a reasonable performance. However, $H_1$ no longer closely followed the evolution under the optimal Hamiltonian. To partially correct for this error we add another term to the Hamiltonian:
\begin{equation}
    \label{eq:corQZ}
    H_{1,improved}=H_1+H_{QZ}.
\end{equation}

 Again, we make use of Hamiltonians for optimal state-transfer to motivate the form of $H_{QZ}$. Finding the optimal Hamiltonian in the presence of an uncontrollable term in the Hamiltonian is known as the quantum Zermello (QZ) problem \cite{Bro15,Brody_2015,Rus14,Rus15}. 

 \begin{figure}
    \centering
\begin{tikzpicture}

\fill[black] (-1,5) circle (0.15cm) node[anchor=south east] {$\ket{\psi_i}$};

\fill[black] (5,4) circle (0.15cm) node[anchor=north west] {$\ket{\psi_f}$};

\draw[myblue,ultra thick,dashed] (5,4) .. controls (4,5) and (4,0) .. (0,2);

\draw[myblue, ultra thick] (-0.1,1.9) -- (0.1,2.1);

\draw[myblue, ultra thick] (-0.1,2.1) -- (0.1,1.9)node[anchor=south west] {$e^{i H_{QW}T} \ket{\psi_f}$};

\draw[mypink, ultra thick](-1,5)--(-2/3,4) node[anchor=south west] {$e^{-i H t} \ket{\psi_i}$};

\draw[mypink, ultra thick, dashed](-2/3,4)--(-1/3,3);
\end{tikzpicture}
    \caption{A cartoon of the evolution of states in the QZ problem for constant $H_{QW}$. In the interaction picture, with background Hamiltonian $H_{QW}$, it appears the final state is moving under the influence of this Hamiltonian. In this frame Eq. \ref{eq:optHam} can then be applied. It then remains to move out of the interaction picture to get Eq.\ \ref{eq:opHQZ}. }
    \label{fig:QZcartoon}
\end{figure}
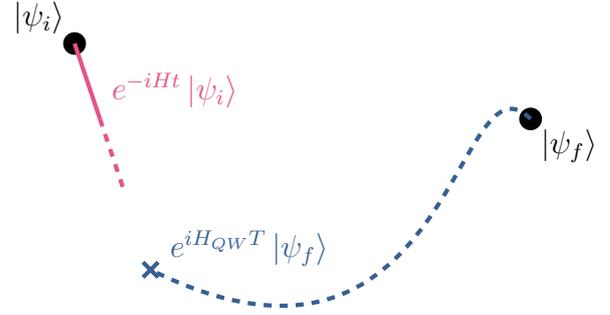

In the rest of this section we expand on the details of the QZ problem. From the exact form of the optimal correction, $H_{cor}$, we then apply a series of approximations so that $H_{cor}$ is time-independent and does not rely on knowledge of $\ket{\psi_f}$. This final Hamiltonian will be $H_{QZ}$ in Eq.\ \ref{eq:corQZ}.

The QZ problem, like the quantum brachistochrone problem, asks what is the Hamiltonian that transfers the system from $\ket{\psi_i}$ to the final state $\ket{\psi_f}$ in the shortest possible time. Unlike the quantum brachistochrone problem, part of the Hamiltonian is uncontrollable. In the case of a constant uncontrollable term, the total Hamiltonian can be written as:
\begin{equation}
    H_{opt | QW}=H_{QW}+H_{cor}(t),
\end{equation}
where $H_{QW}$ is the constant `quantum wind' that cannot be changed and $H(t)$ is the Hamiltonian we are free to vary. Typically, $H_{QW}$ is understood as a noise term  \cite{Lap10,Muk13}. Instead, here we will take $H_{QW}$ to be $H_1$ to provide a favourable quantum wind that $H_{cor}(t)$ can provide an improvement on. 

The optimal form of $H_{cor}(t)$ is (up to some factor) \cite{Brody_2015}:

\begin{multline}
    H_{cor}(t)=-i e^{-iH_{1}t}\left(\ket{\psi_i}\bra{\psi_f}e^{-iH_{1}T}\right.\\  \left. -e^{i H_{1}T}\ket{\psi_f}\bra{\psi_i} \right)e^{iH_{1}t},
    \label{eq:opHQZ}
\end{multline}
where $t$ is the time and $T$ is the final time. The motivation for this equation can be found in Fig.\ \ref{fig:QZcartoon}. This Hamiltonian requires knowledge of the final state, so we introduce a series of approximations to make $H_{cor}(t)$ more amenable for implementation. 

 Since we know that the optimum evolution under $H_1$ tends to be short, we make the assumption that the total time $T$ is small. Therefore, we approximate the optimal correction with $H_{cor}(t)$ with
\begin{multline}
    H_{cor}(0)=-i \left(\ket{\psi_i}\bra{\psi_f}e^{-iH_{1}T} \right. \\
    \left. -e^{i H_{1}T}\ket{\psi_f}\bra{\psi_i} \right).
\end{multline}

Introducing the commutator structure (Sec.\ \ref{sec:hamdes}) with the same pragmatic substitutions as before for $H_1$ gives:
\begin{equation}
   \label{eq:infT}
    H_{QZ}=-i \left[H_i,e^{i H_{1}T}H_fe^{-iH_{1}T} \right],
\end{equation}
where we have introduced the subscript $QZ$ to distinguish this Hamiltonian from $H_{cor}(0)$ prior to the substitutions. Expanding this expression in $T$ gives:

\begin{multline}
    \label{eq:QZexp}
    H_{QZ}=-i \left[H_i, H_f+iT \left[H_{1},H_f\right] \right.\\ 
    \left.-T^2\left[H_{1},\left[H_{1},H_f\right]\right]/2+\mathcal{O}\left(T^3\right)\right].
\end{multline}

From the QZ problem we have motivated the form of the correction $H_{QZ}$ in Eq.\ \ref{eq:corQZ}.  In spite of this we have no guarantee on its performance - to this end we carry out numerical simulations. 

In both its philosophy and structure $H_{QZ}$ is reminiscent of shortcuts to adiabaticity (STA) \cite{Gue19}. In STA the aim is to modify the Hamiltonian in Eq. \ref{eq:QAham}  to reach the adiabatic result in a shorter time, typically by appending to the standard QA Hamiltonian. The approach here is distinctly different  for three key reasons, besides the initial starting point of Hamiltonian. 
\begin{enumerate}
    \item The aim here is to do something distinctly different from $H_1$, not to recover its behaviour in a shorter time.
    \item In the QZ-inspired approach we make use of the excited states, with the aim of finding approximate solutions, as opposed to exact solutions. 
    \item Here we only consider time-independent Hamiltonians.
\end{enumerate}

A clear downside to $H_{QZ}$ is the increased complexity, compared to say QAOA. However, if $H_{QZ}$ is decomposed into a QAOA-style circuit, the single free parameter in $H_{QZ}$ might translate to fewer free parameters in the QAOA circuit, allowing for easier optimisation.

\subsection{Numerical simulations}
\subsubsection{MAX-CUT on two-regular graphs}

\begin{figure}
    \centering
    \begin{subfigure}[t]{0.48\textwidth}
    \centering
    \includegraphics[width=\textwidth]{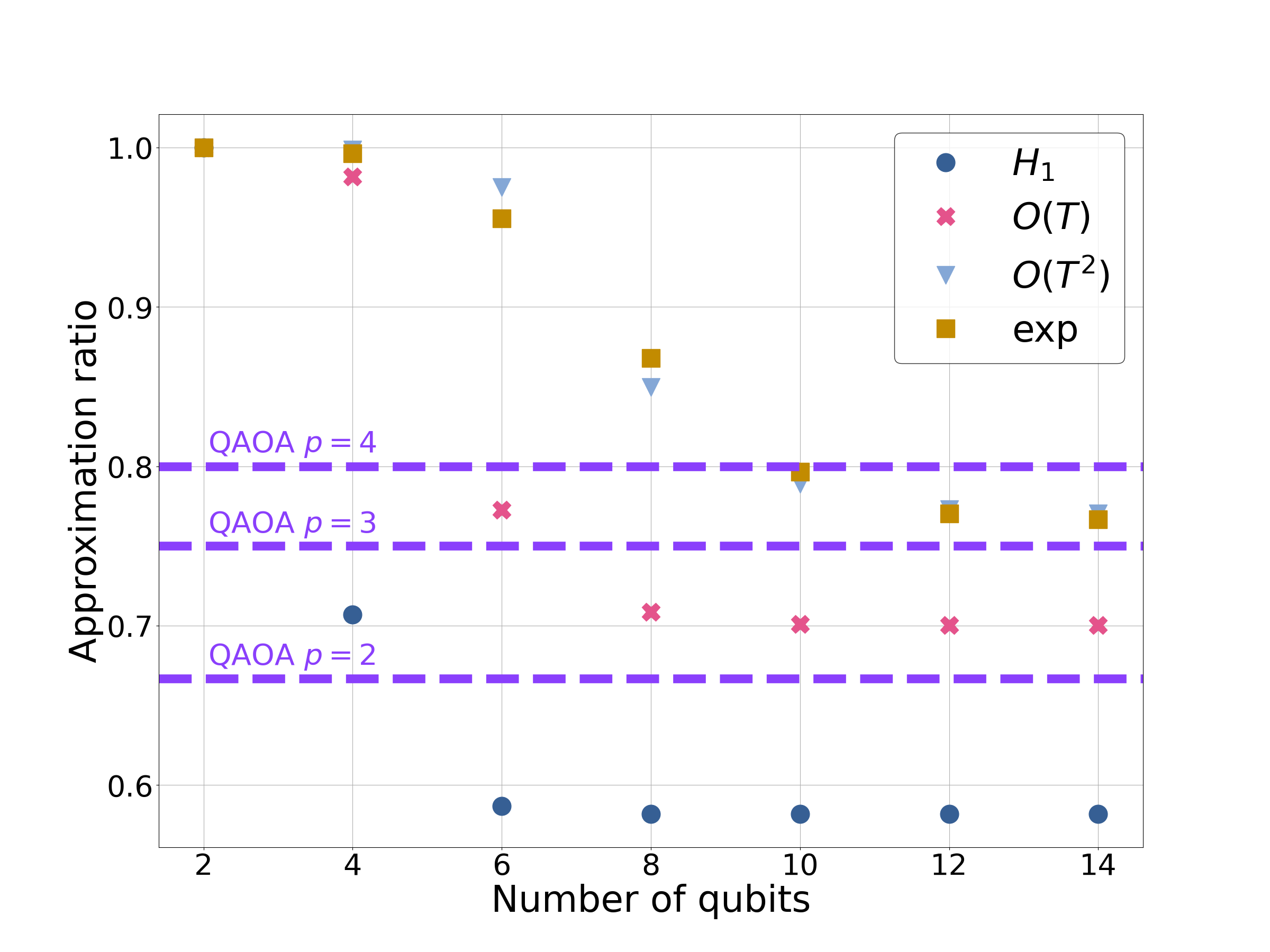}
    \caption{Numerically optimised performance of Eq.\ \ref{eq:QZexp}. Each point has been optimised in the time interval [0,0.3] by considering 3000 divisions. }
    \label{fig:SerPer}
    \end{subfigure}
    \begin{subfigure}[t]{0.48\textwidth}
    \centering
    \includegraphics[width=\textwidth]{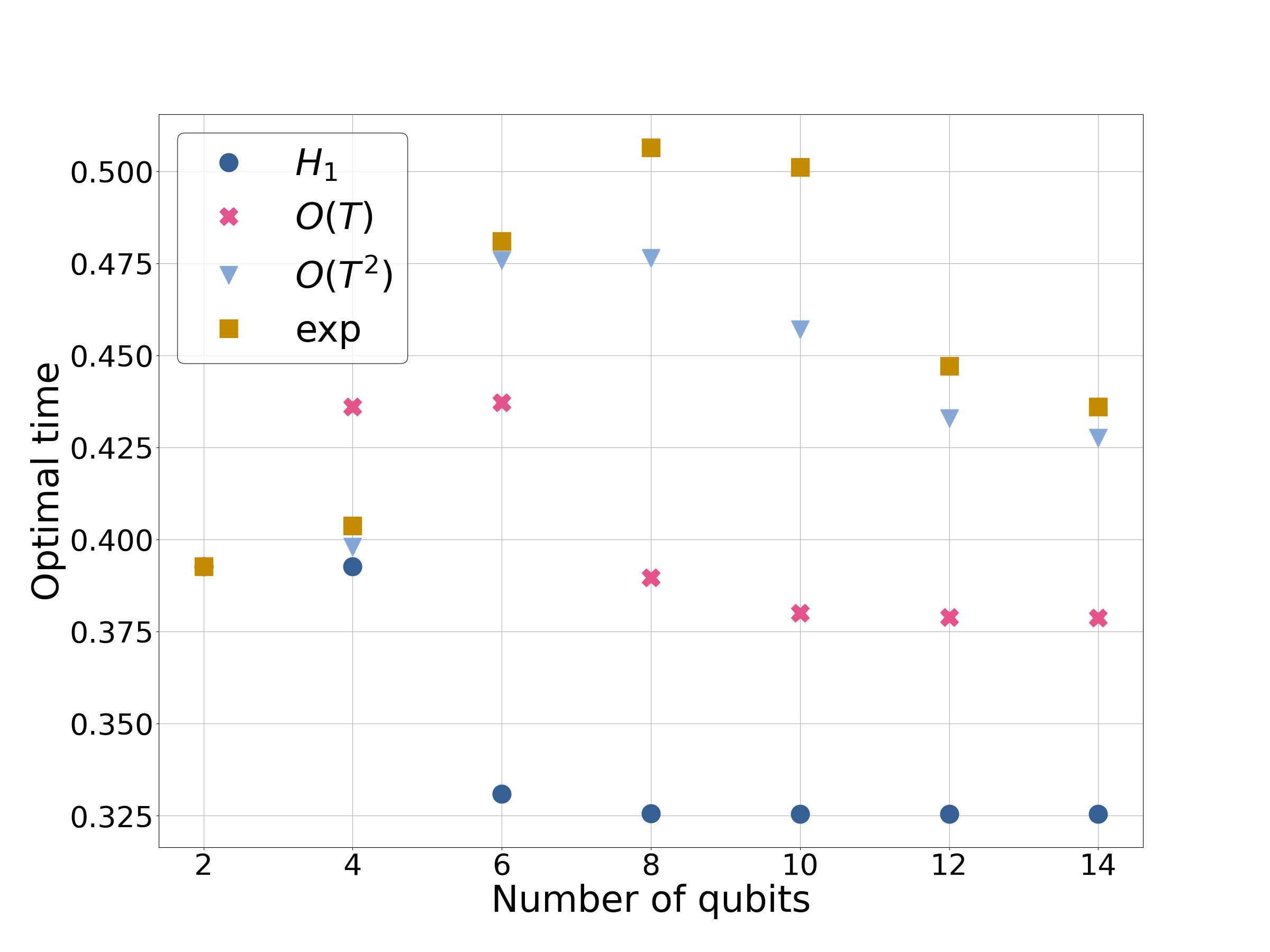}
    \caption{The corresponding optimal times for Fig.\ \ref{fig:SerPer}. The norm of each Hamiltonian, for each problem size has been fixed so Eq.\ \ref{eq:norm} is true, to make comparisons fair.}
    \label{fig:Sertime}
    \end{subfigure}
    \caption{The performance of Eq.\ \ref{eq:QZexp} on two-regular graphs.  The legend shows the order of $T$, with `exp' referring to Eq.\  \ref{eq:infT}. The dashed lines show the asymptotic performance of QAOA.}
\end{figure}

Here we focus on applying Eq.\ \ref{eq:QZexp} up to various orders in $T$ to MAX-CUT on two-regular graphs with an even number of qubits.  We focus on this problem as it is trivial to scale and the performance of QAOA and $H_1$ on this problem is well understood. 

The results for MAX-CUT with two-regular graphs can be seen in Fig.\ \ref{fig:SerPer}. Increasing the expansion in $T$ appears to improve the approximation ratio. But the improvement is capped, shown by the data labelled `exp'. Notably, this approach with a single variational parameter at order $T^2$ is performing better than QAOA $p=3$ (with 6 variational parameters) for 10 qubits. 

The optimal $T$ for the QZ-inspired Hamiltonians can be seen in Fig.\ \ref{fig:Sertime}. Again, the optimal time for each order in $T$ appears to be tending to a constant value, suggesting this approach is still acting in a local fashion. This is consistent with the approximation ratio plateauing. As we can see the QZ-inspired approach is still operating in a rapid fashion.

\subsubsection{MAX-CUT on random graphs}

To complete this section we examine the performance of the QZ-inspired approach (Eq.\ \ref{eq:QZexp}) to the randomly generated instances of MAX-CUT, detailed in Sec.\ \ref{sec:prob}. 

\begin{figure}
    \begin{subfigure}[t]{0.48\textwidth}
    \centering
    \includegraphics[width=\textwidth]{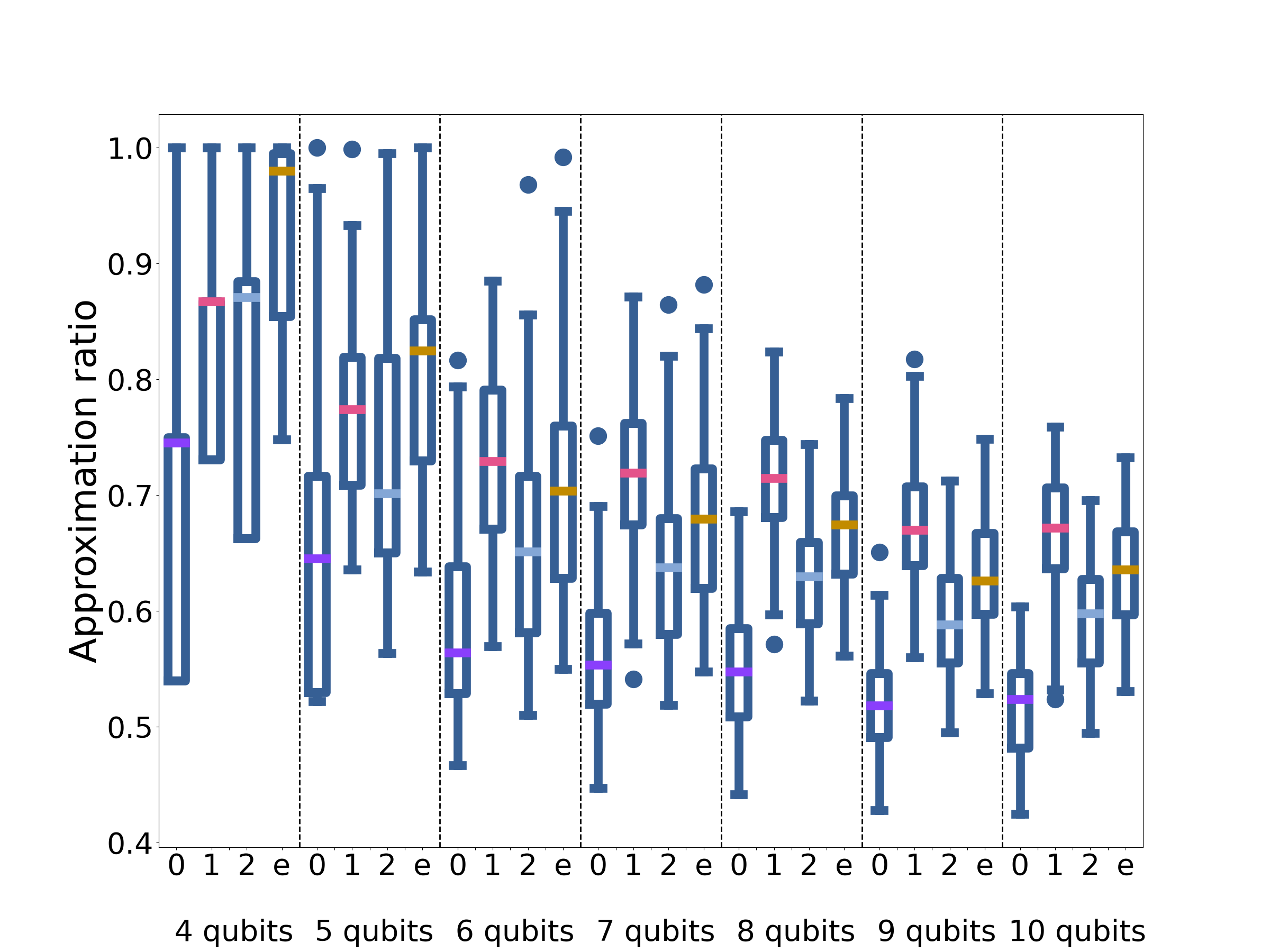}
    \caption{The approximation ratio.}
    \label{fig:qzmc}
    \end{subfigure}
    \begin{subfigure}[t]{0.48\textwidth}
    \centering
    \includegraphics[width=\textwidth]{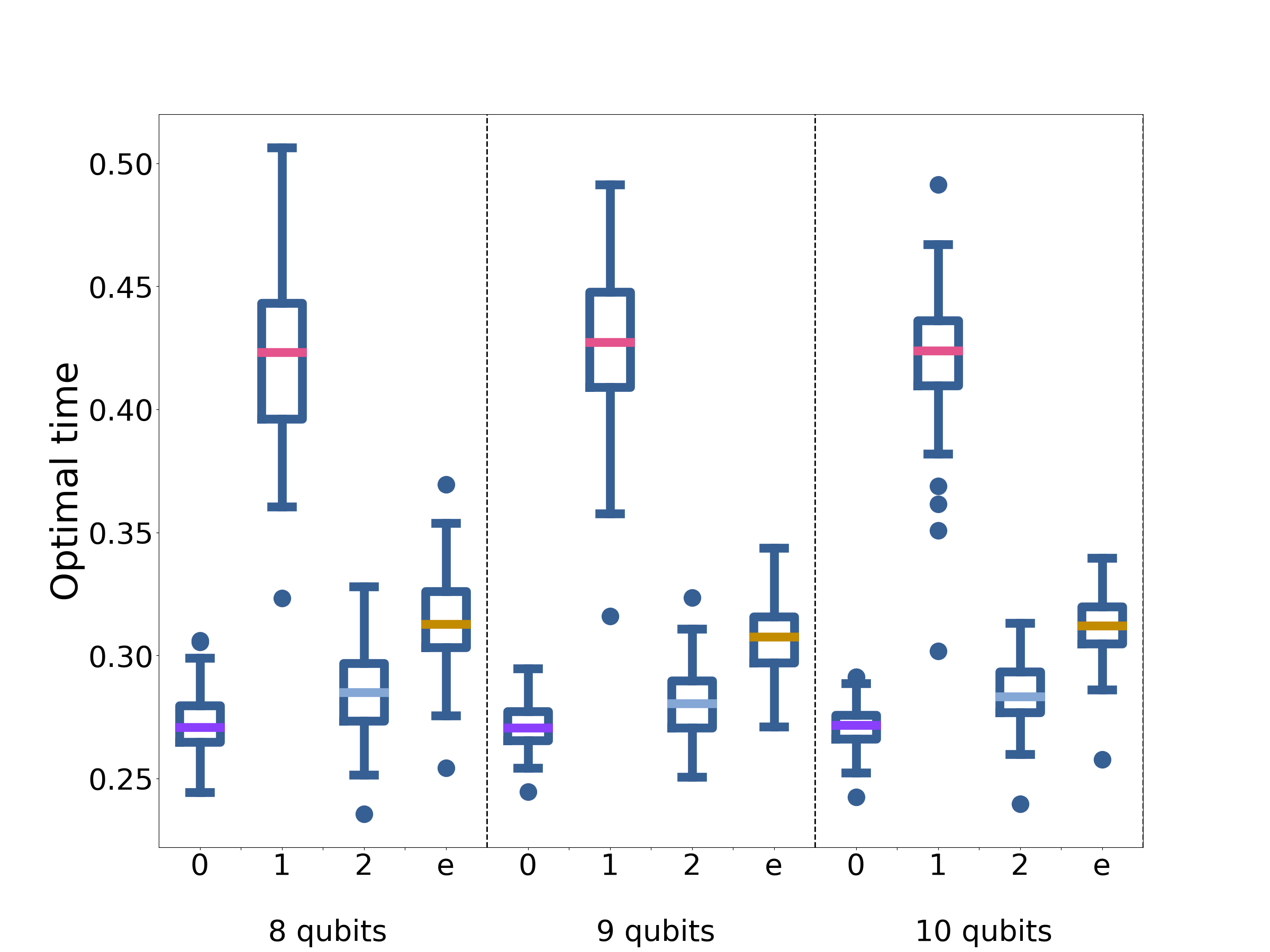}
    \caption{The optimal times for the problem instances seen in Fig.\ref{fig:qzmc}. The norm of each Hamiltonian, for each problem size has been fixed, according to Eq.\ \ref{eq:norm}, ensuring a fair comparison.}
    \label{fig:QZmct}
    \end{subfigure}
    \caption{The performance of the QZ-inspired approach on 100 random MAX-CUT instances. The x-axis label refers to the order of $T$ in the expansion of Eq.\ \ref{eq:QZexp}, with $0$ being $H_1$ and $e$ referring to the full exponential (i.e.\ Eq.\ \ref{eq:infT})}
\end{figure}

The results for different orders in $T$ for the approximation ratio can be seen in Fig.\ \ref{fig:qzmc}. All the QZ-inspired approaches provide an improvement on the original $H_1$ Hamiltonian, indexed by 0 in the figures. However, the performance is not monotonically increasing with the order of the expansion. This is not unusual for a Taylor series of an oscillatory function. Consequently, achieving better approximation ratios is not as simple as increasing the order of $T$. At the same time, this means that it is not necessary to go to high orders in $T$, with very non-local terms, to achieve a significant gain in performance. For example, in going to first order achieves a substantial improvement. 

The optimal times for the QZ-inspired approach can be found in Fig.\ \ref{fig:QZmct}. For clarity we only show the optimal times for the larger problem instances. As with $H_1$ the optimal times are clustered for a given order. The lack of dependence on problem size for optimal times and approximation ratios suggests that the QZ-inspired approach is still optimising locally. Compared to the $H_1$ case, the operators have a larger support. Despite optimising locally, they are optimising less locally than $H_1$, hence the increased performance.

Here we have numerically demonstrated that the QZ-inspired approach can provide an improvement over $H_1$, suggesting how this new design philosophy might be extended. The numerics also suggest that going to first order may provide the best possible advantage. 

\section{Using knowledge of the initial state}
\label{sec:lpa}

As introduced in Sec.\ \ref{sec:hamdes}, in this section we exploit our knowledge of the initial state and evaluate the performance of 
\begin{equation}
    H_{\psi_i}=-i\left[\ket{\psi_i}\bra{\psi_i},f(H_f)\right] 
\end{equation}
within the QA-framework. We take $f(\cdot)$ to be a real function such that:
\begin{equation}
    f(H_f)=\sum_k f(E_k)\ket{E_k}\bra{E_k},
\end{equation}
where $\ket{E_k}$ and $E_k$ are the eigenvectors and associated eigenvalues of $H_f$.

Evolution under $H_{\psi_i}$ can be calculated analytically - the details can be found in Appendix \ref{app:expHam}. By evolving $\ket{\psi_i}$ the state:
\begin{equation}
    \ket{\omega}=\frac{1}{\sqrt{\Tr f^2(H_f)}}\sum_k f(E_k)\ket{E_k},
\end{equation}
can be reached. Indeed $H_{\psi_i}$ will generate linear superpositions of $\ket{\omega}$ and $\ket{\psi_i}$ only.

Assuming that the state $\ket{\omega}$ is prepared, then the probability of finding the ground-state is 
\begin{equation}
    P_{gs}=\frac{g f^2(E_{gs})}{\Tr f^2(H_f)},
\end{equation}
where $g$ is the ground-state degeneracy and $E_{gs}$ the associated energy. If $f(\cdot)$ is the identity, then $\Tr H_f$ scales approximately as $2^n$ and $E_{gs}$ might scale with $n$. Hence the ground state probability will scale as  $\sim n^2/2^n$. Indeed if $f(H_f)=H_f^m$, where $m$ is some  positive integer, then the ground state probability might scale as $\sim n^{2m}/2^n$. This is an improvement over random guessing, but still with exponential scaling. This may be of some practical benefit, depending on the computational cost of calculating $f(H_f)$. 
If $f(\cdot)$ is the projector onto the ground-state then $P_{gs}=1$ (as expected). 

Calculating the expectation of $H_f$ for $\ket{\omega}$ gives:
\begin{equation}
    \langle H_f \rangle =\frac{1}{\Tr f^2(H_f)}\sum_k E_k f^2(E_k).
\end{equation}
Here we can see that $\langle H_f \rangle$ will be dominated by states for which $f^2(E_k)$ is large. If $f(\cdot)$ is the identify this most likely means low energy states and high energy states. Hence, we do not expect a good approximation ratio. This provides some insight into Sec.\ \ref{sec:RoD} where the observed peak in ground-state probability did not coincide with the optimal approximation ratio.

This approach has the potential to provide a modest practical speed-up with a polynomial prefactor on the hardest problems. However, the success of this approach depends on the (unlikely) feasibility of implementing $H_{\psi_i}$ and $f(H_f)$. It does however provide further evidence of the power of commutators for designing algorithms to tackle optimisation problems.

\section{Discussion and Conclusion}

Designing quantum algorithms to tackle combinatorial optimisation problems, especially within the NISQ framework, remains a challenge. Many algorithms have used AQO in their inspiration, such as in the choice of Hamiltonians. In this paper we have explored using optimal Hamiltonians as a guiding design principle. 

With $H_1$, the commutator between $H_i$ and $H_f$, we demonstrated that we can outperform QAOA p=1, with fewer resources. The short run-times which do not appear to scale with problem size suggest that this approach is acting locally. An effective Lieb-Robinson bound prevents the information about the problem propagating instantaneously \cite{Lieb1972,WangZ20}. This helps provide some insights into the performance of $H_1$:
\begin{itemize}
    \item In the local regime, the effective local Hilbert space is smaller than the global Hilbert space, consequently $H_1$ will be a better approximation of the optimal Hamiltonian. This accounts for why we might expect $H_1$ to work better on short run-times.
    \item A local algorithm is unlikely to be able to solve an optimisation problem, as it cannot see the whole graph. It follows that such an approach would have poor scaling of the ground-state probability. 
\end{itemize}

Due to the local nature of $H_1$ we were able to utilise some analytical tricks to assist the numerical assessment of its performance, allowing for some guarantee of the performance of the approach on large problem sizes. The techniques used had already been developed or deployed by the continuous-time quantum computing community in the context of QA/QAOA, indicative of the wide applicability of the tools being developed to assess these algorithms.

Local approaches have clear advantages in NISQ-era computations. The short run-times put fewer demands on the coherence times of the device. The local nature can also help mitigate some errors. If, for example, there is a control misspecification in one part of the Hamiltonian this is unlikely to propagate through the whole system and affect the entire computation. 

Buoyed by the relative success of utilizing Hamiltonians for optimal state-transfer we turned to the quantum Zermello problem to help find improvements. These Hamiltonians compromised a single variational parameter and short run-times, for increased complexity in the Hamiltonian. Again, the saturation of the optimal time suggest these approaches are still operating locally.

The success of this approach, within the NISQ era, will depend on the feasibility of implementing these Hamiltonians. This might be achieved through decomposition into a product formula \cite{Childs_2013} for gate based approaches, resulting in a QAOA like circuit. Alternatively, one could attempt to explicitly engineer the interactions involved. Indeed, for exponentially scaling problems, implementing these Hamiltonians for short times could be less challenging then maintaining coherence for exponentially increasing times.

Although the results of this paper are not fully conclusive, it has shown that by considering Hamiltonians for optimal state-transfer we can develop promising new algorithms. We hope the results presented in this paper will encourage further work into the success of these Hamiltonians. There is scope for taking this work further. This could include changing the choice of $H_i$, exploiting our observation that any stoquastic Hamiltonian can lead to an increase in ground state probability. For $H_f$ we have only explored problems with trivial Ising encodings. There is scope to explore new encodings such as LHZ \cite{Lec15}  or Domain-wall \cite{Cha19} encodings. Such encodings will result in different $H_1$ and presumably distinct dynamics.

\section*{Acknowledgments}
We gratefully acknowledge Filip Wudarski, Glen Mbeng, Henry Chew, Natasha Feinstein, and Sougato Bose for inspiring discussion and helpful comments. This work was supported by the Engineering and Physical Sciences Research Council through the Centre for Doctoral Training in Delivering Quantum Tech-
nologies [grant number EP/S021582/1] and the ESPRC Hub in Quantum Computing and Simulation [grant number EP/T001062/1]. For the purpose of open access, the author has applied a Creative Commons Attribution (CC BY) licence to any Author Accepted Manuscript version arising.

\bibliographystyle{quantum}
\bibliography{main}

\onecolumn\newpage
\appendix

\section{Optimal Hamiltonians for the QA-framework}
\label{sec:opHamQa}
In Sec.\ \ref{sec:hamdes} we saw that the Hamiltonian (up to some constant factor):
\begin{equation}
    \label{eq:ophamap}
    H_{opt}=-i\left(\ket{\psi_i}\bra{\psi_f}-\ket{\psi_f}\bra{\psi_i}\right).
\end{equation}
 transfers the system from $\ket{\psi_i}$ to  $\ket{\psi_f}$ in the shortest possible time.

In standard QA, the time-varying Hamiltonian interpolates between $H_i=-\sum_i X_i$ and an Ising Hamiltonian $H_f$. 
Consider the overlap of $H_i$ with Eq.\ \ref{eq:ophamap}:
\begin{align*}
    \Tr{H_i H_{opt}}=&-i\Tr{H_i\left(\ket{\psi_i}\bra{\psi_f}-\ket{\psi_f}\bra{\psi_i}\right)}\\
    =&-i E_i^{(0)}\Tr{\ket{\psi_i}\bra{\psi_f}-\ket{\psi_f}\bra{\psi_i}}\\
    =&0,
\end{align*}
where $E_i^{(0)}$ is the ground-state energy of $H_i$. The final line follows since in QA, typically $\bra{\psi_i}\ket{\psi_f}=g/\sqrt{2^n}$, where $g$ is the degeneracy of the ground-state. We also choose $\ket{\psi_i}$ and $\ket{\psi_f}$ to have a real overlap in deriving Eq.\ \ref{eq:ophamap}. Similarly, 
\begin{equation*}
    \Tr{H_f H_{opt}}=0.
\end{equation*}

Eq.\ \ref{eq:ophamap} has no overlap with any of the Hamiltonians typically used in QA. More generally, if $M$ is any operator whose eigenstates include $\ket{\psi_i}$ or $\ket{\psi_f}$, then 
\begin{equation*}
    \Tr{M H_{opt}}=0.
\end{equation*}

This means Eq.\ \ref{eq:ophamap} has no overlap with other Hamiltonians besides $H_i$ and $H_f$, including $XX$ terms.

As a final example, consider the optimal Hamiltonian for MAX-CUT on a two-regular graph with four qubits. The Hamiltonian (up to some scaling factor):
\begin{multline}
    \label{eq:opHam4Rod}
    H_4=Z_1Y_2+Y_1Z_2+Z_2Y_3+Y_2Z_3+Z_3Y_4+Y_3Z_4+Z_1Y_4+Y_1Z_4-Z_1Y_3-Y_1Z_3-Z_2Y_4-Y_2Z_4\\
    -Y_1Z_2Z_3Z_4-Z_1Y_2Z_3Z_4-Z_1Z_2Y_3Z_4-Z_1Z_2Z_3Y_4+X_1Y_2Z_3+Y_1X_2Z_4-Y_1X_2Z_3-X_1Y_2Z_4\\
    +Y_1Z_2X_3+X_1Z_2Y_3+X_1Y_3Z_4+Y_1X_3Z_4
    +Y_1Z_2X_4+X_1Z_3Y_4-Y_1Z_3X_4-X_1Z_2Y_4\\
    +Z_1Y_2X_3+X_2Y_3Z_4-Z_1X_2Y_3-Y_2X_3Z_4
    +Z_1Y_2X_4+Y_2Z_3X_4+X_2Z_3Y_4+Z_1X_2Y_4\\
    +Z_2Y_3X_4+Z_1X_3Y_4-Z_1Y_3X_4-Z_2X_3Y_4
    +X_1X_2Y_3Z_4+Y_1X_2X_3Z_4-X_1Y_2X_3Z_4+Y_1Y_2Y_3Z_4\\
    +X_1Y_2Z_3X_4+X_1X_2Z_3Y_4-Y_1X_2Z_3X_4+Y_1Y_2Z_3Y_4
    +Y_1Z_2X_3X_4+X_1Z_2Y_3X_4-X_1Z_2X_3Y_4+Y_1Z_2Y_3Y_4\\
    +Z_1Y_2X_3X_4+Z_1X_2X_3Y_4-Z_1X_2Y_3X_4+Z_1Y_2Y_3Y_4,\\
\end{multline}
solves the problem in the shortest possible time. This Hamiltonian has a huge number of terms, yet none of them are the ones typically used in QA. Note every term involves a $Y$ Pauli. 
 
Although Eq.\ \ref{eq:opHam4Rod} is clearly not implementable, it raises the question of whether QA uses the correct terms in the Hamiltonian to achieve practical speed-up, especially in NISQ devices. This suggests terms like $H_i$, $H_f$ (unless used in a targeted way) might not be the most successful choice of Hamiltonians for fast NISQ algorithms.

\section{Background on the problems considered}
\label{app:prob_plus}
The Hamiltonians proposed in Sec.\ \ref{sec:hamdes} need to be applied to optimisation problems to assess their performance. Algorithms are unlikely to work uniformly well on all problems; for instance QA is believed to work better on problems with tall, thin barriers in the energy landscape \cite{Kat15}. Consequently the apparent performance of a heuristic algorithm will, in general, be dependent on the optimisation problem considered as well as the algorithm. With this caveat in mind this paper makes focuses on the canonical optimisation problems MAX-CUT. To further substantiate the claims made we provide a further numerical study on the Sherrington-Kirkpatrick-inspired problem in Appendix \ref{app:skm}. This section briefly outlines these problems. 

\subsection{MAX-CUT}

MAX-CUT seeks to find the maximum cut of a graph, $G=(V,E)$. A cut separates the nodes of the graph into two disjoint sets. The value of the cut is equal to the number of edges between the two disjoint sets. In general, MAX-CUT is an NP-hard problem \cite{Gar76}. Indeed, finding very good approximations for MAX-CUT is a computationally hard problem \cite{Pap88}.

The corresponding Ising formulation of this problem is:
\begin{equation}
    H_f=\sum_{(i,j) \in E} Z_i Z_j, 
\end{equation}
up to a constant offset (i.e. a term proportional to the identity) and multiplicative factor. In this paper we explore MAX-CUT on a range of graphs.

\subsubsection{Two-regular graphs}

MAX-CUT on two-regular graphs (also known as the Ring of Disagrees or the anti-ferromagnetic ring) is a well studied problem in the context of QA and QAOA \cite{Far00,Wan18,mbe19,Far14}. The problem Hamiltonian
\begin{equation}
    H_f=\sum_{i=1}^n Z_i Z_{i+1}
\end{equation}
with $n+1=1$, consists of nearest-neighbour terms only. The performance of QA and QAOA on this problem has been understood by applying the Jordan-Wigner transformation to map the problem onto free fermions \cite{Far00,Wan18}. In Sec.\ \ref{sec:RoD} we follow the approach laid out by \cite{Wan18} to apply this technique to $H_1$. A further useful tutorial for tackling the Ising chain with the Jordan-Wigner transformations can be found in \cite{Mbe20}.

Alternatively, provided $p$ is sufficiently small, the performance of QAOA can be understood in terms of locality \cite{Far14}. Due to the structure of the ansatz in QAOA (i.e.\ Eq.\ \ref{eq:QAOA}), to find the expectation of a term in $H_f$, such as $Z_iZ_{i+1}$, it is only necessary to consider a subgraph. This subgraph consists of all nodes connected by no more than $p$ edges to a node in the support of the expectation value being calculated. For two regular graphs this is a chain consisting of $2p+2$ nodes. Provided this subgraph is smaller than the problem graph (i.e. the two-regular graph has more than $2p+2$ nodes), then QAOA is operating locally.

For a given $p$, all the subgraphs are identical for a two-regular graph. Hence, the performance of QAOA for this problem depends only on its performance on this subgraph. As a direct consequence of the locality, the approximation ratio of QAOA will not change as the size of the two-regular graph is scaled. By optimising over this subgraph with a classical resource, it is therefore possible to find the optimal time and approximation ratio of QAOA for this problem.

\subsubsection{Three-regular graphs}
\label{sec:prob3reg}
MAX-CUT on three-regular graphs was considered in the original QAOA paper by Farhi et al.\ \cite{Far14}. The local-nature of QAOA allowed them to calculate explicit bounds on the performance of their algorithm. To this end, the graph was broken down into subgraphs to measure local expectation values (i.e., $\langle Z_iZ_{i+1}\rangle$). For QAOA $p=1$, there are three distinct subgraphs. By simulating QAOA for the subgraphs and bounding the proportion of subgraphs in the problem, they calculated a lower bound on the performance.  The small number of relevant subgraphs for three-regular graphs, makes this approach particularly amenable. They found that QAOA $p=1$  will produce a distribution whose average will correspond to at least $0.6924$ times the best-cut. This lower-bound is saturated by triangle-free graphs. These graphs consist of a single subgraph. Therefore, the performance of QAOA $p=1$ is largely dependent of the proportion of edges in this problem that belong to this subgraph. We refer to the performance of a local approach being limited by its performance on one subgraph (or possibly a small handful of subgraphs) as being dominated by this subgraph.

This method was later extended by Braida et al.\  \cite{Bra22} who applied it to QA by using an approach inspired by Lieb-Robinson bounds. By operating QA on short times it can be treated as a local algorithm. By then simulating local sub-graphs (as in the QAOA case) and calculating the bounds outlined in the paper, the worst-case performance for each sub-graph can be calculated.  This approach does not necessarily find a tight bound, but it is nonetheless impressive, with bounds in continuous-time quantum computation a rarity. They show that QA finds at least $0.5933$ times the best cut. The authors conjecture that $0.6963$ times the best cut might be a tighter bound on the performance but are unable to rigorously show that this is the case.

In Sec.\ \ref{sec:3reg} we make use of the approach formulated by Braida et al.\ to prove a lower bound for $H_1$ for this problem.

\subsubsection{Random graph instances}

To generalise the MAX-CUT instances discussed above, we also consider MAX-CUT on randomly generated graphs. These graphs do not have fixed degree. The graphs are generated by selecting an edge between any two nodes with probability $p$. MAX-CUT undergoes a computational phase change for random graphs at $p=1/2$ (harder problems appear for $p>1/2$) \cite{Gam18,Cop03,Pol22}. We set $p=2/3$ for this paper (note that this is no guarantee of hardness for the problem instances considered). 

\subsection{Sherrington-Kirkpatrick inspired model}

In a MAX-CUT problem all the couplers in the graph are set to the same value. Here we introduce a second problem, the Sherrington-Kirkpatrick model (SKM), where this is not the case. The problem is to find the ground-state of 
\begin{equation}
    H_f=\sum_{i,j} J_{i,j} Z_i Z_j 
\end{equation}
where the $J_{i,j}$'s are randomly selected from a normal distribution with mean 0 and variance 1 \cite{She75}. Each qubit is coupled to every other qubit. 

To further distinguish the SKM from the MAX-CUT problems we introduce bias terms to the Hamiltonian,
\begin{equation}
    H_f=\sum_{i,j} J_{i,j} Z_i Z_j+\sum_i^n h_i Z_i,
\end{equation}
where the $h_{i}$'s are also randomly selected from a normal distribution with mean 0 and variance 1. We use the SKM in Appendix \ref{app:skm} to give an indicative idea of the performance of the proposed Hamiltonians on a wider range of problem instances, rather than just MAX-CUT.

\section{Details on the choice of metrics}
\label{app:met}
Once a problem and algorithm have been determined it remains to decide on the metric (or metrics) by which to assess the performance. A common choice is the ground-state probability, $P_{gs}$. This is a reasonable measure for an exact solver. It fails to capture the performance of an approximate solver (a solver that finds \textit{good enough} solutions). A common measure for approximate solvers, such as QAOA, is the approximation ratio. The approximation ratio is a measure on the final distribution produced by the approach. Here we define the approximation ratio to be $\langle H_f \rangle/E_{min}$ where $E_{min}$ is the energy of the ground-state solution and the expectation is with respect to the final state. If the approach finds the ground-state exactly, then $\langle H_f \rangle/E_{min}=1$. Random guessing (for all the problem Hamiltonians considered in Sec.\ \ref{sec:prob} ) has an approximation ratio of $0$. This is distinct from other papers that include terms proportional to the identity in the Hamiltonian. For example, in \cite{Far14} they use terms proportional to the identity meaning random guessing achieves a non-zero approximation-ratio. We make this choice to achieve consistency across different problems. If an algorithm for a specific problem is cited to have a approximation ratio with no explicit reference to time, this refers to an optimised approximation ratio. 

In Appendix \ref{app:skm}, we consider the width of the final energy distribution. Wider distributions may mean the algorithm is harder to optimise in practice. We use\\
$\sigma=\sqrt{\langle H_f^2\rangle-\langle H_f \rangle}/\abs{E_{min}}$ to measure the width of the distribution.

These are by no means all possible metrics to assess the performance of an algorithm. For example time to solution is another popular metric choice \cite{Alb18,Kadowaki_2022}.

\section{Details on numerical work and presentation}
\label{app:num}
\begin{figure}[t]
        \centering
        \includegraphics[width=0.48\textwidth]{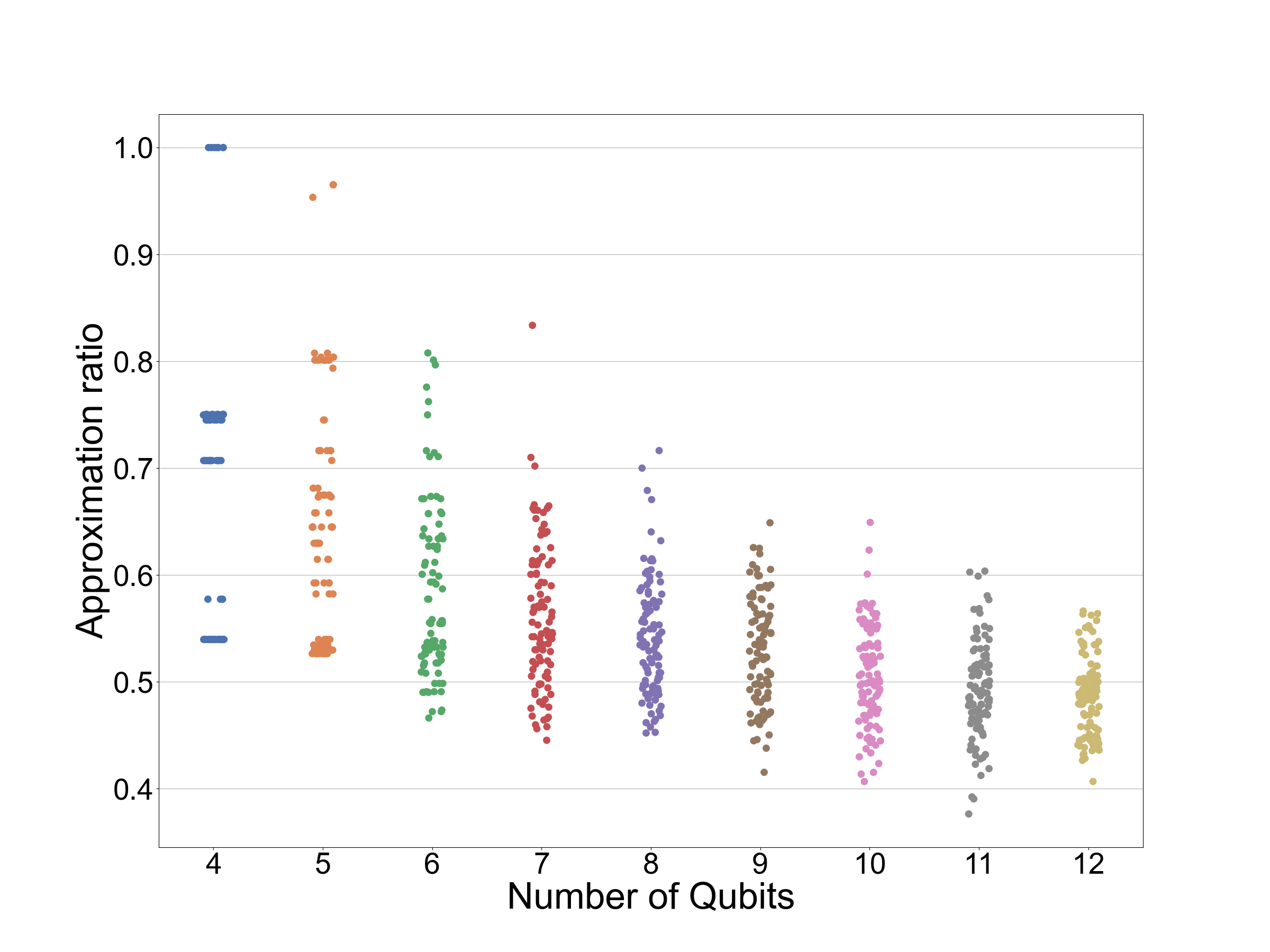}
        \caption{The approximation ratio for MAX-CUT on random graphs (shown as box-plot in Fig.\ \ref{fig:MCRandc}) as a strip-plot, showing the raw data. The horizontal width of the distribution is only to try and minimise overplotting.}
        \label{fig:StripPlot}
\end{figure}

This work makes use of numerical experiments to establish the performance of new approaches. The results are often presented as a box-plot \cite{Hun07, Fre05}.  The central line shows the median. The top and bottom line of the central box shows the lower and upper inter-quartile. Outliers are determined if they are more than 1.5 times more than the inter-quartile range away from the median and denoted by circles. The two caps at the end of the plot show the maximum and minimum data points, excluding outliers. Fig.\ \ref{fig:StripPlot} shows Fig.\ \ref{fig:MCRandc} in terms of the raw-data. The box-plot makes no assumption about the underlying distribution and provides a reasonable representation of the distribution for easy comparison.

As part of the numerics, we need to find the optimum time. For the new methods, this is found by a brute force grid search, dividing the optimal time into 1000 in the time-interval $[0,2\pi]$. For QAOA p=1, the grid was 100 by 100 in the interval $\beta \in [0,\pi]$ and $\gamma \in [0,2\pi]$. By using brute force search we minimise the effect of the classical optimiser on the quantum algorithm.

\section{A further numerical study on a Sherrington-Kirkpatrick inspired model}
\label{app:skm}
To provide further evidence of the applicability of our approach in this section we repeat the numerical experiments performed on MAX-CUT on a  Sherrington-Kirkpatrick inspired model (SKM). The details of the model can be found in Appendix \ref{app:prob_plus}. 

\begin{figure*}
    \centering
    \begin{subfigure}[t]{0.3\textwidth}
        \centering
        \includegraphics[width=\textwidth]{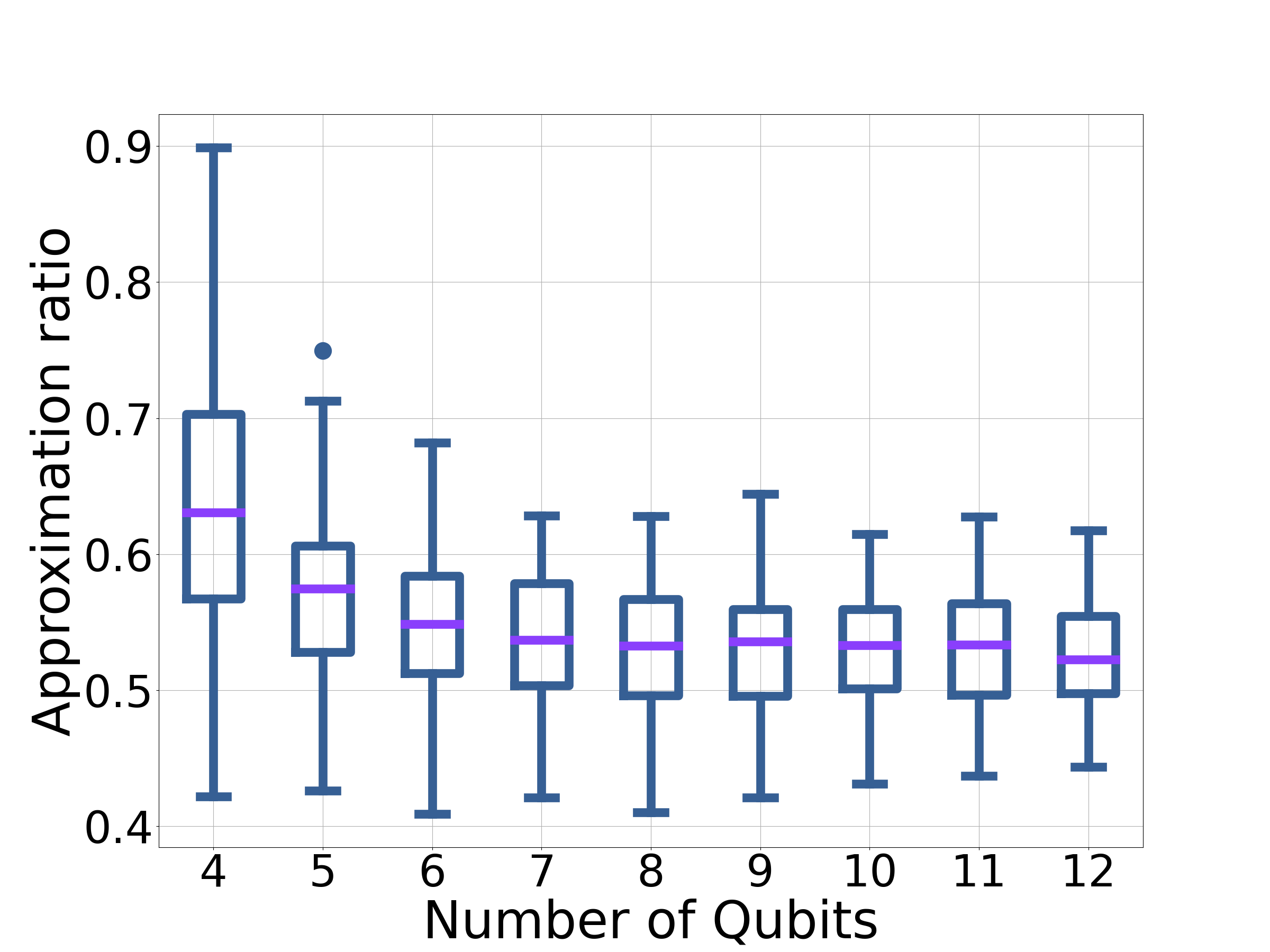}
        \caption{Approximation ratio for the SKM.}
        \label{fig:SKc}
    \end{subfigure}
    \hfill
    \begin{subfigure}[t]{0.3\textwidth}
        \centering
        \includegraphics[width=\textwidth]{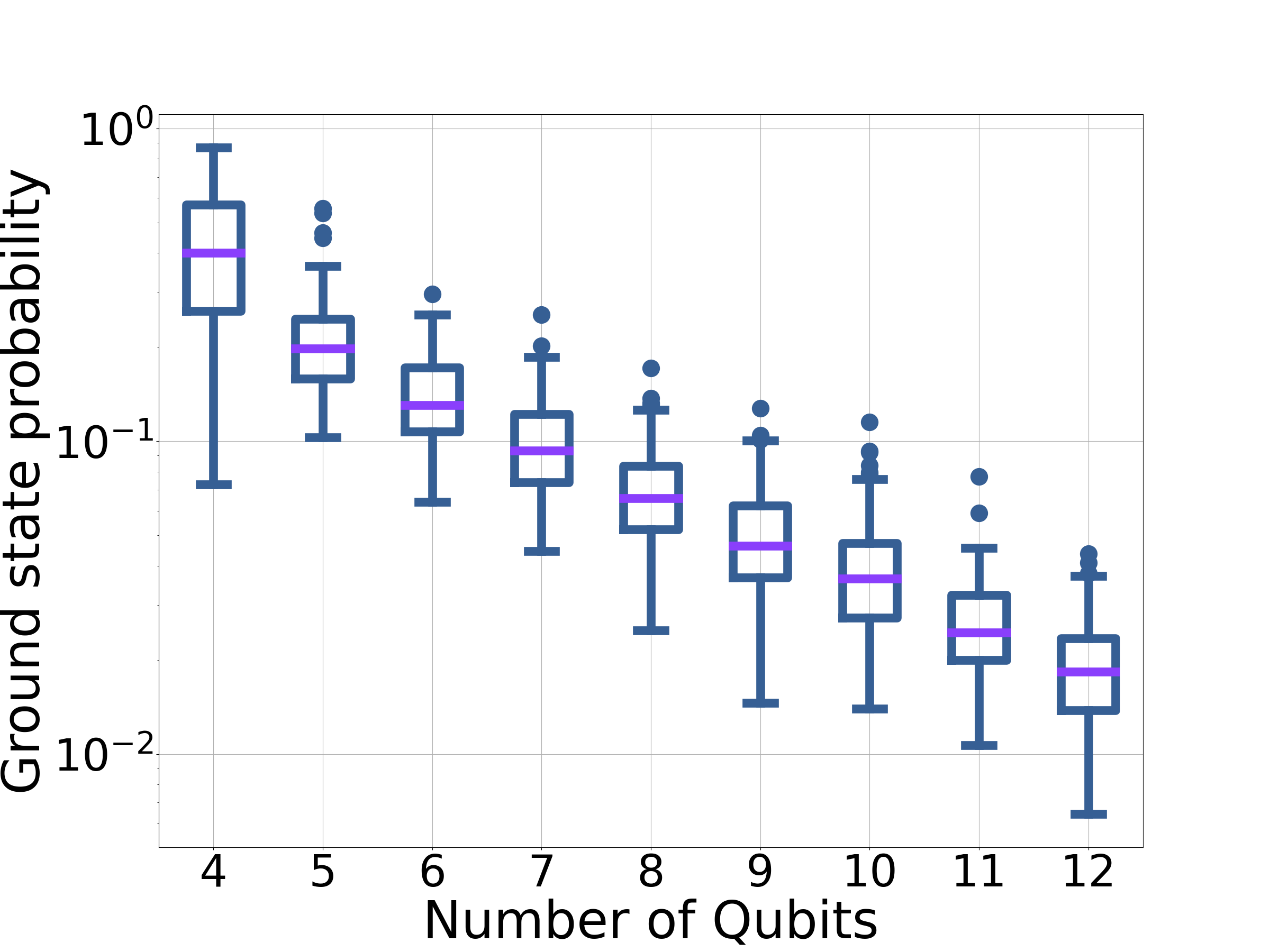}
        \caption{Ground-state probability for the SKM.}
        \label{fig:SKgsp}
    \end{subfigure}
    \hfill
    \begin{subfigure}[t]{0.3\textwidth}
        \centering
        \includegraphics[width=\textwidth]{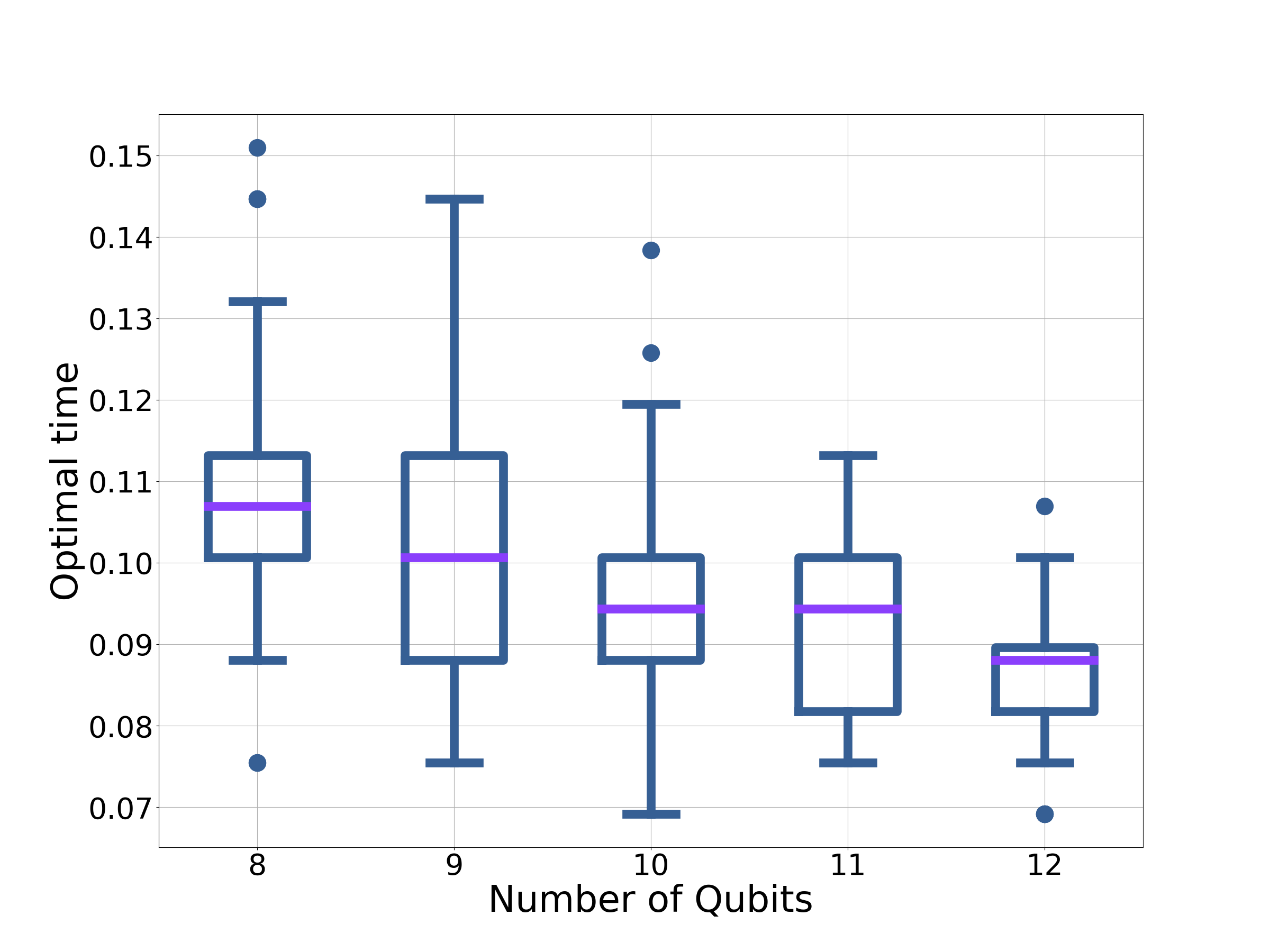}
        \caption{Optimal times for the SKM. The optimal time was found by dividing the interval $[0,2 \pi] $ into 1000 time steps.}
        \label{fig:SKOt}
    \end{subfigure}
    \caption{Performance of $H_1$ on 100 randomly-generated instances of SKM.}
    \label{fig:skmH1}
\end{figure*}

Starting with assessing the performance of $H_1$, Fig.\ \ref{fig:skmH1} shows the performance of $H_1$ on 100 randomly generated instances of the SKM. The approximation ratio appears to have little dependence on the problem size for more than 7 qubits, an indicator that the dynamics under $H_1$ is approximately local for these times. The ground-state probability also appears to decline exponentially (Fig.\ \ref{fig:SKgsp}). The optimal times can be found in Fig.\ \ref{fig:SKOt}. It appears that the optimal time tends to a constant value (or a small range of values), with $T<1$.

\begin{figure}
    \centering
    \begin{subfigure}[t]{0.48\textwidth}
        \centering
        \includegraphics[width=\textwidth]{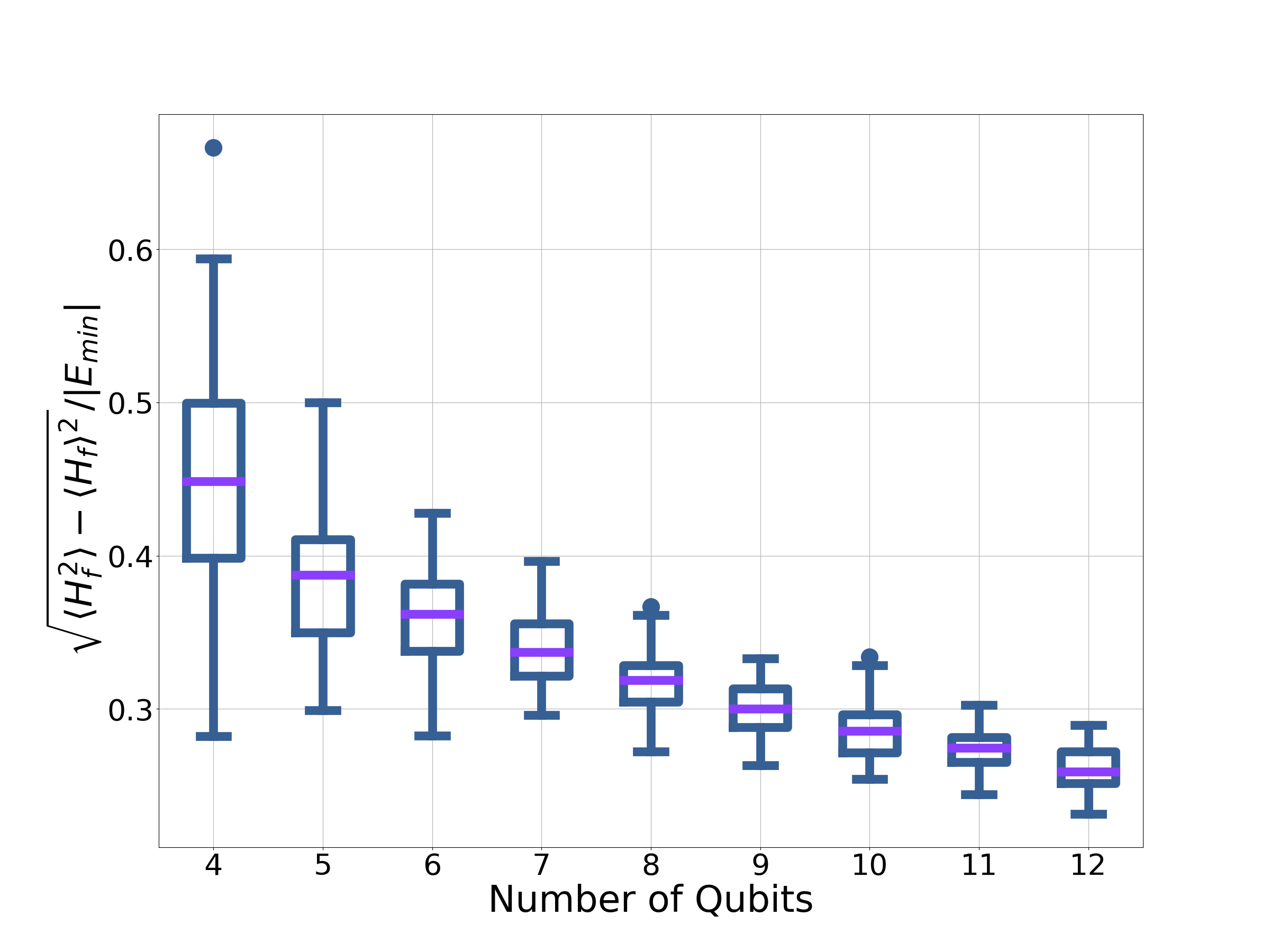}
        \caption{Randomly generated SKM instances.}
        \label{fig:SKVar}
    \end{subfigure}
    \begin{subfigure}[t]{0.48\textwidth}
        \centering
        \includegraphics[width=\textwidth]{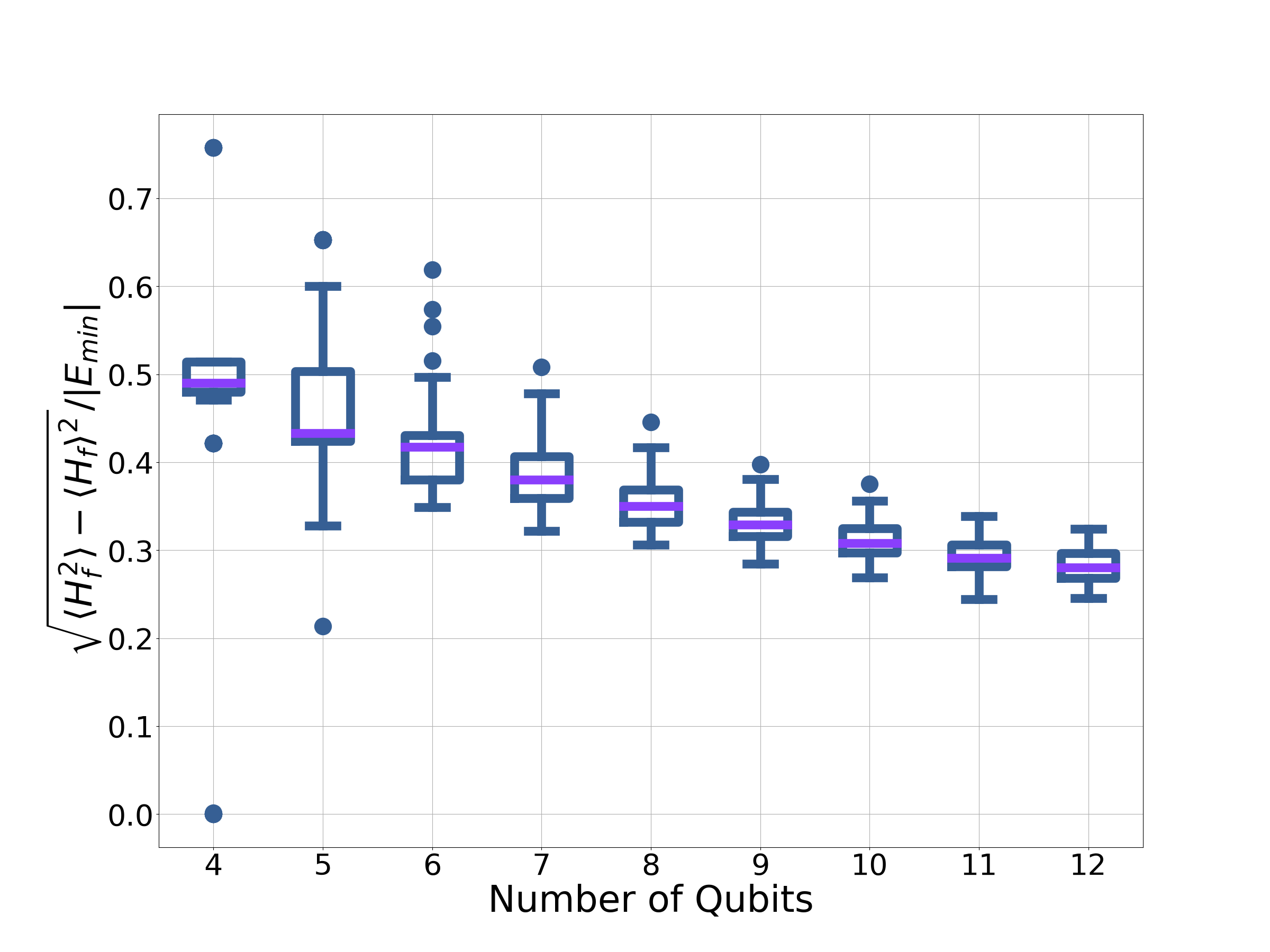}
        \caption{Randomly generated MAX-CUT instances.}
        \label{fig:MCVar}
    \end{subfigure}
\caption{Width of the final distribution, $\sigma$, for randomly generated instances of MAX-CUT and SKM.}
\label{fig:width}
\end{figure}

As mentioned in Appendix \ref{app:met}, knowing the width of the distribution associated with the approximation ratio can also be useful. This is shown for SKM as well as the MAX-CUT instances on randomly generated graphs in Fig.\ \ref{fig:width}. The width, $\sigma$ is non-zero, suggesting the final state is not a computational basis state.

\begin{figure}
    \begin{subfigure}[t]{0.48\textwidth}
        \centering
        \includegraphics[width=\textwidth]{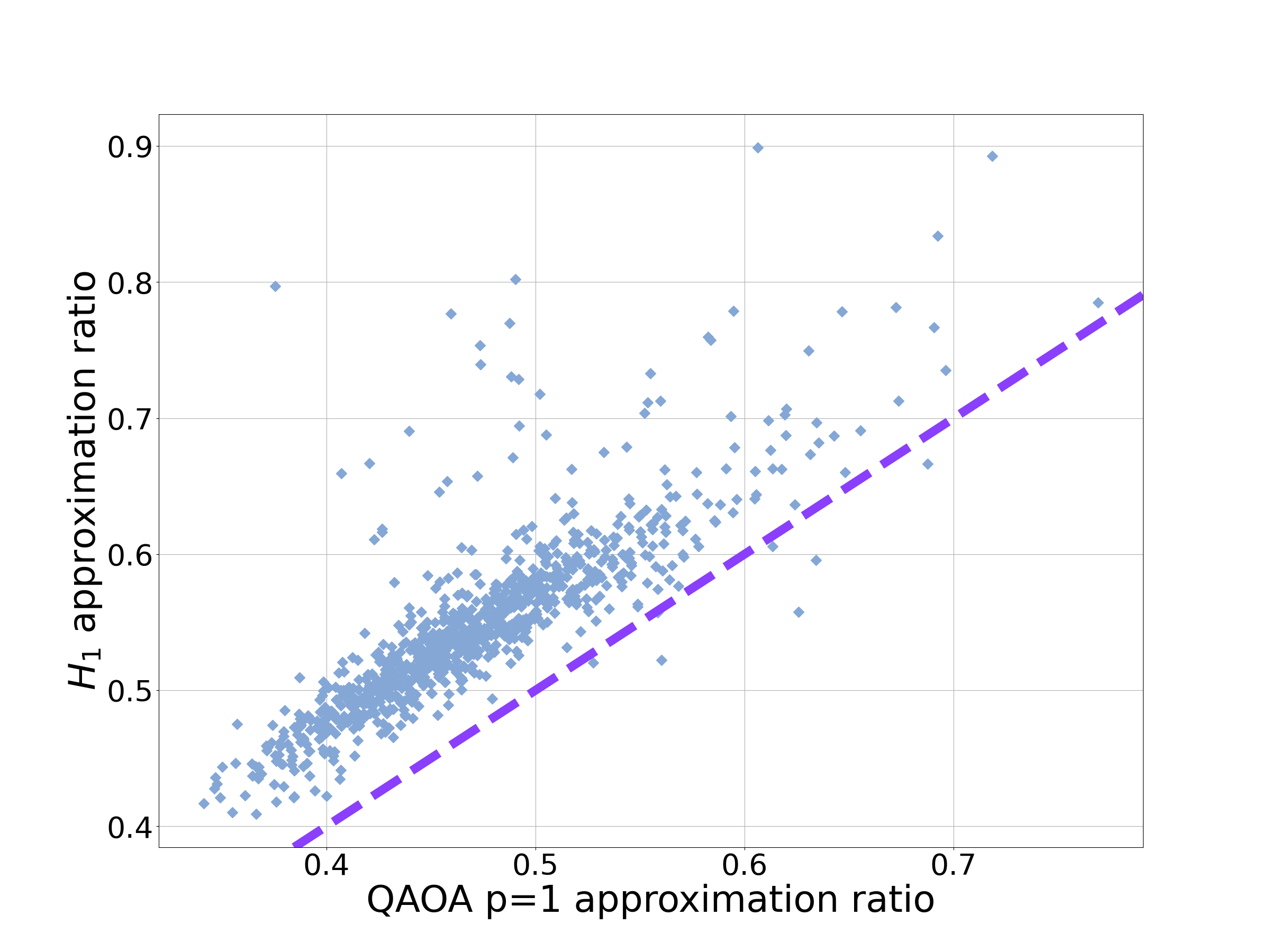}
        \caption{Approximation ratio comparison for the SKM.}
        \label{fig:compskc}
    \end{subfigure}
    \begin{subfigure}[t]{0.48\textwidth}
        \centering
        \includegraphics[width=\textwidth]{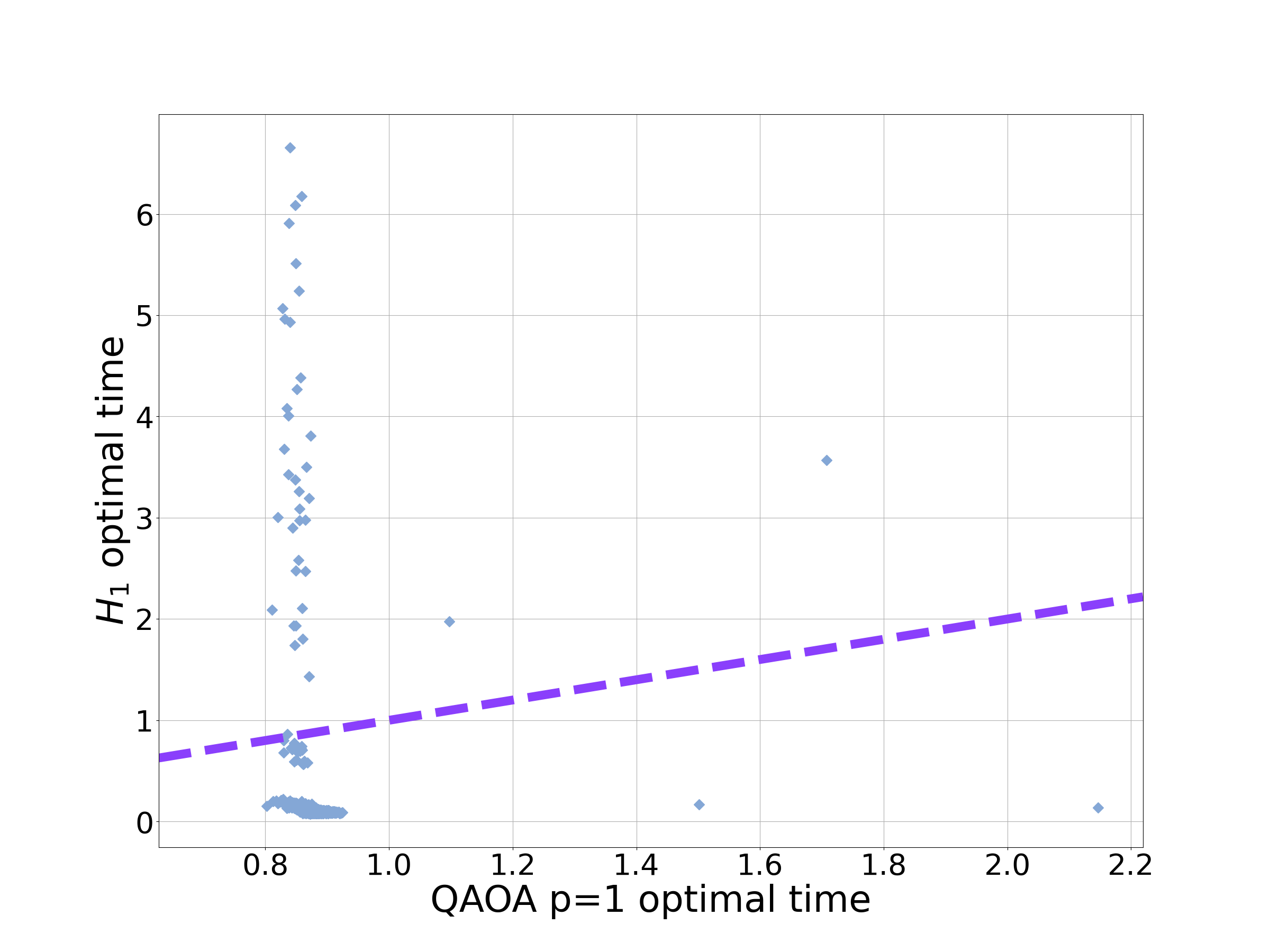}
        \caption{Optimal time comparison for the SKM.}
        \label{fig:compskt}
    \end{subfigure}
    \caption{Comparison of $H_1$ (y-axis on the above plots) with QAOA $p=1$ (x-axis on the above plots). The dashed purple line corresponds to equal performance.}
    \label{fig:compSKMh1QAOA}
\end{figure}

In Fig.\ \ref{fig:compSKMh1QAOA} the performance of $H_1$ is directly compared to QAOA p=1 on 100 instances.  For a handful of problems with the SKM, QAOA $p=1$ outperformed $H_1$, but for the vast majority of problems instances $H_1$ performed better for both approximation ratio and optimal time. In Appendix \ref{app:QAOA_better} we elaborate further on the exceptions.

\begin{figure}
    \begin{subfigure}[t]{0.48\textwidth}
        \centering
        \includegraphics[width=\textwidth]{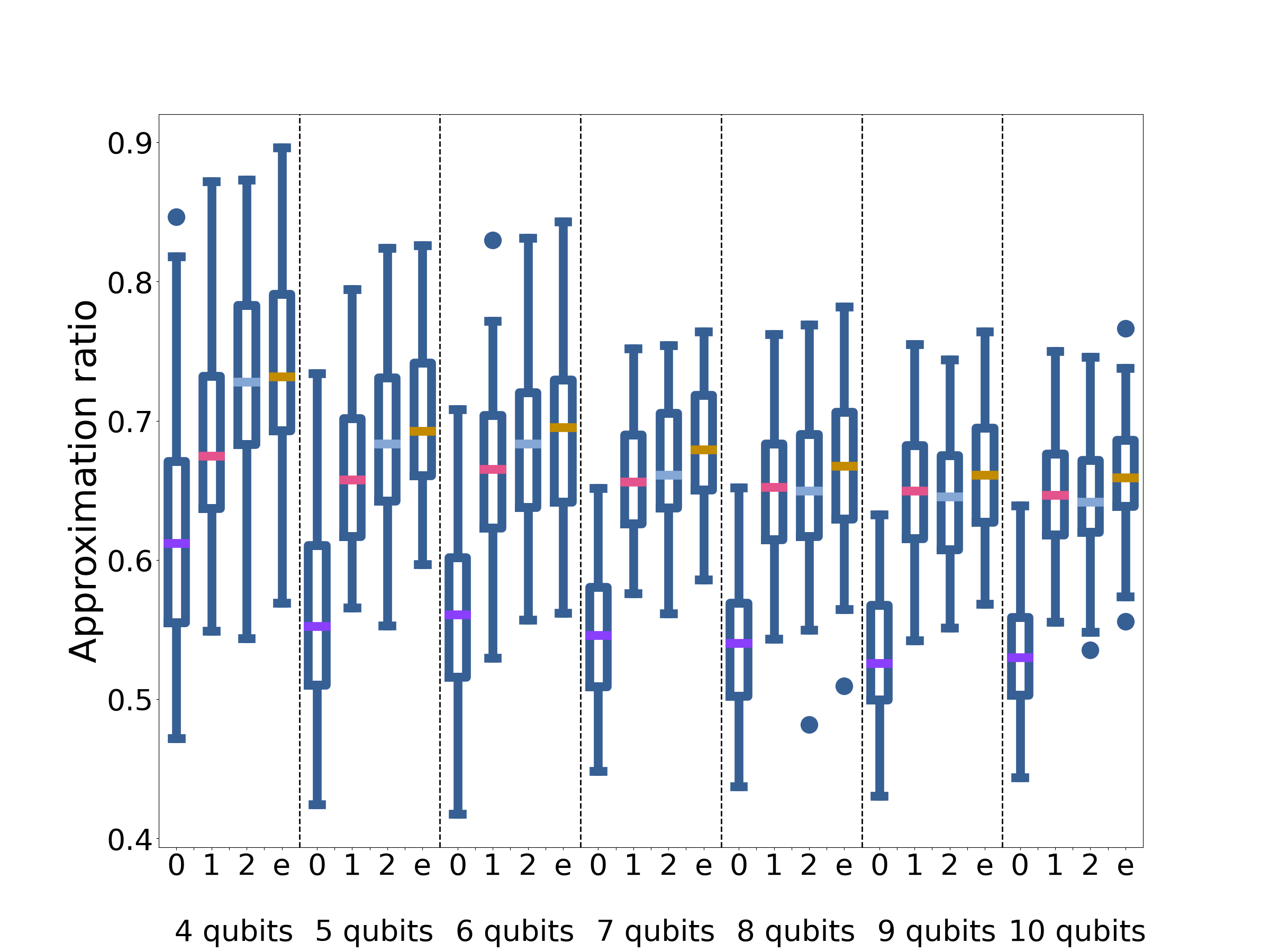}
        \caption{The approximation ratio for the QZ-inspired approach on 100 instances of SKM.}
        \label{fig:qzsk}
    \end{subfigure}
    \begin{subfigure}[t]{0.48\textwidth}
        \centering
        \includegraphics[width=\textwidth]{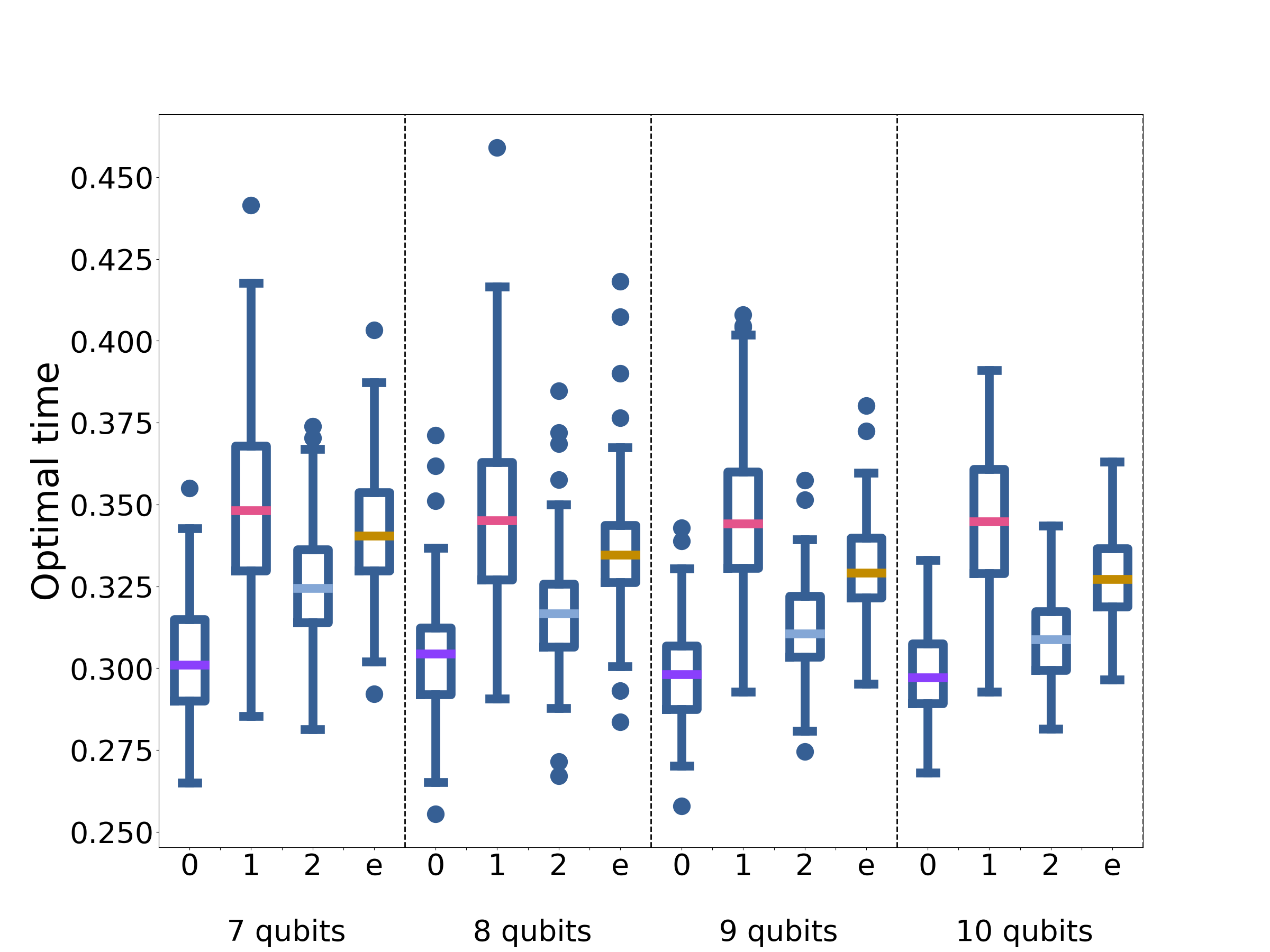}
        \caption{The optimal time for the QZ-inspired approach on 100 instances of the SKM. The norm of each Hamiltonian, for each problem size has been fixed, according to Eq.\ \ref{eq:norm}, ensuring a fair comparison.}
        \label{fig:QZSKt}
    \end{subfigure}
    \caption{Performance on the QZ-inspired approach on 100 instances of the SKM. The x-axis label refers to the order of $T$ in the expansion of Eq.\ \ref{eq:QZexp}, with $0$ being $H_1$ and $e$ referring to the full exponential (i.e.\ Eq.\ \ref{eq:infT}).}
    \label{fig:QZSK}
\end{figure}

Finally, in Fig.\ \ref{fig:QZSK} we assess the performance of the Quantum-Zermello inspired approach for 100 SKM instances. Again, all the QZ-inspired approaches provide an improvement on the original $H_1$ Hamiltonian, indexed by 0 in the figures. Going to first order achieves a substantial improvement as with the MAX-CUT instances. The optimal times for the QZ-inspired approach can be found in Fig.\ \ref{fig:QZSKt} for the SKM. 

In this section we have demonstrated that the approaches inspired by Hamiltonians for optimal state-transfer operate qualitatively similar on SKM as they do on MAX-CUT.

\clearpage
\section{Optimal state-transfer for a single qubit}
\label{app:sq}
This appendix outlines the explicit details for calculating the Hamiltonian for optimal state-transfer for a single qubit. To simplify the calculations, we make use of index notation and the Einstein summation notation convention. For this reason, in this appendix the $k^\text{th}$ Pauli matrix is denoted by $\sigma_k$, with $\sigma_0$ being the identity. 

The first step is to construct traceless Hamiltonians with a trace-norm of one. 
\begin{equation}
    \tilde{H}_i=\frac{H_i-1/2\Tr\left(H_i\right)\sigma_0}{\sqrt{\frac{1}{2}\Tr\left[\left(H_i-1/2\Tr\left(H_i\right)\sigma_0\right)^2\right]}},
\end{equation}
and 
\begin{equation}
    \tilde{H}_f=\frac{H_f-1/2\Tr\left(H_f\right)\sigma_0}{\frac{1}{2}\sqrt{\Tr\left[\left(H_f-1/2\Tr\left(H_f\right)\sigma_0\right)^2\right]}}.
\end{equation}
The Hamiltonians, $\tilde{H}_i$ and $\tilde{H}_f$, have both the same eigenvectors, and ordering in terms of energy, as $H_i$ and $H_f$. Expanding $\tilde{H}_i$ and $\tilde{H}_f$ in terms of Pauli matrices gives: $\tilde{H}_i=\hat{m}\cdot \vec{\sigma}$ and $\tilde{H}_f=\hat{n}\cdot\vec{\sigma}$, where $\vec{\sigma}=(\sigma_x,\sigma_y,\sigma_z)$ and, $\hat{m}$ and $\hat{n}$ are real vectors with Euclidean norm of one. 

The eigenvectors of $\tilde{H}_i$ and $\tilde{H}_f$ correspond to $\pm \hat{m}$ and $\pm \hat{n}$ in the Bloch sphere representation. Ignoring the trivial case, when $\hat{m}$ and $\hat{n}$ are parallel, the two vectors define a plane. The vector $\hat{k}=\hat{m} \cross \hat{n}$ is perpendicular to the plane an, $\hat{k}\cdot\vec{\sigma}$ generates rotations in the $\hat{m}$, $\hat{n}$ plane. Note that:
\begin{align*}
    \frac{1}{2i} \left[\tilde{H}_i,\tilde{H}_f\right] 
    &=\frac{1}{2i} \left[m_i \sigma_i,n_j \sigma_j\right]\\
    &=\frac{1}{2i} m_i n_j \left[\sigma_i,\sigma_j\right]\\
    &= m_i n_j \epsilon_{ijk}\sigma_k\\
    &=\epsilon_{kij} m_i n_j \sigma_k\\
    &=\left(\hat{m}\cross\hat{n}\right)\cdot \vec{\sigma}, 
\end{align*}
where $\epsilon_{ijk}$ is the Levi-Civita tensor \cite{Ril06}.

It is clear that $\left[\tilde{H}_i,\tilde{H}_f\right]/2i$ generates a rotation in the plane spanned by $\hat{m}$ and $\hat{n}$. This result also follows trivially from Eq.\ \ref{eq:optHam} for the single qubit case. The second step is to calculate the angle, $\theta$, between the respective ground states (which is the same as the angle between the excited states). This can be deduced from the overlap between $\tilde{H}_i$ and $\tilde{H}_f$:
\begin{align*}
    \frac{1}{2} \Tr(\tilde{H}_i \tilde{H}_f) &= \frac{1}{2}\Tr(m_i n_j \sigma_i \sigma_j)\\
    &=\frac{1}{2}m_i n_j\Tr(\delta_{ij}I+i\epsilon_{ijk}\sigma_k)\\
    &=m_i n_i\\
    &=\cos(\theta).
\end{align*}

Relating this angle to a time can be done by using the Anandan-Aharonov relation:
\begin{equation}
    \frac{d\theta}{dt}=2 \delta E (t),
\end{equation}
where $\delta E(t) = \sqrt{\bra{\psi(t)}H(t)^2\ket{\psi(t)}-\bra{\psi(t)}H(t)\ket{\psi(t)}^2}$. The Hamiltonian being considered is constant in time. Thus $\delta E (t)$ can be calculated using the initial state (i.e., $\delta E(0) := \delta E$). Hence, we can deduce the time of evolution as:
\begin{equation}
    T=\frac{\arccos{\left[\Tr\left(\tilde{H}_i \tilde{H}_f\right)/2\right]}}{2 \delta E}.
\end{equation}

Finally, the Hamiltonians, $H_0$ and $H_1$, can be replaced by the original Hamiltonians. The time-optimal Hamiltonian for a single-qubit is:
\begin{equation}
    H_{1}=\frac{1}{2i}\left[H_i,H_f\right].
\end{equation}
The corresponding time is:
\begin{equation}
    T=\frac{\arccos{\left[\frac{1}{2}\Tr\left(\tilde{H}_i \tilde{H}_f\right)\right]}}{2 \delta E},
\end{equation}
where $\delta E$ is the uncertainty in energy corresponding to $H_{1}$, and $\tilde{H}_i$, $\tilde{H}_f$ are as defined earlier in this appendix.

It remains an open question as to how much of this geometric intuition in three dimensions can be mapped onto higher dimensional problems. Promisingly, the final operations (commutator, trace) are well defined outside of three dimensions.

\section{Mapping the evolution to free fermions for MAX-CUT on two-regular graphs}
\label{app:ferm}

In this appendix we outline how to utilize the Jordan-Wigner transformation to analyse the performance of $H_1$ for MAX-CUT on two-regular graphs with $n$ qubits. The end goal is to map $H_1=\sum_j Y_jZ_{j+1}+Z_jY_{j+1}$, $H_i=-\sum_j X_j$, and $H_f=\sum_j Z_jZ_{j+1}$ onto fermionic operators. Applying a Fourier Transform then decouples the Hamiltonians into pseudo-spins that can be easily studied. The notation and method closely follows the calculation presented in \cite{Wan18}. 

The first step is to introduce spin-raising and -lowering operators:
\begin{align}
    S_j^{+}=\frac{1}{2}\left(Y_j+iZ_j\right),\\
    S_j^{-}=\frac{1}{2}\left(Y_j-iZ_j\right).
\end{align}

In terms of these new operators:
\begin{align}
    H_i=&\sum_j I-2S_j^+S_j^-,\\
    H_f=&\sum_j S_j^+S_{j+1}^-+S_j^-S^+_{j+1}-S_j^+S_{j+1}^+-S_j^-S_{j+1}^-,\\
    H_1=&2i\sum_j S_j^-S_{j+1}^--S_j^+S_{j+1}^+.
\end{align}

The Jordan-Wigner transformation can now be applied to map these spin operators onto fermionic operators, $a_j$ and $a_j^\dagger$, where:
\begin{align}
    a_j&=S_j^-e^{-i\phi_j},\\
    a_j^\dagger&=S_j^+e^{i\phi_j},
\end{align}
and $\phi_j=\pi\sum_{j'<j}a^\dagger_{j'}a_{j'}$. The fermionic operators obey the standard anti-commutation relationship for fermionic operators (i.e., $\{a^\dagger_j,a_k\}=\delta_{j,k}$). In terms of the fermionic operators the Hamiltonians are:
\begin{equation}
    H_i=\sum_{j=1}^n I-2a_j^\dagger a_j,
\end{equation}

\begin{equation}
    H_f=\sum_{j=1}^{n-1} a^\dagger_ja_{j+1}-a_ja^\dagger_{j+1}-a_j^\dagger a_{j+1}^\dagger+a_ja_{j+1}+G\left(-a_n^\dagger a_1+a_na_1^\dagger +a_n^\dagger a_1^\dagger-a_na_1\right),
\end{equation}
and
\begin{equation}
    H_1=-2i\sum_{j=1}^{n-1}a^\dagger_ja^\dagger_{j+1}+a_ja_{j+1}+2iG\left(a_n^\dagger a_1^\dagger+a_n a_1\right),
\end{equation}
where $G=e^{i\pi\sum_{j=1}^n a^\dagger_ja_j}$. For even $n$, $G=1$ (anti-periodic boundary conditions - ABC) and for odd $n$, $G=-1$ (periodic boundary conditions - PBC). 

Apply a Fourier Transform with appropriate $p$, such that $e^{ipn}=1$ for PBC and $e^{ipn}=-1$ for ABC, such that,
\begin{equation}
    c_p=\frac{1}{\sqrt{n}}\sum_je^{ipj}a_j.
\end{equation}

The Hamiltonians in this new basis are:
\begin{equation}
    H_i=\sum_{k=0}^{n-1}I-2c_k^\dagger c_k
\end{equation}

\begin{equation}
    H_f=2\sum_{k=0}^{\lfloor \frac{n-1}{2} \rfloor} \cos{\theta_k} \left(c_k^\dagger c_{k}+c_{-k}^\dagger c_{-k}\right)+i \sin{\theta_k} \left(c_kc_{-k}+c_k^\dagger c_{-k}^\dagger\right) +H_{f,0}
\end{equation}
and
\begin{equation}
    H_1=4 \sum_{k=0}^{\lfloor \frac{n-1}{2} \rfloor} \sin{\theta_k} \left(c_k^\dagger c_{-k}^\dagger-c_kc_{-k}\right),
\end{equation}
 where for odd $n$:
 \begin{align*}
     & \theta_k=\frac{2 \pi k}{n}\\
     & H_{f,0}=-2c_0^\dagger c_0\\
     & c_{-k}=c_{n-k},
 \end{align*}
 and for even $n$:
 \begin{align*}
     & \theta_k=\frac{\left(2 k +1\right)\pi }{n}\\
     & H_{f,0}=0\\
     & c_{-k}=c_{n-1-k}.
 \end{align*}

These Hamiltonians couple the vacuum state, $\ket{\emptyset}$, with doubly-excited states with opposite momentum, e.g., $c_k^\dagger c_{-k}^\dagger \ket{\emptyset}$. Therefore, we can express the Hamiltonians as pseudo-spins,
\begin{equation}
    H_*=\sum_{k=0}^{\lfloor \frac{n-1}{2} \rfloor}H_{*,k},
\end{equation}
with $*=i,f,1$ and:
\begin{align}
    H_{i,k}&=-2Z\\
    H_{f,k}&=2 \cos \theta_k Z - 2 \sin \theta_k Y\\
    H_{1,k}&=4 \sin \theta_k X,
\end{align}
with the exception of $H_{*,0}$ for $*=i,f$, for odd n, which is half of the above expressions. The observation that $H_{1,k}$ generates rotations in the plane spanned by the eigenvectors of $H_{f,k}$ and $H_{i,k}$ provides further evidence that $H_1$ is capturing some of $H_{opt}$.

The initial state for each pseudo-spin is the ground-state of $-Z$. Calculating the evolution for each pseudo-spin gives:
\begin{equation}
    \langle H_f \rangle=\sum_{k=0}^{\lfloor \frac{n-1}{2} \rfloor} F_k,
\end{equation}

with $F_0=1$ for odd n, otherwise:
\begin{equation}
    F_k=2 \cos{\theta_k} \cos \left(8 \sin \theta_k t\right)-2\sin{\theta_k}\sin{\left(8 \sin{\theta_k}t\right)}.
\end{equation}

The ground state probability is given by:
\begin{equation}
    P_{gs}=\prod_{k=0}^{\lfloor \frac{n-1}{2} \rfloor} G_k,
\end{equation}
again $G_0=1$ for odd n, otherwise: 
\begin{equation}
    G_k=\frac{1}{2}\left(1-\cos{\theta_k} \cos \left(8 \sin \theta_k t\right)+\sin{\theta_k}\sin{\left(8 \sin{\theta_k}t\right)}\right).
\end{equation}

\begin{figure}
    \begin{subfigure}[t]{0.48\textwidth}
    \centering
        \includegraphics[width=\textwidth]{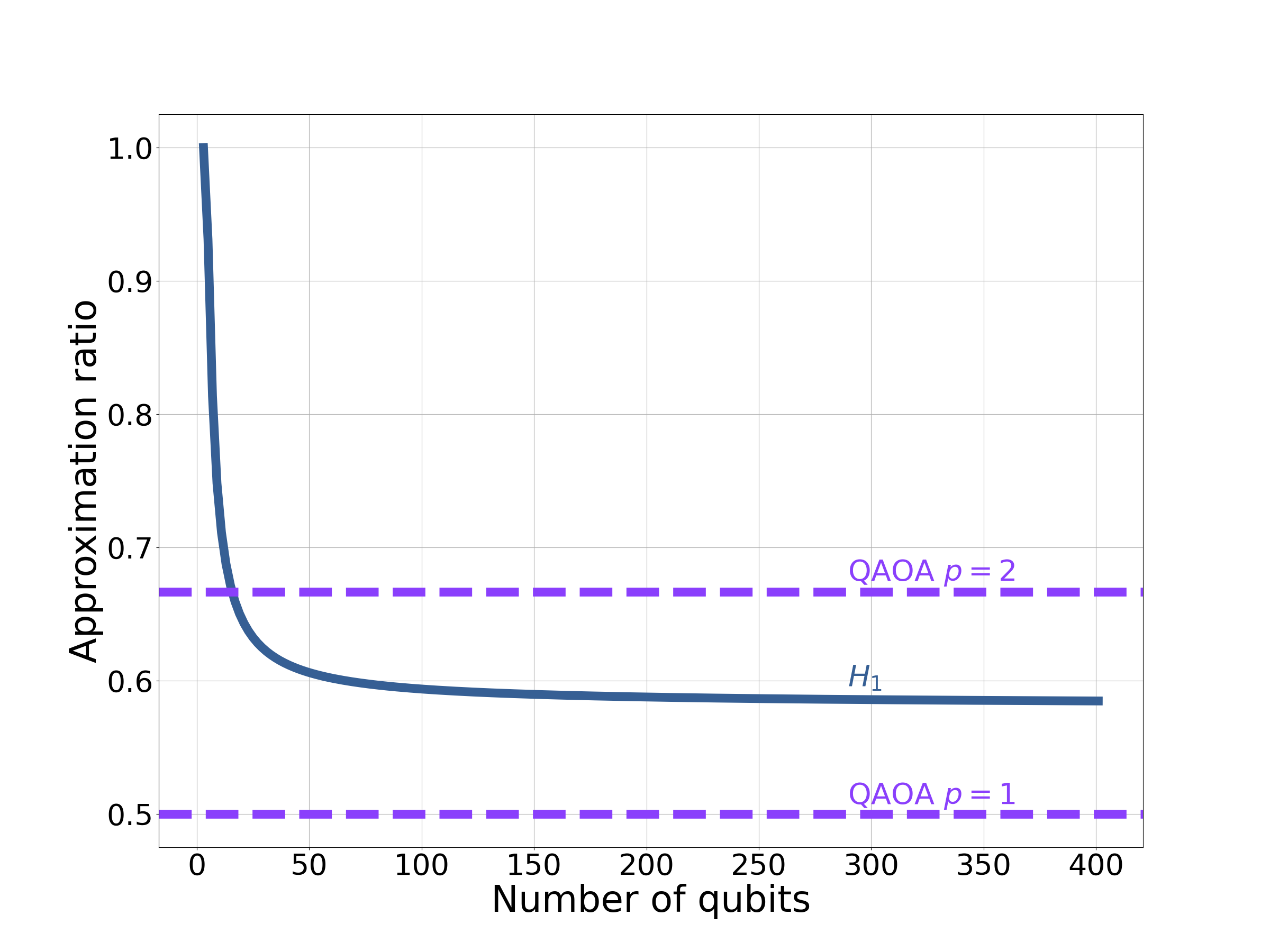}
        \caption{The approximation ratio for $H_1$ on MAX-CUT with two-regular graphs, compared to the performance of QAOA. The approximation ratio freezes out at $0.5830$ - marginally higher than the even case (see Fig.\ \ref{fig:RodeC}). The time also freezes out at $0.2301$ (not shown).}
        \label{fig:rodOdd}
    \end{subfigure}
    \begin{subfigure}[t]{0.48\textwidth}
        \centering
        \includegraphics[width=\textwidth]{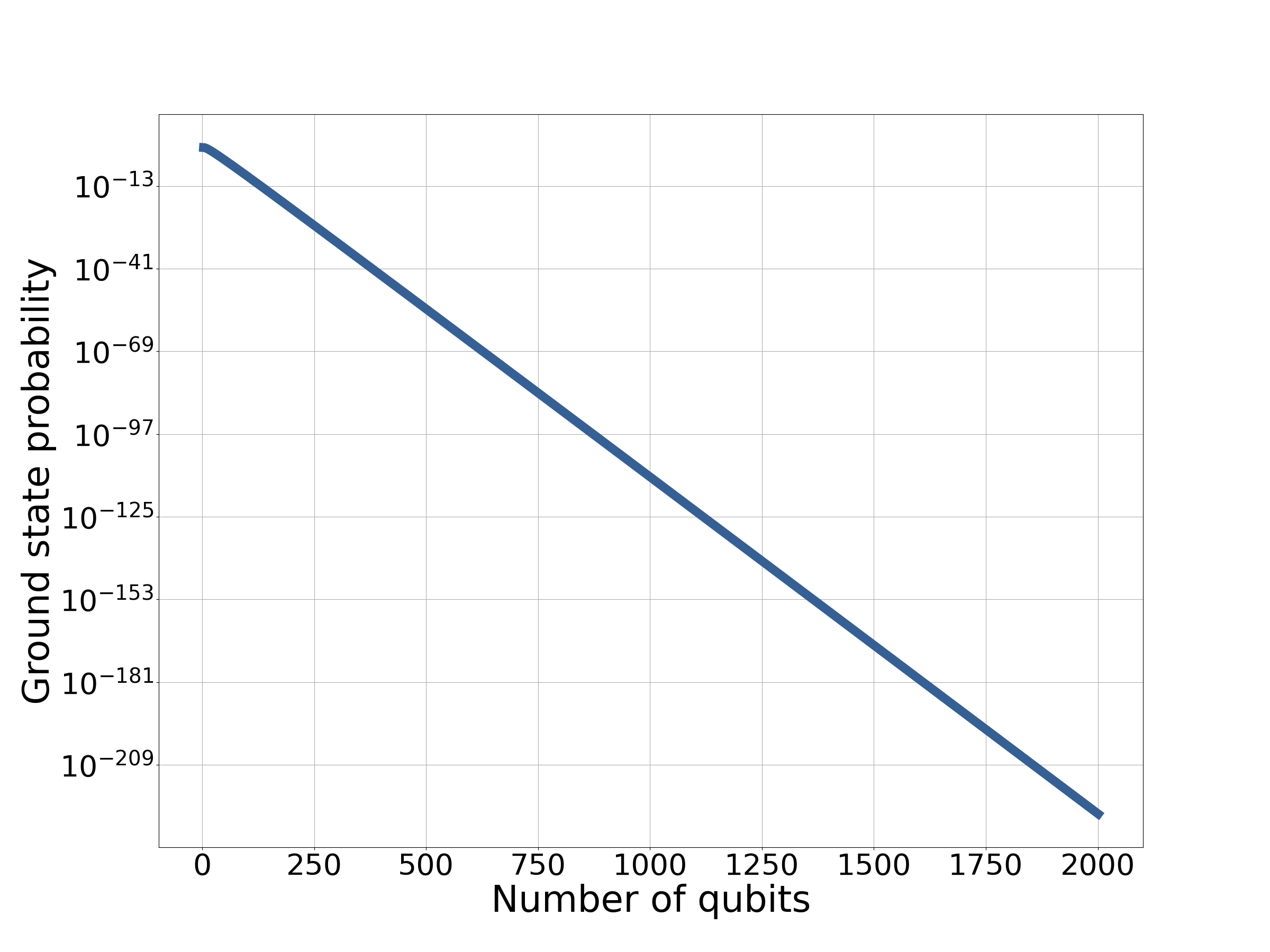}
        \caption{The ground-state probability for different problem sizes under the evolution of $H_1$. The system is measured at times that maximises the approximation ratio. Like the even qubit case (Fig.\ \ref{fig:RoDegsp}), the ground-state probability scales exponentially. }
        \label{fig:rodOddscal}
    \end{subfigure}
    \caption{The performance of $H_1$ on MAX-CUT with two-regular graphs on an odd number of qubits}
\end{figure}

This completes the analytical work used to analyse the performance of $H_1$ for MAX-CUT on two-regular graphs. It remains to find the optimal time to minimise $\langle H_f \rangle$ - this was done numerically. Here we include the plots for an odd number of qubits (Fig.\ \ref{fig:rodOdd} and Fig.\ \ref{fig:rodOddscal}).  

\section{Lieb-Robinson inspired bound}
\label{app:LRB}

In this section we outline how the Lieb-Robinson inspired bound (LRB) from \cite{Bra22} was applied to analysing the performance of $H_1$ on three-regular graphs. 

First, we demonstrate how the bound was derived in \cite{Bra22}. The goal of this approach is to estimate the expectation value of a local observable $O_L$ by simulating part of the system. For this to be a useful estimate it is necessary to quantify the error in doing do, that is to calculate:
\begin{equation}
    \epsilon=\abs{\bra{\psi_i}U^\dagger(t)O_LU(t)\ket{\psi_i}-\bra{\psi_i}U_L^\dagger(t)O_LU_L(t)\ket{\psi_i}},
\end{equation}
where $\ket{\psi_i}$ is the initial state of the system, $U(t)$ is the global unitary evolution of the system and $U_L(t)$ the local unitary we wish to simulate. The error is bounded by
\begin{equation}
    \epsilon\leq \| U^\dagger(t)O_LU(t)-U_L^\dagger(t)O_LU_L(t)\|,
\end{equation}
where $\|\cdot\|$ is the matrix-norm. Adapting the proof from \cite{Bra22} for the case of evolution under a global time-independent Hamiltonian $H$ and local time-independent Hamiltonian $H_L$ gives:
\begin{equation*}
    \left\| U^\dagger(t)O_LU(t)-U_L^\dagger(t)O_LU_L(t)\right\|
    =\left\|\int_0^t ds \frac{d}{ds}
    \left(U^\dagger(s)U_L(s)U_L^\dagger(t)O_LU_L(t)U_L^\dagger(s)U(s)\right) \right\|
\end{equation*}

Substitute in the Schr\"odinger equation for the time derivatives to get: 
\begin{equation*}
    =\left\| \int_0^t ds\
    U^\dagger(s)\left(H-H_L\right)U_L(s)\tilde{O}_L(t)U_L^\dagger(s)U(s) 
     + U^\dagger(s) U_L(s)\tilde{O}_L(t)
     U_L^\dagger(s)\left(H_L-H\right)U(s)\right\|\\,
\end{equation*}
where $\tilde{O}_L(t)=U_L^\dagger(t)O_LU_L(t)$. Tidying this up with $\Delta H=H-H_L$ gives:
\begin{equation}
     =\left\|\int_0^t ds\ U^\dagger(s) \left[\Delta H,U_L(s)\tilde{O}_LU(s)\right] U(s)\right\|
\end{equation}

Using the triangle-inequality,
\begin{equation*}
    \epsilon \leq \int_0^t ds \left\|\left[\Delta H,U_L(s)\tilde{O}_LU(s)\right] \right\|
\end{equation*}

This is essentially the expression (and proof) given in \cite{Bra22}. For Hamiltonians constant in time this can further be tidied up to:
\begin{equation}
\label{eq:LRB}
    \epsilon \leq \int_0^t du \left\|\left[\Delta H,e^{i H_l u}O_Le^{-i H_l u}\right] \right\|
\end{equation}

 The right-hand-side term in the commutator is a local operator depending only on the local system which we wish to simulate, while $\Delta H$ only includes terms in $H$ but not $H_L$. Therefore, in the context of simulating qubits, the only terms in $\Delta H$ that do not commute through are the ones that couple qubits from the local system to qubits outside the system being simulated. 
 
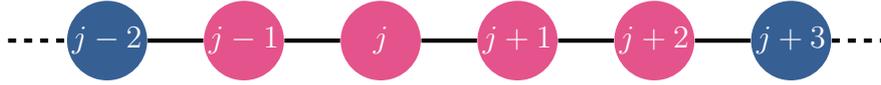
\begin{figure*}
\centering
\begin{tikzpicture}[scale=1.8, auto,swap]
\foreach \pos/\name in { {(-1,0)/j-1}, {(0,0)/j}, {(1,0)/j+1}, {(2,0)/j+2}}
        \node[vertex] (\name) at \pos {$\name$};
        
\foreach \pos/\name in { {(-2,0)/j-2}, {(3,0)/j+3}}
        \node[vertexb] (\name) at \pos {$\name$};

\foreach \pos/\name in { {(-3,0)/j-3}, {(4,0)/j+4}}
        \node[vertexblank] (\name) at \pos {$\name$};
\foreach \source/ \dest  in {j-1/j, j/j+1, j+1/j+2}
        \path[edge] (\source) -- node {}(\dest);
        
\foreach \source/ \dest  in { j-2/j-1, j+2/j+3}
        \path[edge] (\source) -- node {}(\dest);
        
\foreach \source/ \dest  in { j-3/j-2, j+3/j+4}
        \path[dedge] (\source) -- node {}(\dest);

\end{tikzpicture}
\caption{A cartoon for calculating the LRB on MAX-CUT with two-regular graphs. Each node represents a qubit, and each edge the interactions between them. To estimate a local expectation-value, say $Z_jZ_{j+1}$, the local subgraph in pink is simulated. Bounding the error requires $\Delta H$, corresponding to all the interactions leaving the subgraph. In this case this corresponds to all the interactions which connect the blue qubits to the pink ones.}
\label{fig:loc_graph_RD}
\end{figure*}

\begin{figure}
     \centering
     \includegraphics[width=0.48\textwidth]{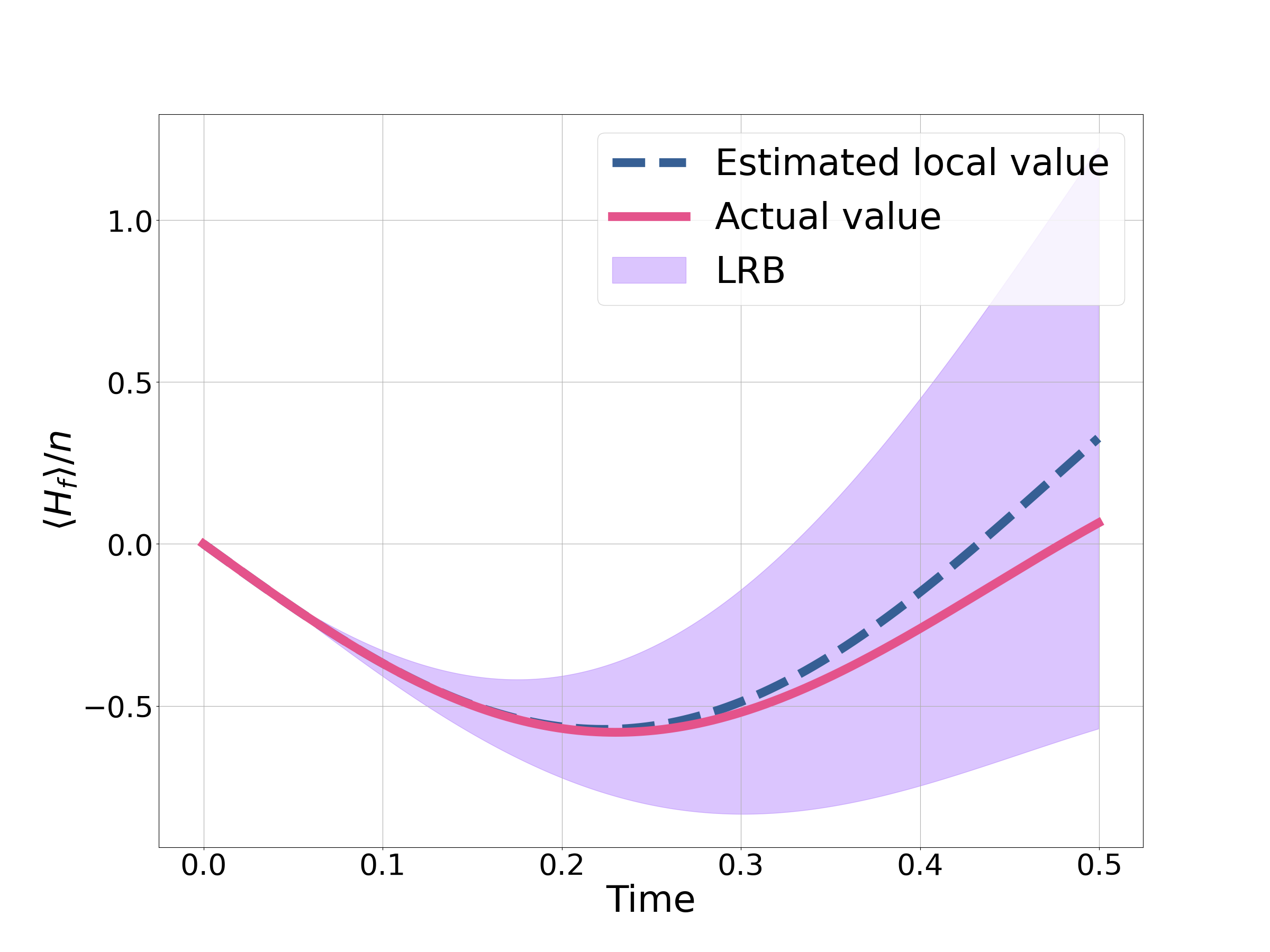}
     \caption{LRB applied to MAX-CUT on two-regular graphs. The dashed blue line shows the local estimate of $\langle Z_j Z_{j+1} \rangle$ and the shaded violet region the allowed region from the LRB. The red line shows the actual value for 400 qubits. They dynamics of $H_1$ closely resembles the local simulation. }
     \label{fig:RoDLRB}
 \end{figure}
 
At this point it is useful to examine a simple system. Take MAX-CUT on two-regular graphs with $H_1$.  We wish to estimate $\langle Z_jZ_{j+1}\rangle$, so we estimate the value by simulating the evolution under $H_1$ for the interactions between the pink qubits shown in Fig.\ \ref{fig:loc_graph_RD}. Choosing a larger subgraph should make the computation more accurate but will also result in a more difficult computation.  Then using Eq.\ \ref{eq:LRB} we can calculate a bound on the error. Here $\Delta H$ corresponds to the interactions between the pink qubits and the blue qubits. Calculating the local estimate with LRB gives Fig.\ \ref{fig:RoDLRB}. The figure also shows the result for the MAX-CUT for a two-regular graph with 400 qubits.

The simulation shows that the bound is not very tight and, except at very short times, is not meaningful. It also demonstrates that $H_1$ behaves in a local fashion, with the full simulation (through the Jordan-Wigner transformation) closely matching the estimate from the local simulation. The LRB would be the lowest upper-bound shown in Fig.\ \ref{fig:RoDLRB} 
 
Having established the idea behind the LRB, we now apply it to find the performance of $H_1$ on three-regular graphs. To do this we use 
\begin{equation}
    H_f=\sum_{(i,j)\in E}Z_iZ_j
\end{equation}
for the graph (or subgraph) $G=(V,E)$ being considered. The minimum value of $\langle H_f \rangle$ is then mapped onto the length of the cut for easier comparison with QAOA and QA.
 
The performance of $H_1$ was determined by looking at the three local sub-graphs that can be found in \cite{Far14,Bra22}. The error was calculated, taking the worst-case scenario, where each sub-graph has the maximum number of interactions exiting the sub-graph. Using the relative ratios of the sub-graphs \cite{Far14}, a worst case performance can be calculated. The numerical details can be found in Tab.\ \ref{tab:LRB}.

\begin{table}
\centering
\begin{tabular}{c | c c c }

& Subgraph 1 & Subgraph 2 & Subgraph 3 \\ 
\hline 
& 

\begin{tikzpicture}[scale=1, auto,swap]
\node[vertexblank] (n) at (-0.5, 1.8) {};
\node[vertex_mt] (a) at (-1,0) {$i$};
\node[vertex_mt] (b) at (1,0) {$j$};

\foreach \pos/\name in {{(0,1)/c}, {(0,-1)/d}}
        \node[vertex_mt] (\name) at \pos {$*$};

\foreach \source/ \dest  in {a/b, a/c, a/d, c/b, d/b}
        \path[edge] (\source) -- node {}(\dest);

\end{tikzpicture}

& 
\begin{tikzpicture}[scale=1, auto,swap]
\vspace{0.3 mm}
\node[vertex_mt] (a) at (-1,0) {$i$};
\node[vertex_mt] (b) at (1,0) {$j$};
\foreach \pos/\name in {{(0,1)/c}, {(-2,-1)/d},{(2,-1)/e}}
        \node[vertex_mt] (\name) at \pos {$*$};

\foreach \source/ \dest  in {a/b, a/c, a/d, c/b, e/b}
        \path[edge] (\source) -- node {}(\dest);

\end{tikzpicture}
& 
\begin{tikzpicture}[scale=1, auto,swap]
\vspace{0.3 mm}
\node[vertex_mt] (c) at (-0.5,0) {$i$};
\node[vertex_mt] (d) at (0.5,0) {$j$};
\foreach \pos/\name in { {(-1.5,1)/a}, {(-1.5,-1)/b}, {(1.5,1)/e},{(1.5,-1)/f}}
        \node[vertex_mt] (\name) at \pos {$*$};

\foreach \source/ \dest  in {a/c, b/c, c/d, d/e, d/f}
        \path[edge] (\source) -- node {}(\dest);

\end{tikzpicture} \\ 
Local estimate of $Z_iZ_j$ & -0.2056 & -0.2676 & -0.3377 \\ 
Upper estimate from LRB & -0.1333 & -0.1652 & -0.2007\\

Cut value & 0.5666 & 0.5826 & 0.6003 \\

\end{tabular}
\caption{Numerical details for the LRB applied to $H_1$ on MAX-CUT with three-regular graphs. Each column shows, from top to bottom, the local subgraph being simulated; the local estimate of $Z_i Z_j$ to be minimised; the corresponding worst case from the LRB; and the corresponding cut value for this worst case. All of these values are taken at the optimised time of $0.093$.}
\label{tab:LRB}
\end{table}

\begin{figure}
     \centering
     \includegraphics[width=0.48\textwidth]{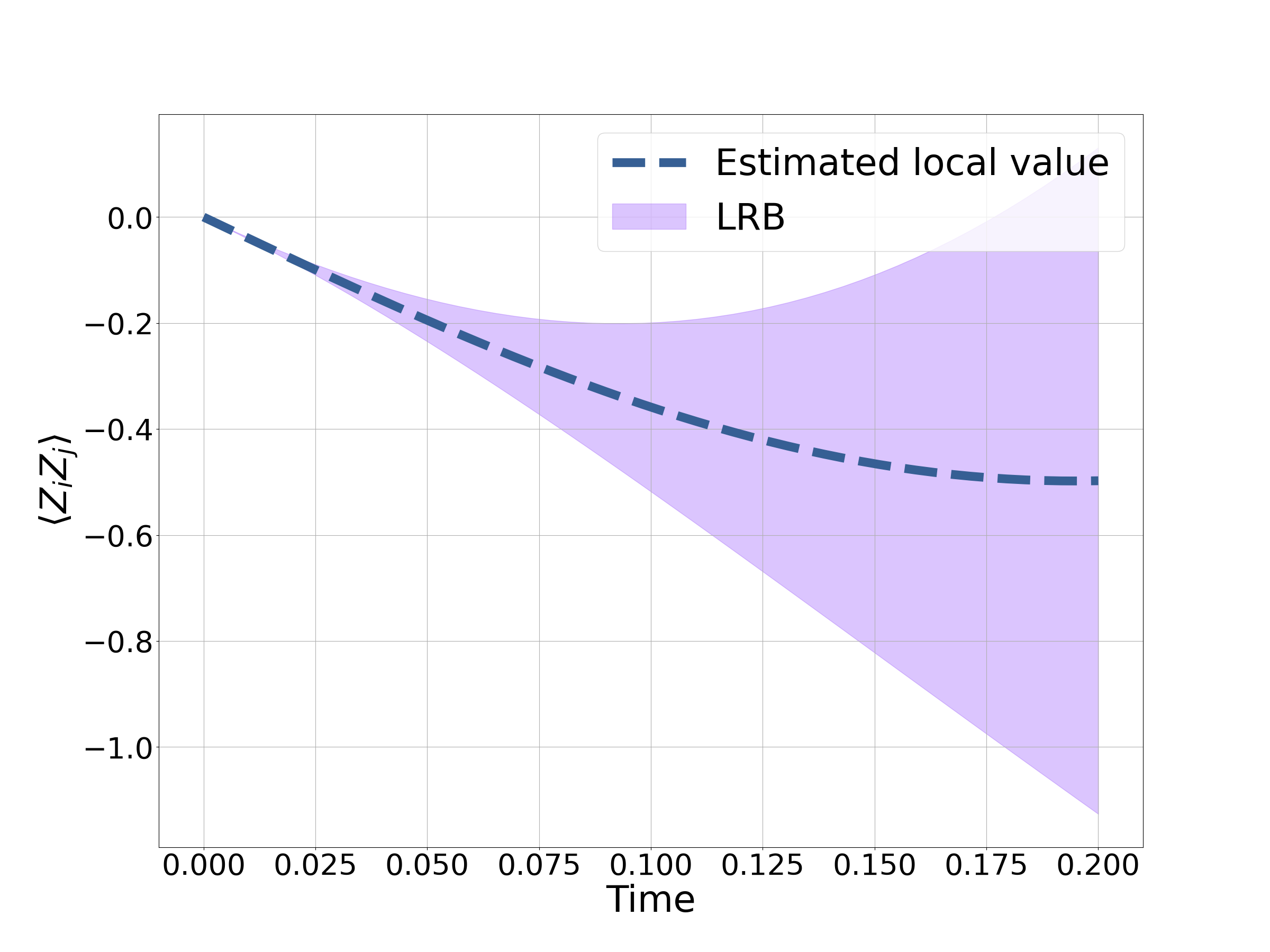}
     \caption{The Lieb-Robinson inspired bound for subgraph 3 in Table \ref{tab:LRB}. This subgraph dominates the worst-case bound. The LRB has a minimum at around a time of  $0.1$ while the locally estimated value has a minimum at around a time of $0.2$.}
     \label{fig:sg3LRB}
\end{figure}

Similarly to QAOA and QA \cite{Bra22}, the LRB approach suggests that $H_1$ struggles the most with triangle-free graphs. The LRB, with local estimate, for the triangle-free subgraph can be seen in Fig.\ \ref{fig:sg3LRB}. 

The LRB is taken at a time of 0.093, while the minimum for the local estimate occurs at around 0.2. Hence the LRB is sampling far from what is optimal for the local graph. Since we know the bound is not very tight, it is reasonable to assume that the worst-case performance of $H_1$ is actually better than the LRB and occurs at a later time (around $0.2$).

\section{Direct comparison of QAOA p=1 with three-regular graph}
\label{app:regQAOA}

\begin{figure}
    \centering
    \begin{subfigure}[t]{0.48\textwidth}
        \centering
        \includegraphics[width=\textwidth]{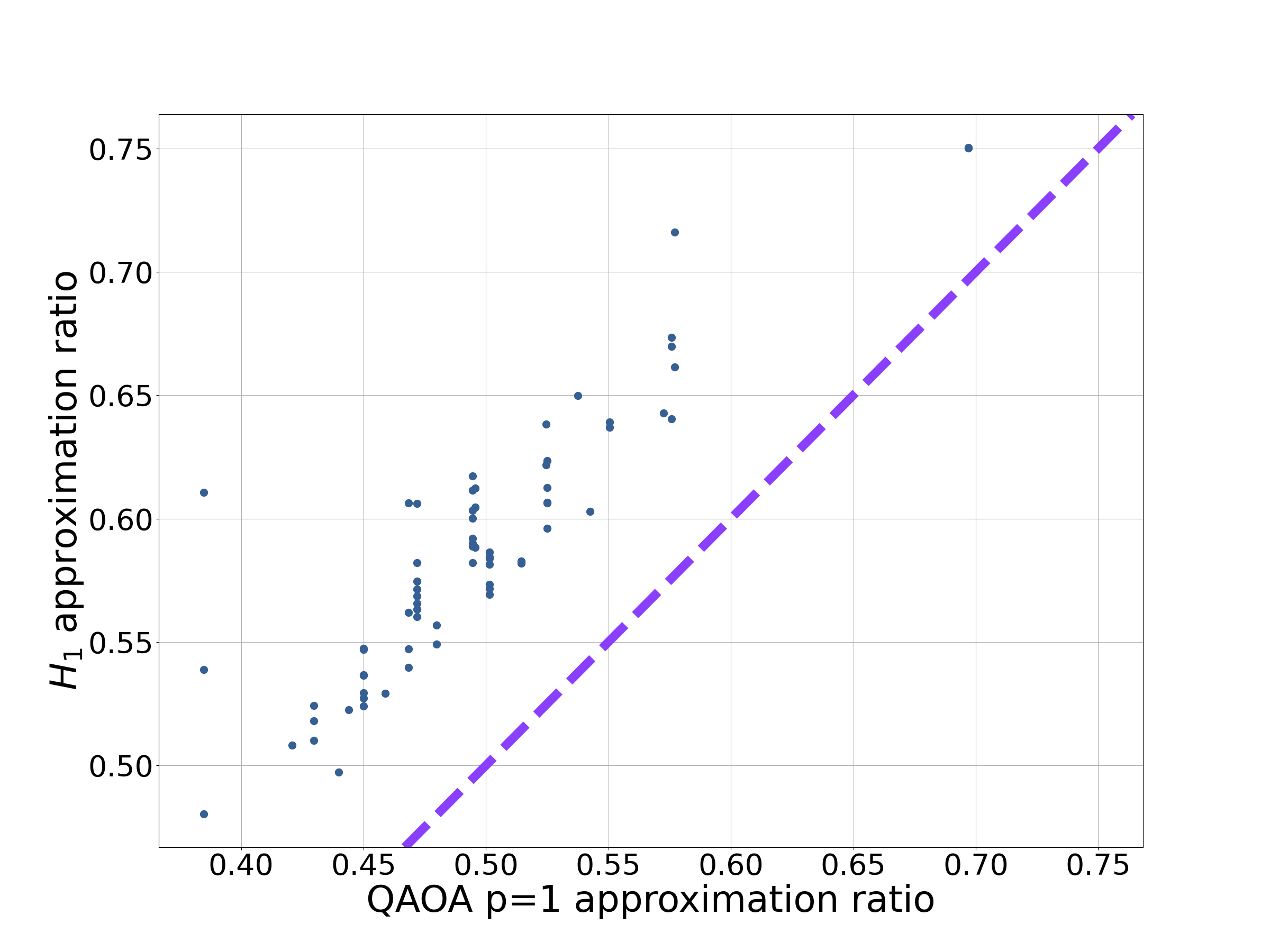}
        \caption{Approximation ratio}
        \label{fig:comp3regc}
    \end{subfigure}
    \begin{subfigure}[t]{0.48\textwidth}
        \centering
        \includegraphics[width=\textwidth]{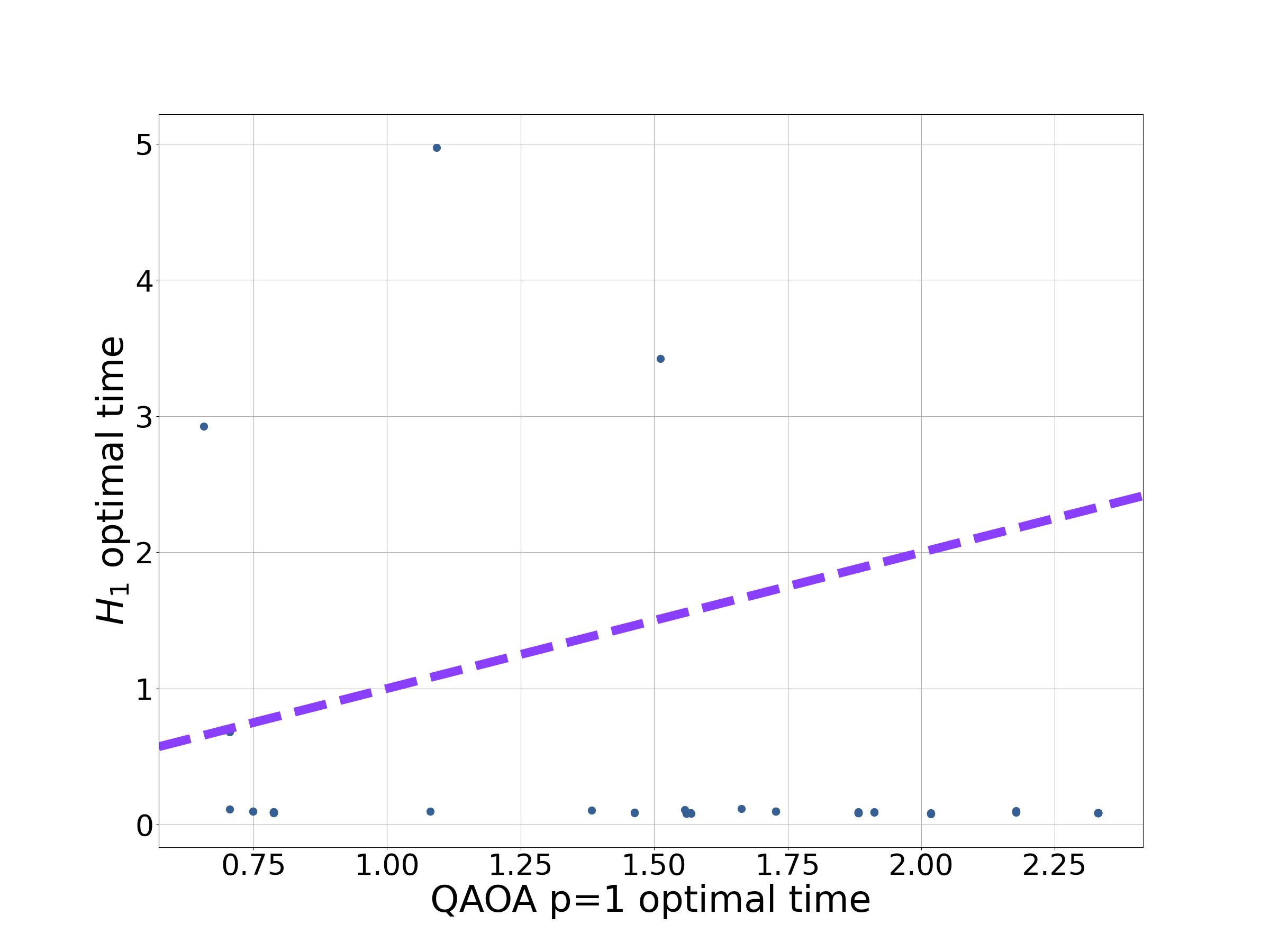}
        \caption{Optimal time comparison for MAX-CUT}
        \label{fig:comp3regt}
    \end{subfigure}
    \caption{Comparison of $H_1$ (y-axis on the above plots) with QAOA $p=1$ (x-axis on the above plots) for three-regular graphs. The dashed purple line shows QAOA and $H_1$ performing the same.}
    \label{fig:comp3h1QAOA}
\end{figure}

In Sec.\ \ref{sec:3reg} we compared $H_1$ to the lower bound of QAOA p=1. In Fig.\ \ref{fig:comp3h1QAOA} we make a direct comparison for the problem instances. For all instances $H_1$ provides a better approximation ratio and in the vast majority of instances in a shorter optimal time. Details on the outliers can be found in Appendix \ref{app:QAOA_better}.

\newpage
\section{Instances for which QAOA p=1 outperforms \texorpdfstring{$H_1$}{H1}}
\label{app:QAOA_better}

\begin{table}
\centering
\begin{tabular}{c c c c }

Graph 1 & Graph 2 & Graph 3 & Graph 4 \\ 
\hline

\begin{tikzpicture}[scale=0.75, auto,swap]
\node[vertex_small] (a) at (-1,-1) {};
\node[vertex_small] (b) at (-1,1) {};
\node[vertex_small] (c) at (1,-1) {};
\node[vertex_small] (d) at (1,1) {};

\foreach \source/ \dest  in {a/b, a/c, a/d, b/c, b/d, c/d}
        \path[edge] (\source) -- node {}(\dest);

\end{tikzpicture}

& 
\begin{tikzpicture}[scale=0.5, auto,swap]

\node[vertex_small] (a) at (-1,-1) {};
\node[vertex_small] (b) at (-1,1) {};
\node[vertex_small] (c) at (1,1) {};
\node[vertex_small] (d) at (1,-1) {};
\node[vertex_small] (e) at (0,-2) {};
\node[vertex_small] (f) at (0,2) {};

\foreach \source/ \dest  in {a/b, b/d,c/d, a/c, a/e,d/e,e/f,b/f,c/f}
        \path[edge] (\source) -- node {}(\dest);

\end{tikzpicture}
& 
\begin{tikzpicture}[scale=1, auto,swap]

\node[vertex_small] (a) at (-1,-1){};
\node[vertex_small] (b) at (-1,0){};
\node[vertex_small] (c) at (0,0) {};
\node[vertex_small] (d) at (0,-1){};

\node[vertex_small] (e) at (-1/2,-1/2) {};
\node[vertex_small] (f) at (-1/2,1/2) {};
\node[vertex_small] (g) at (1/2,1/2) {};
\node[vertex_small] (h) at (1/2,-1/2) {};

\foreach \source/ \dest  in {a/b, b/c, c/d, d/a, e/f, f/g, g/h, h/e, a/e, b/f,c/g,d/h}
        \path[edge] (\source) -- node {}(\dest);

\end{tikzpicture}
&
\begin{tikzpicture}[scale=1/2, auto,swap]
\node[vertex_small] (a) at (-1,-1) {};
\node[vertex_small] (b) at (-1,1) {};
\node[vertex_small] (c) at (1,-1) {};
\node[vertex_small] (d) at (1,1) {};

\node[vertex_small] (e) at (1,2) {};
\node[vertex_small] (f) at (1,4) {};
\node[vertex_small] (g) at (3,2) {};
\node[vertex_small] (h) at (3,4) {};

\foreach \source/ \dest  in {a/b, a/c, a/d, b/c, b/d, c/d, e/f,e/g,e/h,f/g,f/h,g/h}
        \path[edge] (\source) -- node {}(\dest);

\end{tikzpicture}
\\

\end{tabular}
\caption{The exceptions for the three-regular graphs.}
\label{tab:3regexcepgraphs}
\end{table}

\begin{figure}
    \centering
    \begin{subfigure}{0.48\textwidth}
        \centering
        \includegraphics[width=\textwidth]{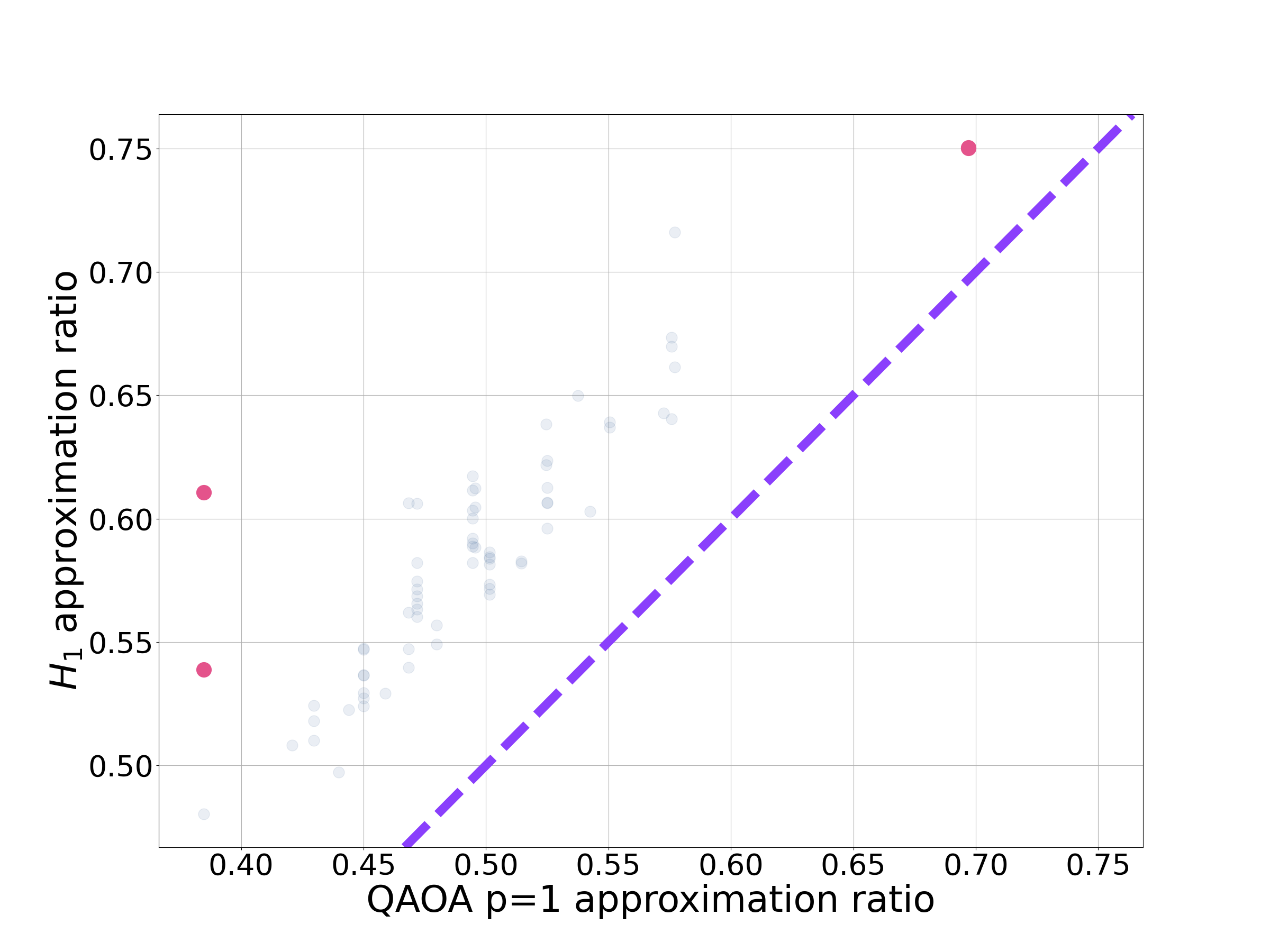}
        \caption{Approximation ratio}
    \end{subfigure}
    \begin{subfigure}{0.48\textwidth}
        \centering
        \includegraphics[width=\textwidth]{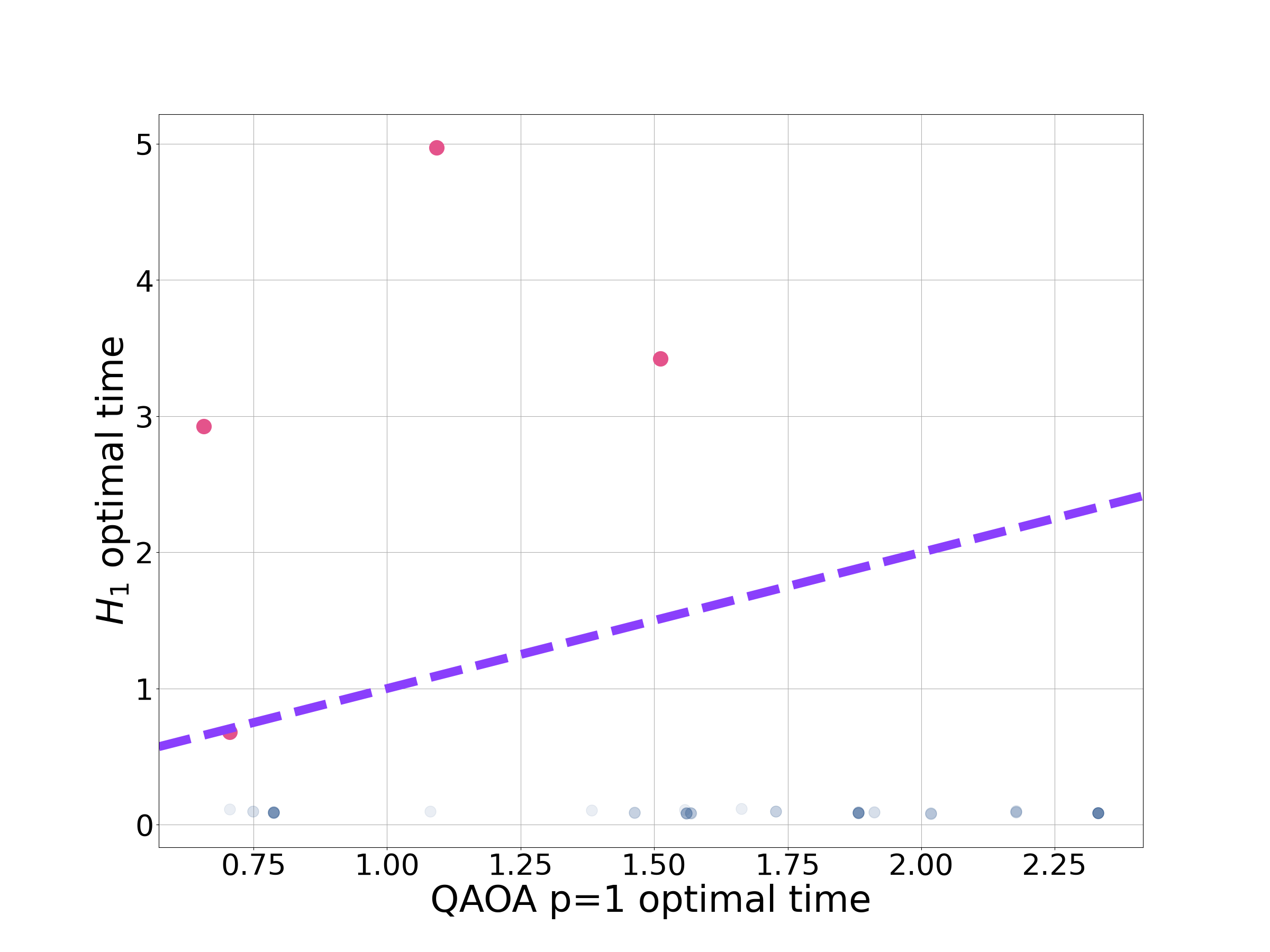}
        \caption{Optimal time.}
    \end{subfigure}
    \caption{The performance of $H_1$ compared to QAOA p=1 on MAX-CUT with three-regular graphs. The instances with atypical run-times for $H_1$ are highlighted in pink. The rest of the data has been faded for clarity and to show how the remaining data is clustered, with darker regions corresponding to more data points. The dashed purple line shows both approaches performing the same.}
    \label{fig:3reghighlight}
\end{figure}

The aim of this section is to explore the instances where QAOA p=1 has a shorter run-time than $H_1$ and/or provides a better approximation ratio, as shown in Fig.\ \ref{fig:NumPer}. The first thing to note is that these tend to be the exception, rather than the rule.

\begin{figure}
    \centering
    \begin{subfigure}{0.48\textwidth}
        \centering
        \includegraphics[width=\textwidth]{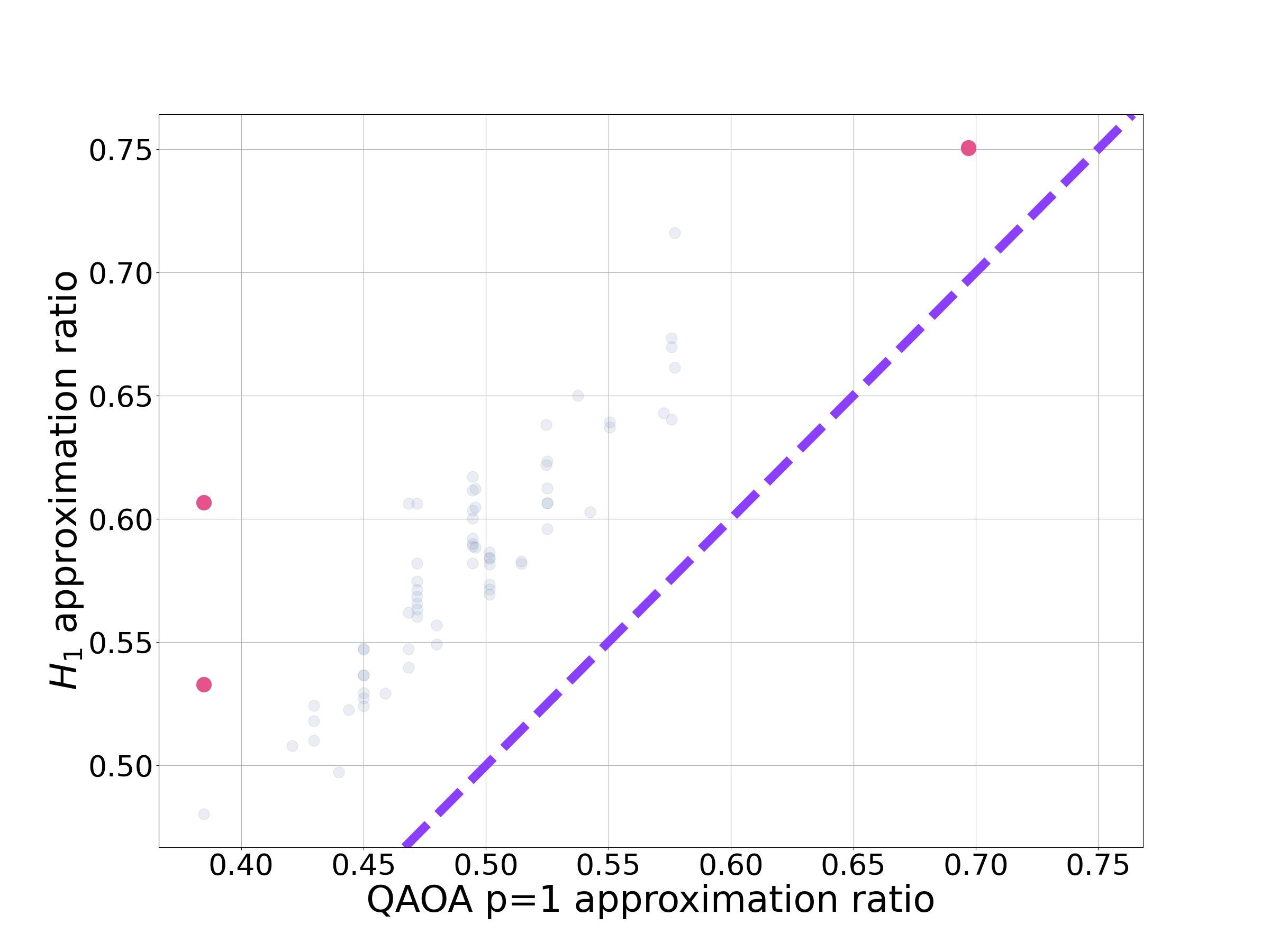}
        \caption{Approximation ratio}
    \end{subfigure}
    \begin{subfigure}{0.48\textwidth}
        \centering
        \includegraphics[width=\textwidth]{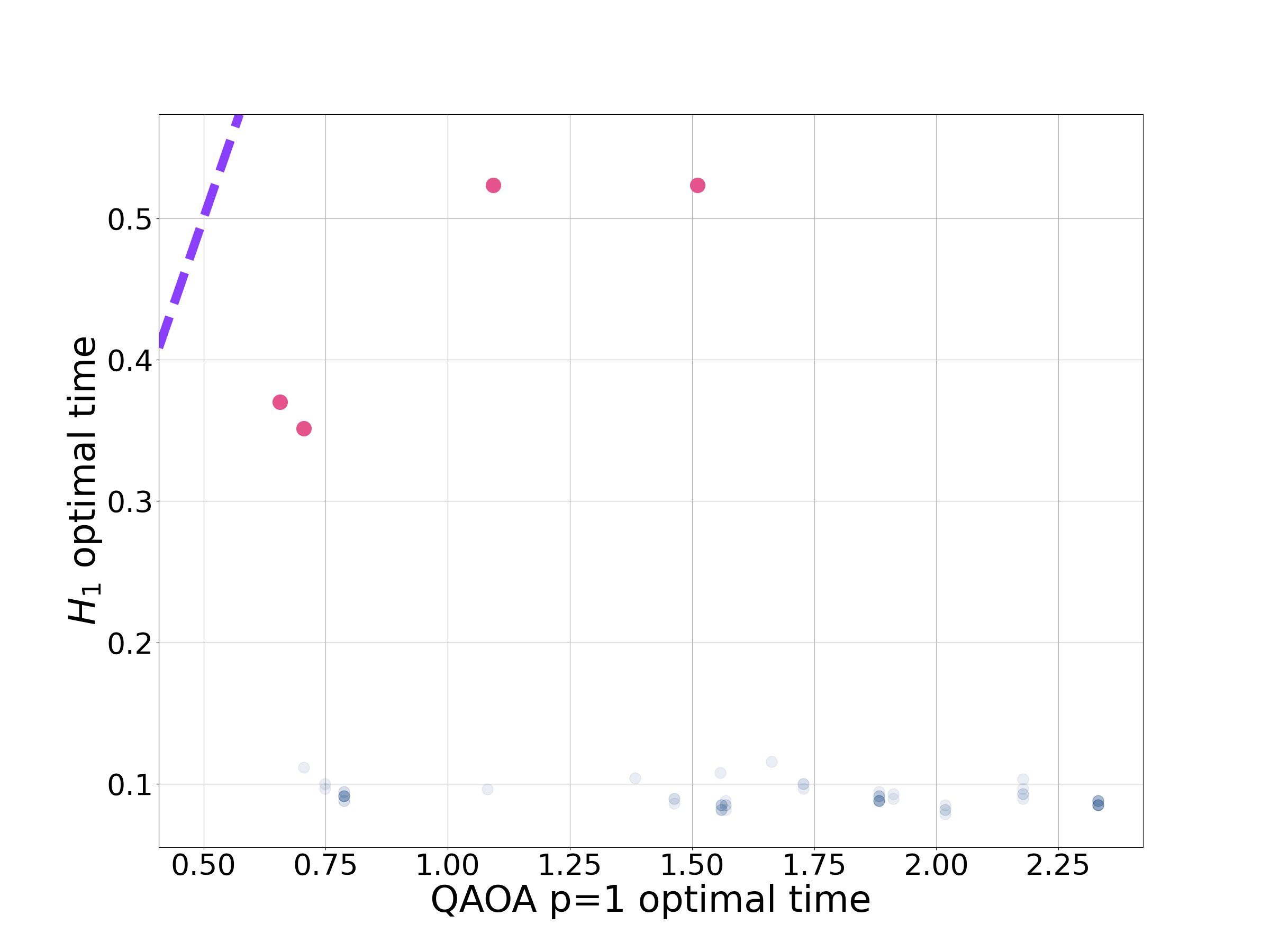}
        \caption{Optimal time.}
    \end{subfigure}
    \caption{The performance of $H_1$ compared to QAOA p=1 on MAX-CUT with three-regular graphs. The highlighted pink dots show the instances with atypical $H_1$ run-times, optimised to give the best possible approximation ratio with shorter run-times than QAOA p=1. The new run-times are shown in the lower plot, with the updated approximation ratio plotted in the upper plot. The dashed purple line shows both approaches performing the same.}
    \label{fig:3reghighlight_cor}
\end{figure}

Looking first at the MAX-CUT instances for three-regular graphs, $H_1$ always provided a better approximation ratio than QAOA p=1, typically in a much shorter time. There are four instances (two with the same approximation ratio) where QAOA has a similar or shorter run-time as highlighted in Fig.\ \ref{fig:3reghighlight}. The corresponding graphs are shown in Tab.\ \ref{tab:3regexcepgraphs}. The first thing to note is that these problems are small, the largest being 8 qubits, despite problem sizes up to 12 qubits being considered. Graph 1 is a complete graph with four nodes and Graph 4 is two copies of this graph. Both Graph 2 and 3 (a cube) consist of a large number of small loops. Due to the relatively high degree of connectivity in these problems, it is likely $H_1$ is no longer operating in the local regime as with the rest of the three-regular problems. In Fig.\  \ref{fig:3reghighlight_cor} we investigate operating $H_1$ suboptimally, optimising only over run-times shorter than QAOA p=1 for these four problems. The result is a negligible decrease in performance. Hence, even for these instances for which $H_1$ has atypical optimal times, it is possible for $H_1$ to provide a better approximation ratio in a shorter time than QAOA p=1.


\begin{figure}
    \centering
    \begin{subfigure}{0.48\textwidth}
        \centering
        \includegraphics[width=\textwidth]{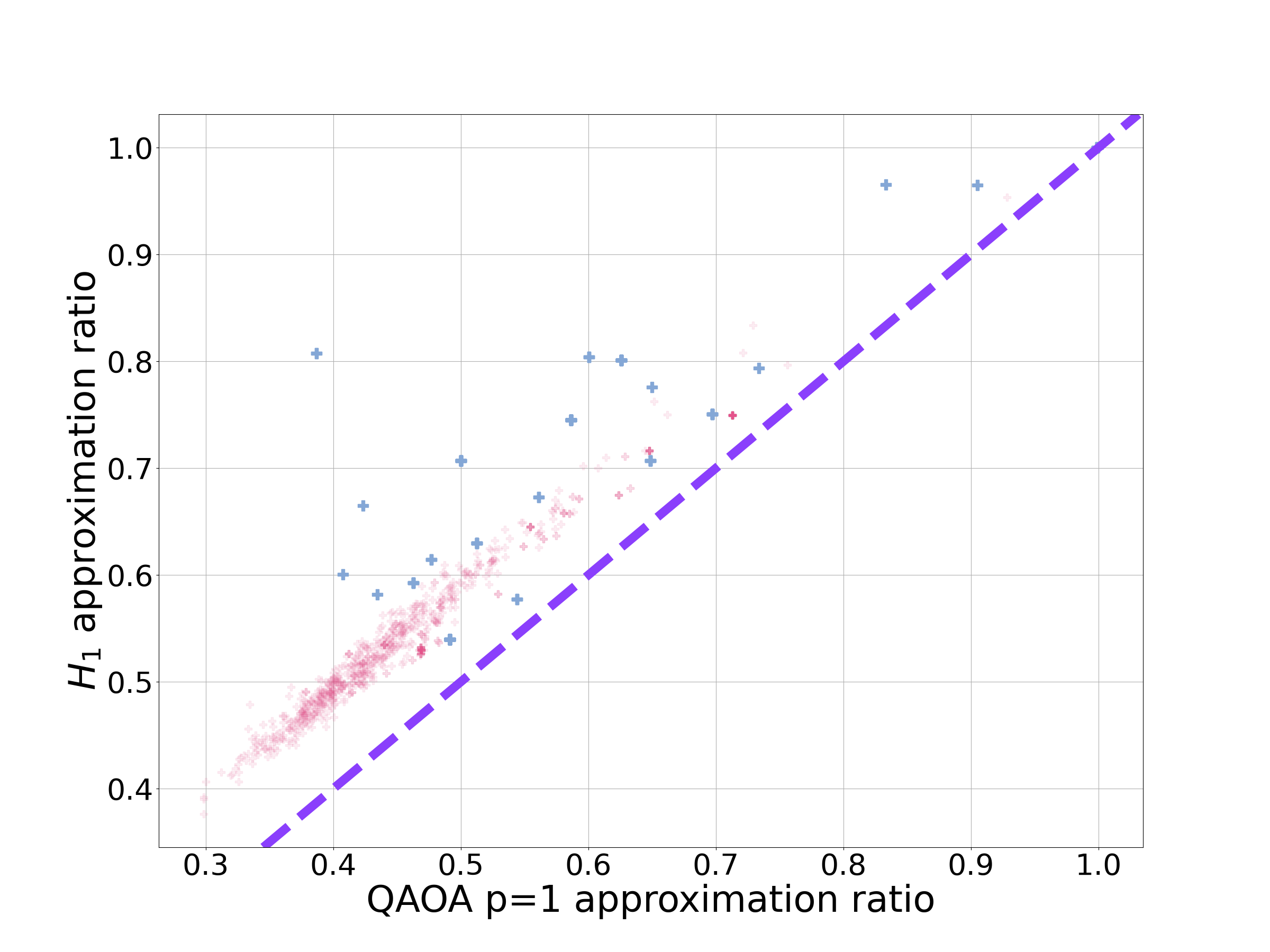}
        \caption{Approximation ratio}
    \end{subfigure}
    \begin{subfigure}{0.48\textwidth}
        \centering
        \includegraphics[width=\textwidth]{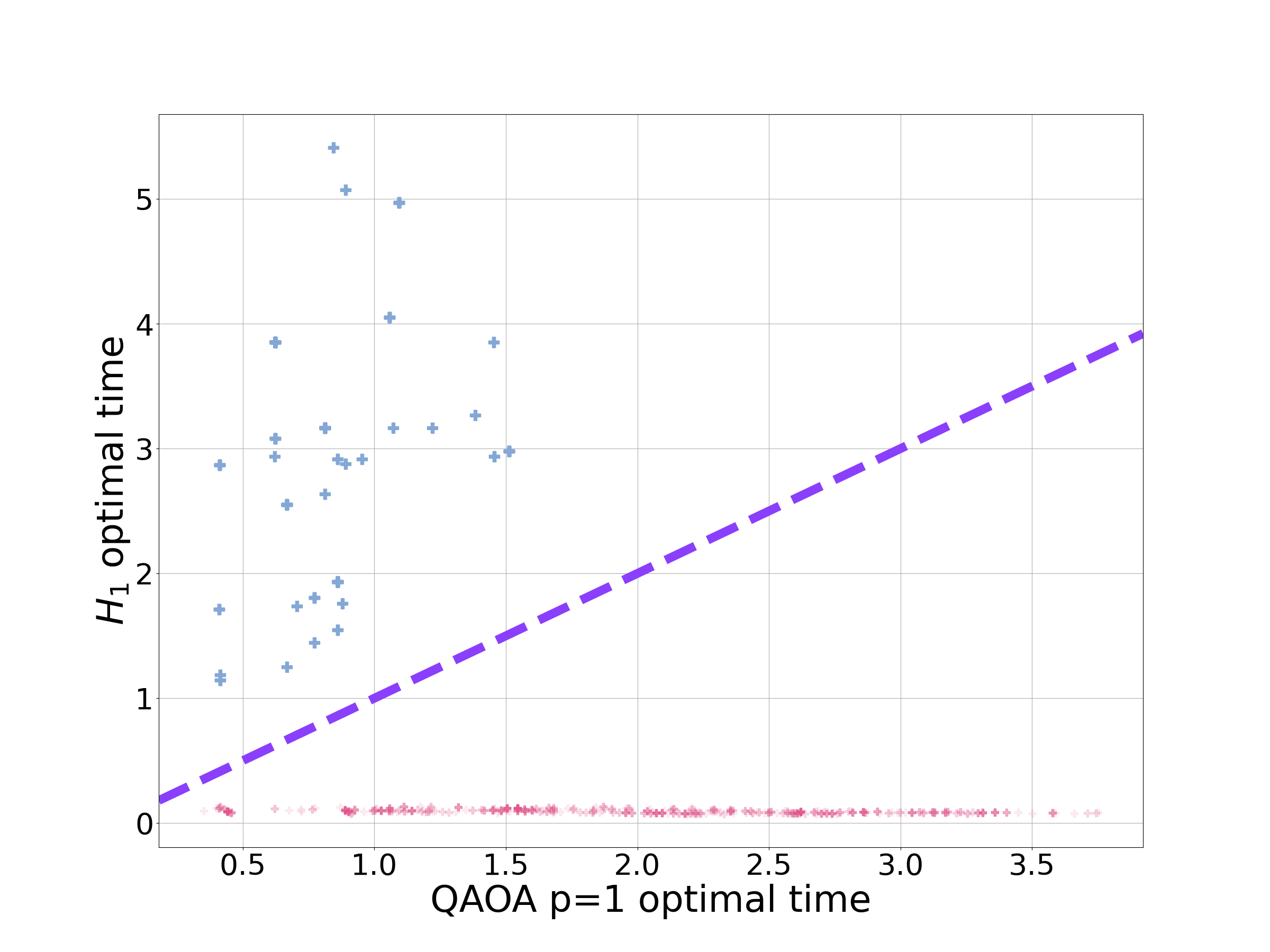}
        \caption{Optimal time.}
    \end{subfigure}
    \caption{The performance of $H_1$ compared to QAOA p=1 on MAX-CUT with randomly generated graphs. The instances with atypical run-times for $H_1$ are highlighted in blue. The rest of the data has been faded for clarity and to show how the remaining data is clustered, with darker regions corresponding to more data points. The dashed purple line shows both approaches performing the same.}
    \label{fig:randhighlight}
\end{figure}

\begin{figure}
    \centering
    \begin{subfigure}{0.48\textwidth}
        \centering
        \includegraphics[width=\textwidth]{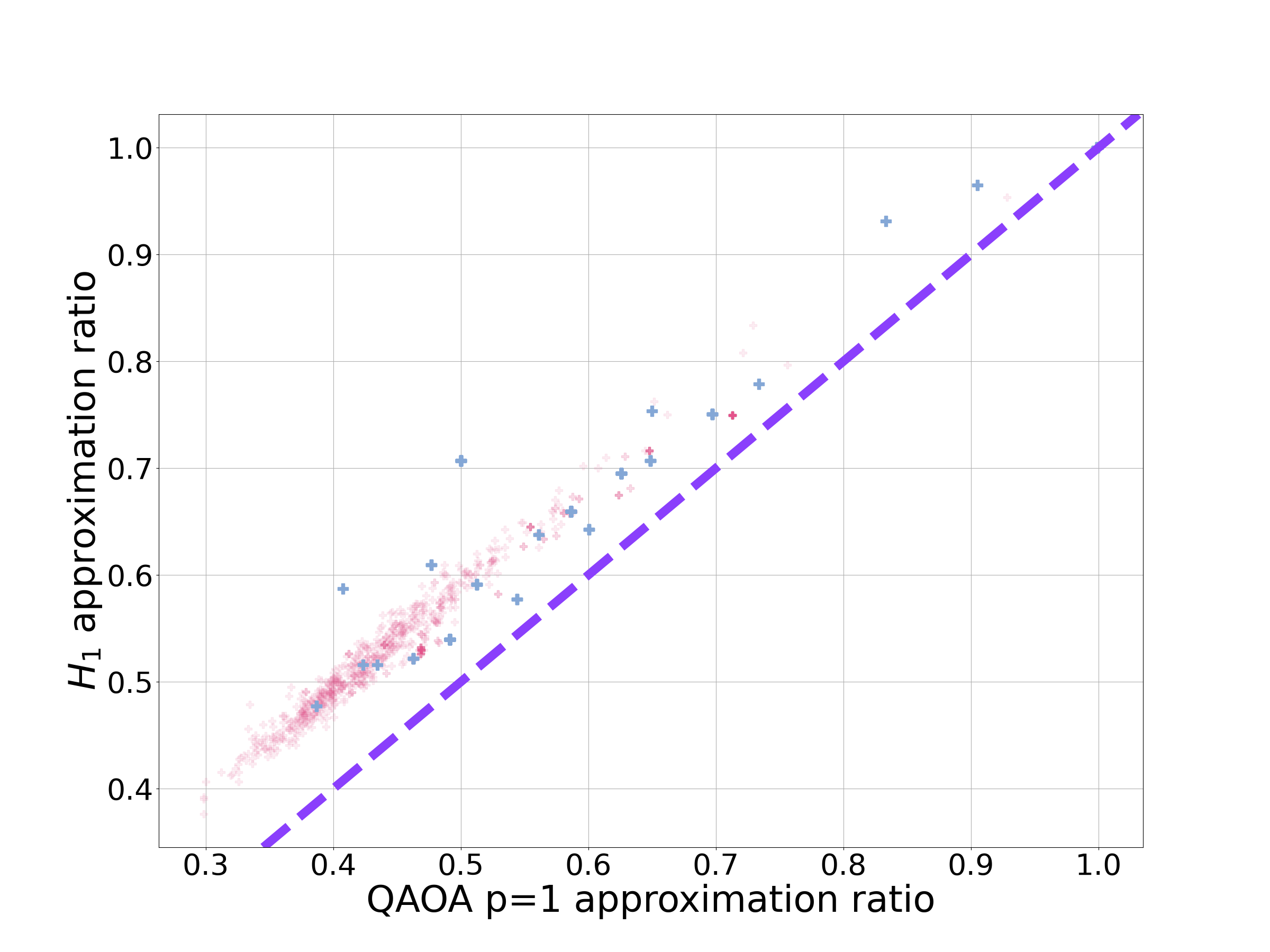}
        \caption{Approximation ratio}
    \end{subfigure}
    \begin{subfigure}{0.48\textwidth}
        \centering
        \includegraphics[width=\textwidth]{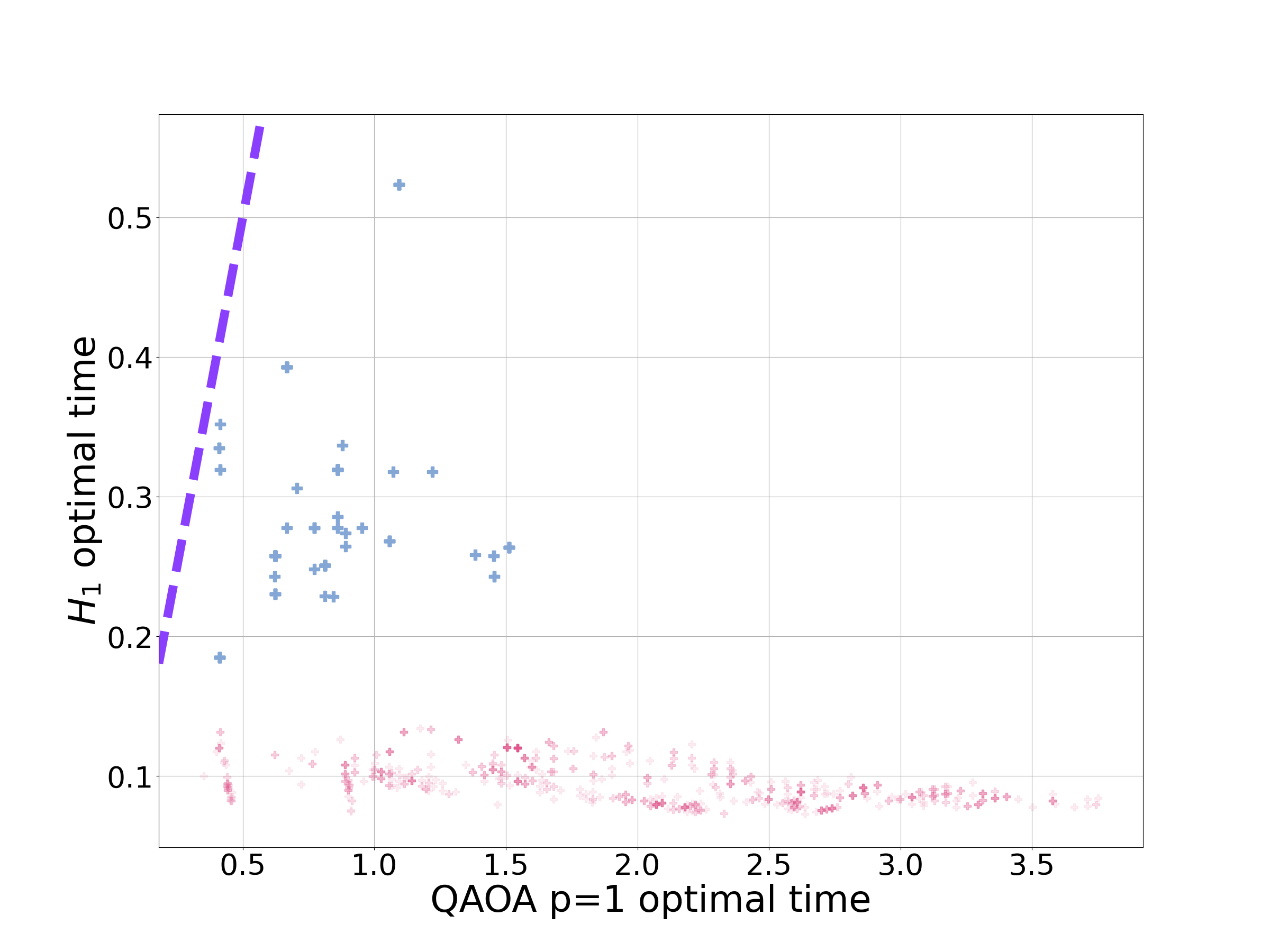}
        \caption{Optimal time.}
    \end{subfigure}
    \caption{The performance of $H_1$ compared to QAOA p=1 on MAX-CUT with randomly generated graphs. The instances with atypical run-times for $H_1$ are highlighted in blue, these have been optimised to give the best possible approximation ratio with a corresponding run-time smaller than the QAOA p=1 optimal time. The new run-times are shown in the lower plot, with the updated approximation ratio plotted in the upper plot. The rest of the data has been faded for clarity. The dashed purple line shows both approaches performing the same.}
    \label{fig:randhighlight_cor}
\end{figure}

As with the three-regular graphs, $H_1$ always gave a better approximation ratio than QAOA p=1 on MAX-CUT with the randomly generated graphs (highlighted in Fig.\ \ref{fig:randhighlight}). Similarly, we can look at the instances with atypical optimal times for $H_1$. The story is similar to before, with only 116 out of 900 instances having run-times longer than QAOA. None of these problem instances consisted of more than 7 qubits (despite simulations going up to 12 qubits). Again we conclude that these atypical optimal times are likely a small problem-setting phenomenon. We can also look for run-times of $H_1$ that provide a better approximation ratio in a shorter time than QAOA p=1 for these problems. The results are shown in Fig.\ \ref{fig:randhighlight_cor} which shows the performance and new run-times of $H_1$ compared to QAOA p=1. For all problem instances it is possible to operate $H_1$ with a shorter run-time than QAOA p=1 and provide a better approximation ratio. Unlike the previous discussion with three-regular graphs, the change in approximation ratio with these new run-times is not negligible for some of the problem instances.

\begin{figure}
    \begin{subfigure}{0.48\textwidth}
        \centering
        \includegraphics[width=\textwidth]{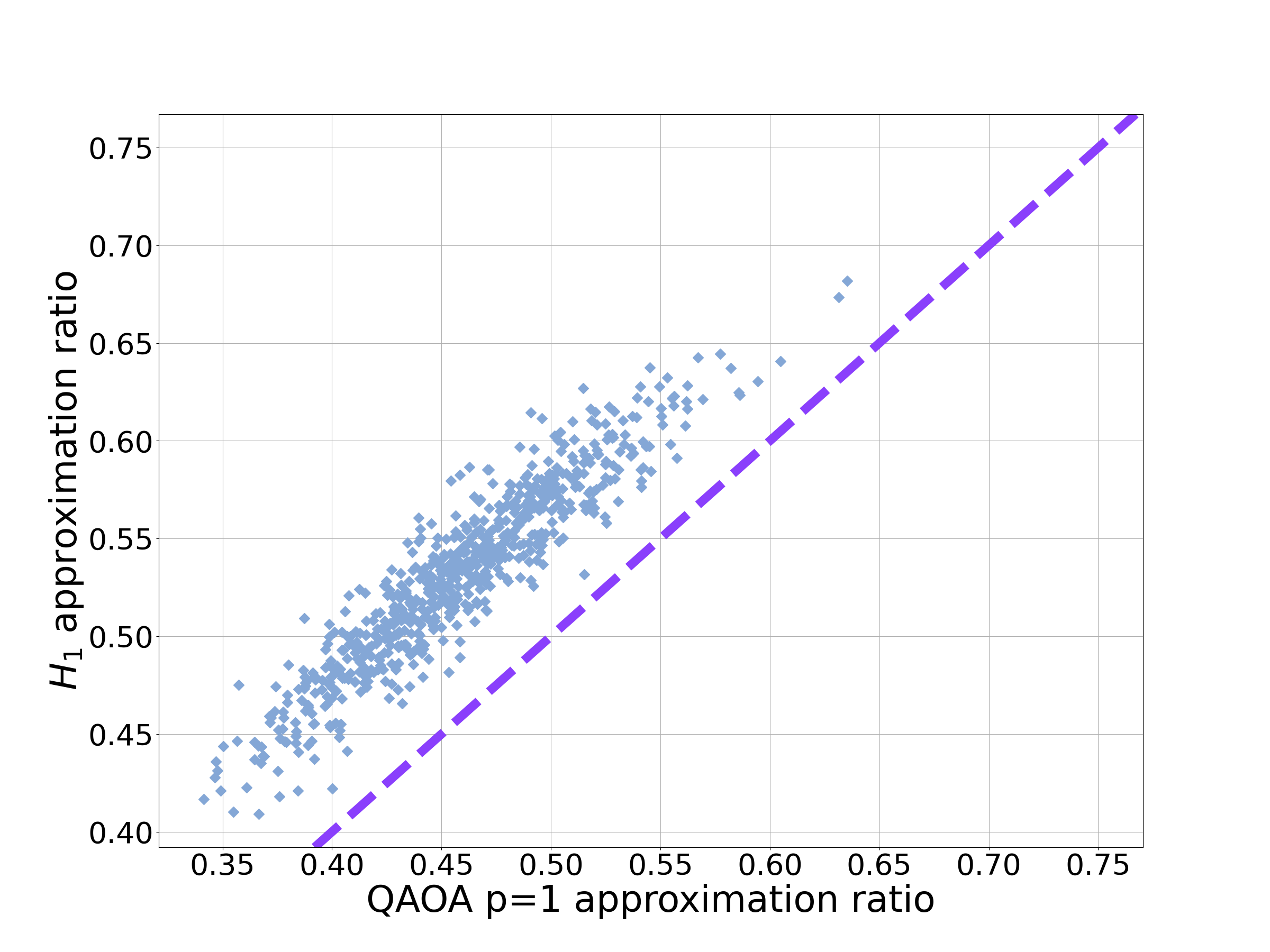}
        \caption{Approximation ratio}
    \end{subfigure}
    \begin{subfigure}{0.48\textwidth}
        \centering
        \includegraphics[width=\textwidth]{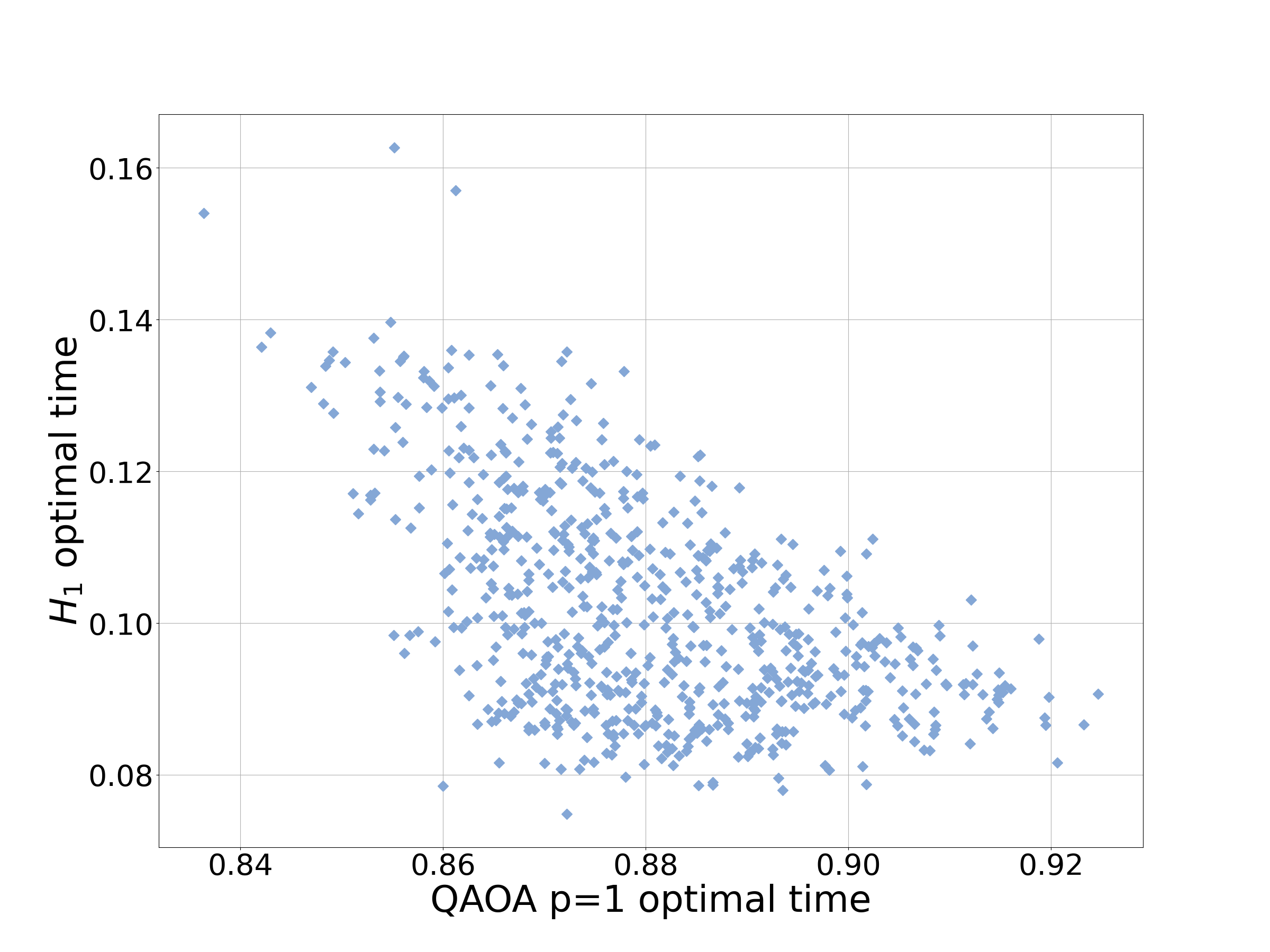}
        \caption{Optimal time.}
    \end{subfigure}
    \caption{The performance of $H_1$ compared to QAOA p=1 on the SKM. Here we neglect smaller problem instances, plotting problem sizes between 6 and 12 qubits.}
    \label{fig:SKMhighlight}
\end{figure}

Finally, we turn to the SKM instances where QAOA p=1 was able to provide a better approximation ratio than $H_1$ for 7 problem instances (out of 900). Five of these instances are 4-qubit problems, the remaining two are 5-qubit problems. 
In Fig.\ \ref{fig:SKMhighlight} we have plotted the SKM data, including only those instances with problem sizes between 6 and 12 qubits. As we can see by ignoring small problem sizes from the data set, the behaviour is  more predictable, with both the approximation ratio and the optimal time more clustered, largely independent of problem size. 

In summary, we have demonstrated numerically that QAOA p=1 might have some advantages over $H_1$ on small, highly connected problems. In general these problems are unlikely to be of any practical interest. Indeed for the MAX-CUT instances it was possible to operate $H_1$ suboptimally so that it still outperformed QAOA p=1 with a shorter run-time. \hfill\\ \vspace{6cm}

\section{Details for using knowledge of the initial state}
\label{app:expHam}
In this appendix we provide the details for $H_{\psi_i}=-i\left[\ket{\psi_i}\bra{\psi_i},f(H_f)\right]$. First rewriting $H_{\psi_i}$ in terms of the eigenstates of $H_f$, $\ket{E_k}$s with associated eigenenergies $E_k$:

\begin{align*}
    H_{\psi_i}&=-i\left[\ket{\psi_i}\bra{\psi_i},f(H_f)\right]\\
    &=-i\left[\ket{\psi_i}\bra{\psi_i},\sum_k f(E_k)\ket{E_k}\bra{E_k}\right]\\
    &=-i\left(\ket{\psi_i}\sum_k f(E_k)\bra{\psi_i}\ket{E_k}\bra{E_k} -\sum_k f(E_k)\bra{E_k}\ket{\psi_i}\ket{E_k}\bra{\psi_i}\right).\\
\end{align*}

Let
\begin{equation}
    \ket{\omega}=\frac{\sum_k f(E_k)\bra{E_k}\ket{\psi_i}\ket{E_k}}{\sqrt{\sum_k f^2(E_k) \abs{\bra{E_k}\ket{\psi_i}}^2}}.
\end{equation}

Apply the Gram-Schmidt procedure to generate an orthonormal basis $\{\ket{\psi_i},\ket{\omega^\perp}\}$, spanning the same space as $\{\ket{\psi_i},\ket{\omega}\}$.
\begin{equation*}
    \ket{\omega^\perp}\propto \ket{\omega}-\bra{\psi_i}\ket{\omega}\ket{\psi_i}
\end{equation*}
Rearranging for $\ket{\omega}$ gives:
\begin{equation}
    \ket{\omega}=\sqrt{1-\abs{\bra{\psi_i}\ket{\omega}}^2}\ket{\omega^\perp} +\bra{\psi_i}\ket{\omega}\ket{\psi_i},
\end{equation}
substituting this into $H_{\psi_i}$ gives:
\begin{equation}
    H_{\psi_i}=-i\beta\left\{ \ket{\psi_i}\left(\bra{\omega^\perp}+\bra{\omega}\ket{\psi_i}\bra{\psi_i}\right) -\left(\ket{\omega^\perp}+\bra{\psi_i}\ket{\omega}\ket{\psi_i}\right)\bra{\psi_i}\right\},
\end{equation}
where
\begin{equation}
    \beta=\sqrt{\left(1-\abs{\bra{\psi_i}\ket{\omega}}^2\right) \left(\sum_k f^2(E_k)\abs{\bra{E_k}\ket{\psi_i}}^2\right)}.
\end{equation}

Using the freedom in the global-phase of the wave-function, choose $\bra{\omega}\ket{\psi_i}$ to be real, giving:
\begin{equation}
    \label{eq:ham2d}
    H_{\psi_i}=-i\beta\left(\ket{\psi_i}{\bra{\omega^{\perp}}}-\ket{\omega^\perp}\bra{\psi_i}\right).
\end{equation}

From Eq.\ \ref{eq:ham2d} it is clear that $H_{\psi_i}$ evolves $\ket{\psi_i}$ to linear superpositions of $\ket{\psi_i}$ and $\ket{\omega^\perp}$. Again by using the Anandan-Aharonov relation, we can determine the time to generate $\ket{\omega}$ and $\ket{\omega^\perp}$. Calculating $\delta E$:
\begin{align*}
    \langle H_{\psi_i} \rangle&=-i \beta \bra{\psi_i} \left(\ket{\psi_i}{\bra{\omega^{\perp}}}-\ket{\omega^\perp}\bra{\psi_i}\right) \ket{\psi_i}\\
    &=0\\
    \langle H_{\psi_i}^2 \rangle &=\beta^2 \bra{\psi_i} G^2 \ket{\psi_i}\\
    &=\beta^2\\
    \delta E &=\sqrt{\langle H_{lp}^2\rangle-\langle H_{lp} \rangle^2}\\
    &=\beta.
\end{align*}

The distance between $\ket{\psi_i}$ and $\ket{\omega^\perp}$ is 
\begin{align*}
    \theta_{\omega^\perp}&=2\arccos{\bra{\psi_i}\ket{\omega^\perp}}\\
    &=\pi.
\end{align*}

The distance between $\ket{\psi_i}$ and $\ket{\omega^\perp}$ is
\begin{align*}
    \theta_{\omega}&=2\arccos{\abs{\bra{\omega}\ket{\psi_i}}}.
\end{align*}

The time then to evolve $\ket{\psi_i}$ to $\ket{\omega^\perp}$ is $t_{\omega^\perp}=\pi/2 \beta$ and the time to evolve to $\ket{\omega}$ is $t_\omega=\arccos{\abs{\bra{\omega}\ket{\psi_i}}}/\beta$.

We can explicitly verify this by exponentiating $H_{\psi_i}$. Focusing on $G=-i\left(\ket{\psi_i}{\bra{\omega^{\perp}}}-\ket{\omega^\perp}\bra{\psi_i}\right)$, then
\begin{align*}
    G^2&=-\left(\ket{\psi_i}{\bra{\omega^{\perp}}}-\ket{\omega^\perp}\bra{\psi_i}\right)\left(\ket{\psi_i}{\bra{\omega^{\perp}}}-\ket{\omega^\perp}\bra{\psi_i}\right)\\
    &=\left(\ket{\psi_i}\bra{\psi_i}+\ket{\omega^\perp}\bra{\omega^\perp}\right).
\end{align*}
This is a projector. Calculating $G^3$ gives:
\begin{align*}
    G^3&=-i\left(\ket{\psi_i}\bra{\psi_i}+\ket{\omega^\perp}\bra{\omega^\perp}\right)\left(\ket{\psi_i}{\bra{\omega^{\perp}}}-\ket{\omega^\perp}\bra{\psi_i}\right)\\
    &=-i\left(\ket{\psi_i}{\bra{\omega^{\perp}}}-\ket{\omega^\perp}\bra{\psi_i}\right)\\
    &=G.
\end{align*}

Let $H$ be a Hamiltonian, where $H^2=P$, $P$ being a projector. Then substituting $H$ into the power-series for the exponential gives:
\begin{align*}
    e^{-i H t}&=\sum_{k=0}^\infty \frac{\left(-i H t\right)^k}{k!}\\
    &=\sum_{\text{even k}}^\infty \frac{\left(-i H t\right)^k}{k!}+\sum_{\text{odd k}}^\infty \frac{\left(-i H t\right)^k}{k!}\\
    &=\sum_{k=0}^\infty \frac{\left(-i H t\right)^{2k}}{\left(2k\right)!}+\sum_{k=0}^\infty \frac{\left(-i H t\right)^{2k+1}}{(2k+1)!}\\
    &=I+\sum_{k=1}^\infty \frac{\left(-1\right)^kt^{2k}}{\left(2k\right)!}P-i\sum_{k=0}^\infty \frac{\left(-1\right)^k t^{2k+1}}{(2k+1)!}PH\\
    &=I+\left(\cos(t)-1\right)P-i \sin(t) PH
\end{align*}
Therefore,
\begin{equation}
    e^{-iH_{lp}t}=I+\left(\cos(\beta t)-1\right)G^2-i\sin \left(\beta t\right) G
\end{equation}
and $\ket{\psi(t)}=e^{-iH_{lp}t}\ket{\psi_i}$,
\begin{equation}
    \ket{\psi(t)}=\cos(\beta t)\ket{\psi_i}+\sin \left(\beta t\right) \ket{\omega^\perp}.
\end{equation}

Checking the times from the Anandan-Aharonov relationship gives:
\begin{align*}
    \ket{\psi(\pi/2\beta)}&=\cos(\pi/2)\ket{\psi_i}+\sin \left(\pi/2\right) \ket{\omega^\perp}\\
    &=\ket{\omega^\perp}
\end{align*}
and
\begin{align*}
    &\ket{\psi(\arccos{\abs{\bra{\psi_i}\ket{\omega}}}/\beta)}\\
    &=\abs{\bra{\psi_i}\ket{\omega}}\ket{\psi_i}+ \sqrt{1-\abs{\bra{\psi_i}\ket{\omega}}^2} \ket{\omega^\perp}\\
    &=\ket{\omega}.
\end{align*}

Up to now we have kept the conversation in this section fairly general. Now we apply $H_{lp}$ to the QA-framework with $\ket{\psi_i}=\ket{+}$ and $\ket{E_k}$s corresponding to computational basis states, so $\bra{+}\ket{E_k}=1/\sqrt{2^n}$. Simplifying $\ket{\omega}$ gives:
\begin{align*}
    \ket{\omega}&=\frac{1}{\sqrt{\sum_k f^2(E_k) \abs{\bra{E_k}\ket{+}}^2}}\sum_k f(E_k) \bra{E_k}\ket{+}\ket{E_k}\\
    &=\frac{1}{\sqrt{\sum_k f^2(E_k)}}\sum_k f(E_k)\ket{E_k}\\
    &=\frac{1}{\sqrt{\Tr{f^2(H_f)}}}\sum_k f(E_k)\ket{E_k}
\end{align*}
and the overlap with the initial state is:
\begin{align*}
    \bra{+}\ket{\omega}&=\frac{\sqrt{2^n}}{\sqrt{\sum_k f^2(E_k)}}\sum_k\frac{1}{\sqrt{2^n}}f(E_k)\bra{+}\ket{E_k}\\
    &=\frac{1}{\sqrt{\sum_k f^2(E_k)}}\sum_k \frac{1}{\sqrt{2^n}} f(E_k)\\
    &=\frac{1}{\sqrt{2^n}}\frac{\Tr{f(H_f)}}{\sqrt{\Tr{f^2(H_f)}}}.
\end{align*}

To summarise, the Hamiltonian $H_{lp}=-i[\ket{+}\bra{+},f(H_f)]$ evolves $\ket{+}$ to 
\begin{equation}
    \ket{\omega}=\frac{1}{\sqrt{\Tr f^2(H_f)}}\sum_k f(E_k)\ket{E_k},
\end{equation}
in a time
\begin{equation}
    T=\frac{\sqrt{2^n}\arccos{\abs{\bra{+}\ket{\omega}}}}{\sqrt{\Tr{f^2(H_f)}\left(1-\abs{\bra{+}\ket{\omega}}^2\right)}},
\end{equation}
where
\begin{equation}
    \bra{+}\ket{\omega}=\frac{1}{\sqrt{2^n}}\frac{\Tr{f(H_f)}}{\sqrt{\Tr{f^2(H_f)}}}.
\end{equation}

\end{document}